\documentclass{jfm}
\usepackage{graphicx}
\usepackage{epstopdf, epsfig}
\usepackage{hyperref}

\shorttitle{Laminar shockwave/boundary-layer interactions for duct flows}
\shortauthor{D.J Lusher and N.D. Sandham}

\title{The effect of flow confinement on laminar shockwave/boundary-layer interactions}

\author{David J. Lusher\aff{1}
  \corresp{\email{d.lusher@soton.ac.uk}}
 \and Neil D. Sandham\aff{1}}

\affiliation{\aff{1}Aerodynamics and Flight Mechanics Group, University of Southampton, Highfield Campus, Southampton SO17 1BJ, United Kingdom}

\begin{document}

\maketitle

\begin{abstract}
Numerical work on shockwave/boundary-layer interactions (SBLIs) to date has largely focused on span-periodic quasi-2D configurations that neglect the influence lateral confinement has on the core flow. The present study is concerned with the effect of flow confinement on Mach 2 laminar SBLI in rectangular ducts. An oblique shock generated by a $2^{\circ}$ wedge forms a conical swept SBLI with sidewall boundary layers before reflecting from the bottom wall of the domain. Multiple large regions of flow-reversal are observed on the sidewalls, bottom wall and at the corner intersection. The main interaction is found to be strongly three-dimensional and highly dependent on the geometry of the duct. Comparison to quasi-2D span-periodic simulations showed sidewalls strengthen the interaction by 31\% for the baseline configuration with an aspect ratio of one. The length of the shock generator and subsequent trailing edge expansion fan position was shown to be a critical parameter in determining the central separation length. By shortening the length of the shock generator, control of the interaction and suppression of the central interaction is demonstrated. Parametric studies of shock strength and duct aspect ratio were performed to find limiting behaviours. For the largest aspect ratio of four, three-dimensionality was visible across 30\% of the span width away from the wall. Topological features of the three-dimensional separation are identified and shown to be consistent with `owl-like' separations of the first kind. The reflection of the initial conical swept SBLIs is found to be the most significant factor determining the flow structures downstream of the main interaction.
\end{abstract}



\section{Introduction}

\subsection{Shockwave/boundary-layer interactions}
Shockwave/boundary-layer interactions (SBLI) play an important role in the study of high-speed compressible gas dynamics. The ubiquity of SBLIs in aeronautical flows of practical interest is well established (\cite{Dolling2001}, \cite{Gaitonde2015}) posing considerable challenges for high-speed aircraft design. SBLIs can occur in both internal and external flow configurations, comprised of a complex coupling between inviscid and viscous effects. An incident shock in the internal case can interact with multiple surfaces and result in a complex and dynamic shock system. Flow separation and unsteadiness is a major concern in applications such as supersonic engine intakes, where non-uniform flow entering the compressor can lead to variable heat transfer rates and pressure losses. Detrimental effects include reduced engine efficiency and an increase in the structural fatigue of components. In more severe cases, SBLIs can lead to a full unstart of the engine. The adverse pressure gradient applied by an impinging shock causes a thickening of the target boundary layer, and for sufficiently strong shocks, a separation of the flow will occur. For a given strength of incident shock, the susceptibility of the boundary layer to separate is largely dependent on the upstream state of the boundary layer \citep{babinsky_harvey_2011}. Turbulent boundary layers are most capable of resisting flow separation: the higher mixing rates effectively energise the boundary layer and stave off stagnation by transferring low-momentum fluid away from the wall. Laminar boundary layers separate far more easily than their turbulent counterparts, with flow separation observed for shocks weaker than required for incipient separation of a turbulent boundary layer. 

Current concerns over the environmental impact of aircraft has contributed to a renewed interest in laminar aerodynamics, taking advantage of the lower skin-friction drag of laminar boundary layers. The present work focuses on numerical simulation of laminar SBLI for internally confined rectangular duct flows, which are applicable to supersonic engine intakes. An initial oblique shockwave interacts with boundary layers on both the sidewalls and bottom wall of the duct, resulting in multiple regions of three-dimensional reverse flow. While many real-world applications will be fully turbulent, laminar solutions provide useful comparisons to wind tunnel experiments where small-scale models are investigated at lower Reynolds numbers. Furthermore, shock and expansion wave patterns are easier to distinguish in the absence of turbulence and the mechanism of transition can be investigated in laminar SBLI. Laminar flows can be used as a basis for stability analysis to gain insight into the mechanism of transition to turbulence. As noted by the laminar-transitional SBLI work of \cite{Giepman2016}, experimental techniques such as particle image velocimetry (PIV) can suffer from seeding issues with laminar boundary layers. Numerical simulations are well placed to complement the existing experimental literature, offering additional insight into the complex flow features of laterally confined SBLI. The next section gives an overview of previous studies on laminar SBLI and laterally confined SBLI in ducts.

\subsection{Previous studies}
\subsubsection{Laminar and transitional shockwave/boundary-layer interactions}

Two-dimensional laminar SBLI is a largely well understood phenomena and has historically been treated by a range of both theoretical and numerical approaches \citep{Adamson1980}. An important numerical study on laminar oblique-SBLI was carried out by \cite{Katzer1989}, based on the earlier experiments of \cite{hakkinen1959}. Laminar SBLI were simulated over a flat plate for a range of Mach numbers from 1.4 to 3.4, with the results agreeing well with predictions from free interaction theory. The length of the separation bubble was found to be linearly dependent on the incident shock strength. A combined numerical/experimental study on two-dimensional laminar SBLI was performed by \cite{Degrez1987} at Mach 2.15. It was reported that experimental configurations with an aspect ratio greater than 2.5 were required to achieve two-dimensional behaviour of the SBLI. More recent work has focused on instability and the transition mechanisms, often motivated by the widely reported low-frequency unsteadiness present in turbulent SBLI \citep{Clemens2014}. A numerical study using the same conditions as \cite{Katzer1989} at Mach 2 was carried out by \cite{Sivasubramanian2015} with and without upstream disturbances. The disturbances were found to be strongly amplified by the laminar separation bubble and at higher shock strengths the flow transitioned to turbulence downstream of the bubble. Both high and low frequency unsteadiness was observed. The high frequency component was attributed to vortical shedding  at the reattachment point during the breakdown.

A central theme of SBLI research has been whether the unsteadiness in turbulent SBLI is caused by structures in the upstream boundary layer or due to a downstream influence intrinsic to the system. \cite{Sansica2016} simulated a Mach 1.5 laminar SBLI forced with a pair of unstable oblique modes at the inlet. The introduction of unstable modes led to a transition to turbulence downstream of the reattachment point and an associated low-frequency unsteadiness. The study demonstrated a low-frequency response of shock induced separation even in the absence of upstream turbulence. In the hypersonic regime \cite{Dwivedi2017} performed direct numerical simulation (DNS) of a Mach 5.92 laminar SBLI. Above a critical shock angle the flow became three-dimensional and unsteady, with the downstream region being found to support significant growth of perturbations starting at the reattachment point. Further work with the same flow conditions \citep{Hildebrand2018_1,Hildebrand2018_2} studied transient growth of disturbances and the instability mechanism within the laminar recirculation bubble itself. Above a critical shock angle a self-sustaining process was identified using global stability analysis. The instability was attributed to streamwise vortices created within the recirculation bubble that redistribute momentum normal to the wall and develop into elongated streaks downstream of reattachment. 

Recent experimental studies on laminar-transitional SBLI include those of \cite{Giepman2015} and \cite{Giepman2018}, in which a range of shock impingement locations were investigated for Mach numbers between $M=\left[1.6, 2.3\right]$. All experiments were performed with high-resolution PIV in a wind tunnel with a partial-span shock generator. For the laminar impingement locations long, triangular recirculation bubbles were observed, with a linear dependence of shock strength on the distance between the separation point and the top of the bubble. The largest separation bubbles were recorded for the purely laminar interactions, while a significant shortening of the separation length was observed when the boundary layer was in a transitional state. The dependence on the upstream boundary layer state has led to studies on optimal tripping methods to obtain a transitional state close to the SBLI. The experiments of \cite{Giepman2016} and the complimentary numerical study by \cite{Quadros2018} investigated tripped transition of laminar SBLI at $M=1.7$. Both cases confirmed that for a given shock strength the size of the recirculation bubble was highly dependent on the incoming boundary layer state. The experimental work showed that the recirculation region could be removed entirely by placing a trip close to the interaction. Although this grants control of the separation, the trade-off is a substantially thicker boundary layer and increased skin-friction drag downstream of the interaction.

\subsubsection{Confinement effects for shockwave/boundary-layer interactions}
Despite the progress in understanding SBLI, the infinite-span (quasi-2D) assumption persists in much of the numerical literature as a way of reducing computational complexity. For internally bounded flows this is not a valid assumption as lateral confinement leads to multiple boundary layers for the shock to interact with. The modified interaction may be highly three-dimensional and strongly influenced by the geometry of the duct. Numerical studies of confined turbulent SBLI include \cite{Garnier2009}, \cite{Bermejo-Moreno2014} and \cite{wang_sandham_hu_liu_2015}. In each case the presence of sidewalls resulted in strong three-dimensionality and a significant strengthening of the central interaction. The wall-modelled large-eddy simulations (LES) of \cite{Bermejo-Moreno2014} studied turbulent SBLI with comparison to experimental PIV data for rectangular ducts with a $20^{\circ}$ flow deflection. It was observed that the structure and location of the internal shock system was heavily modified compared to span-periodic simulations. Furthermore Mach stems were observed at the primary interaction for the case strengthened by sidewalls, a feature not present in the span-periodic simulations. \cite{wang_sandham_hu_liu_2015} performed LES at Mach 2.7 with a flow deflection of $9^{\circ}$. An upstream shift of the separation and reattachment points was observed as the aspect ratio was decreased from four to one. The same reduction in aspect ratio led to a 30\% increase in centreline separation length compared to quasi-2D predictions. Three-dimensional flow features near the main interaction included corner compression waves, secondary sidewall shocks and strong attached transverse flow between the central and corner separations. The main factors responsible for the modified interaction were the swept sidewall SBLI and aspect ratio.

Considerable attention has been given to three-dimensional corner effects experimentally in recent years owing to their prevalence in supersonic intake applications. Duct SBLI for normal shocks have been investigated by \cite{Bruce2011} and \cite{Burton2012} among others. Oblique duct SBLI studies include \cite{Eagle2011}, \cite{Eagle2014} and \cite{Morajkar2016}. An open question is to determine the importance of corner separations in relation to the main interaction and how modifications to the corner flow results in divergence from quasi-2D predictions. Much of the work has focused on identifying compression waves generated by the flow deflection in the corner. Oil-streak images and pressure-sensitive paint have been used to infer the impact of corner compressions and their ability to modify other parts of the flow. \cite{Xiang2019} is a recent example of work in this area at Mach 2.5, adding corner blockages to shrink the duct cross section and obtain exaggerated corner separations. It was observed that the central separation was sensitive to variations in the onset and magnitude of the corner separation. A mechanism was proposed to predict the central separation based on the crossing point of the inferred corner compression waves near the bottom wall. For increased corner separations the topology of the central interaction was seen to transition between the `owl-like' first and second states introduced by \cite{1984ZFl}. The transition to the secondary owl-like topology is indicative of increased three-dimensionality of the separated region. It was argued that corner compression waves crossing on the centreline before the interaction region led to reduced separation, while a crossing point within the interaction resulted in larger separations.

Differences also exist between experimental configurations, one notable feature being the effect of sidewall gaps for partial-span shock generators. \cite{Grossman2017} investigated the effect of duct geometry and the sidewall gap on a Mach 2 SBLI with a $12^{\circ}$ flow deflection. The central separation bubble length was sensitive to the size of the sidewall gap, with reduced three-dimensionality and smaller separations seen for larger gaps. Furthermore, the impingement location of the trailing edge expansion fan was observed to be a critical parameter when determining the size of the central separation. Shifting the expansion fan downstream led to an increase in both the strength and streamwise extent of the separation. A follow up study \citep{Grossman2018} expanded on these themes in the context of regular-irregular transition of SBLI, where for a fixed initial flow deflection Mach reflections were observed for certain aspect ratios. The streamwise separation length was found to be linearly dependent on the distance between the main SBLI and the impingement point of the trailing expansion fan. The increase in separation length was shown to be linked primarily to an upstream shift of the separation line.

\subsection{Aims and outline of the paper}
The aim of this work is to investigate the effect of confinement on laminar SBLI in rectangular ducts. The paper is organised as follows: Section \ref{sec:numerical_method} outlines the governing equations and numerical methods to be applied. Section \ref{sec:problem_specification} specifies the physical problem and computational domain. In section \ref{sec:laminar_grid_study} a grid refinement study is performed to demonstrate grid independence. Section \ref{sec:ramp_length} examines the effect that the shock generator length has on the central separation size for both two and three-dimensional flows. Section \ref{sec:duct_flow} discusses the baseline configuration, highlighting the main flow features and making comparisons to quasi-2D predictions. The topology of the laminar SBLI is shown in section \ref{sec:laminar_topology}, analysing both the global shock structures and critical points found in near-wall streamlines. Qualitative comparisons are made to previous turbulent studies to assess whether similarities can be drawn to flow structures in the laminar case. Parametric effects of duct aspect ratio and incident shock strength are given in sections \ref{sec:aspect_ratio} and \ref{sec:shock_strength} respectively. Section \ref{sec:long_domain_control} uses the trailing expansion fan effect of section \ref{sec:ramp_length} to demonstrate its use as a flow control method.

\section{Numerical method}\label{sec:numerical_method}
\subsection{Governing equations}
The governing equations for all simulations in this work are the dimensionless compressible Navier-Stokes equations for a Newtonian fluid. Applying conservation of mass, momentum and energy in three spatial directions $x_i$ $\left(i=1, 2, 3\right)$ results in a system of five partial differential equations given by 
\begin{equation}\label{ns_eqn1}
\frac{\partial \rho}{\partial t} + \frac{\partial}{\partial x_k} \left(\rho u_k \right) = 0,
\end{equation}
\begin{equation}\label{ns_eqn2}
\frac{\partial}{\partial t}\left(\rho u_i\right) + \frac{\partial}{\partial x_k} \left(\rho u_i u_k + p \delta_{ik} - \tau_{ik}\right) = 0,
\end{equation}
\begin{equation}\label{ns_eqn3}
\frac{\partial}{\partial t}\left(\rho E\right) + \frac{\partial}{\partial x_k} \left(\rho u_k \left(E + \frac{p}{\rho}\right) + q_k - u_i \tau_{ik}\right) = 0,
\end{equation}
with Fourier's heat flux $q_k$ and viscous stress tensor $\tau_{ij}$ defined as 
\begin{equation}\label{heat_flux}
q_k = \frac{-\mu}{\left(\gamma - 1\right) M_{\infty}^{2} Pr Re}\frac{\partial T}{\partial x_k},
\end{equation}
\begin{equation}\label{stress_tensor}
\tau_{ik} = \frac{\mu}{Re} \left(\frac{\partial u_i}{\partial x_k} + \frac{\partial u_k}{\partial x_i} - \frac{2}{3}\frac{\partial u_j}{\partial x_j} \delta_{ik}\right).
\end{equation}
Throughout this work the coordinates $x_i$ $\left(i=1, 2, 3\right)$ are referred to as $\left(x,y,z\right)$ for the streamwise, bottom wall-normal and spanwise directions respectively, with corresponding velocity components $\left(u, v, w\right)$. The equations are non-dimensionalized by freestream velocity, density and temperature $\left(U^{*}_{\infty}, \rho^{*}_{\infty}, T^{*}_{\infty}\right)$, with a characteristic length based on the displacement thickness $\delta^{*}$ of the boundary layer imposed at the inlet. Further details of the boundary layer initialization are given in section \ref{sec:inlet_profile}. Freestream Mach number, Prandtl number and ratio of specific heat capacity for air are taken to be $M_{\infty} = 2$, $Pr = 0.72$ and $\gamma = 1.4$ respectively. Reynolds number based on the inlet displacement thickness is set as $Re_{\delta^{*}} = 750$ throughout. The dynamic viscosity $\mu\left(T\right)$ is computed by Sutherland's law 
\begin{equation}\label{sutherland}
\mu\left(T\right) = T^{\frac{3}{2}} \left(\frac{1+\frac{T_{s}}{T_{\infty}} } {T + \frac{T_{s}}{T_{\infty}}}\right),
\end{equation}
with reference and Sutherland temperatures taken to be $T_{\infty} = 288.0 \textrm{K}$ and $T_s = 110.4 \textrm{K}$. For an ideal Newtonian fluid pressure can be calculated through the equation of state such that
\begin{equation}\label{pressure_eqn}
p = \left(\gamma - 1\right) \left(\rho E - \frac{1}{2} \rho u_i u_i\right) = \frac{1}{\gamma M^{2}_{\infty}} \rho T.
\end{equation}
Throughout this work wall-normal skin friction $C_f$ is calculated as 
\begin{equation}\label{skin_friction_eqn}
C_f = \frac{\tau_w}{\frac{1}{2} \rho_{\infty} U^{2}_{\infty}},
\end{equation}
for a wall shear stress
\begin{equation}\label{cf_definition}
\tau_w = \mu {\frac{\partial u}{\partial y}\vline}_{y=0} \quad \textrm{or} \quad \tau_w = \mu {\frac{\partial u}{\partial z}\vline}_{z=0, L_z}
\end{equation}
depending on whether the quantity is being evaluated on the bottom wall $(y=0)$ or sidewalls $(z=0, L_z)$ of the domain.

\subsection{Discretisation schemes}
The high-order finite-difference code OpenSBLI \citep{JACOBS201712}, \citep{LUSHER201817} is used to perform the simulations, which uses the stencil-based Oxford Parallel Structured software (OPS) embedded Domain-specific language (eDSL) \citep{Reguly:2014:ODS:2691166.2691173} for parallelisation. Validation of the OpenSBLI code for laminar shockwave/boundary-layer interactions was shown in \cite{LUSHER201817} for a 2D version of the present case. Spatial discretisation is performed by a $5\textrm{th}$ order Weighted Essentially Non-Oscillatory (WENO) scheme, specifically the improved WENO-Z scheme introduced by \citet{Borges2008}. WENO schemes are a robust and well established method for numerical shock capturing, \cite{Gross2016} is an example of WENO being applied to a wide range of laminar-transitional SBLI with comparison to experiments. The WENO reconstruction is performed in characteristic space to minimise oscillations and uses the local Lax-Friedrich flux-splitting method. Viscous, heat flux and metric terms are computed by standard $4\textrm{th}$ order central differencing, replaced at domain boundaries by the $4\textrm{th}$ order boundary scheme of \cite{Carpenter1998}. To minimise memory usage a low-storage explicit $3\textrm{rd}$ order Runge-Kutta scheme is used for time advancement, in the form provided by \cite{carpenter_kennedy_1994}.

\section{Problem specification and computational domain}\label{sec:problem_specification}
\subsection{Domain specification and physical parameters}\label{sec:inlet_profile}
\begin{figure}
\begin{centering}
  \includegraphics[width=0.75\textwidth]{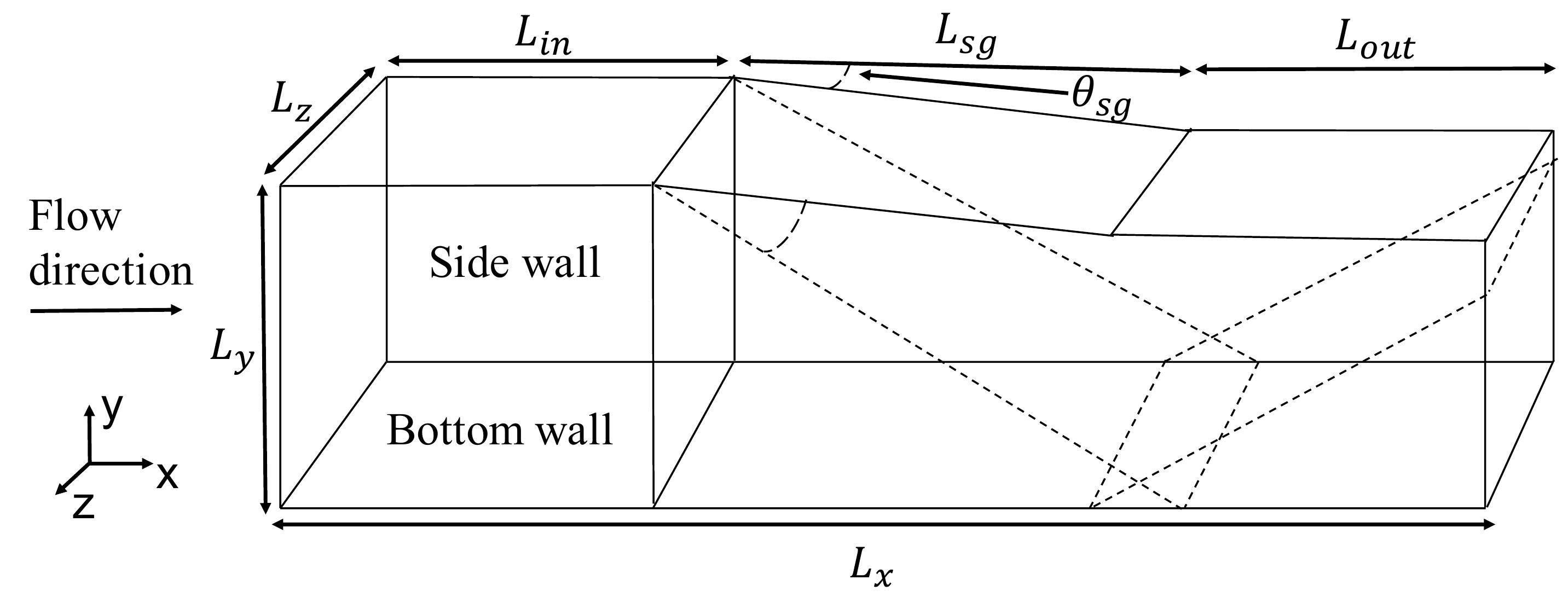}
  \caption{Schematic of the computational domain. An oblique shockwave is generated by deflecting the oncoming flow with a ramp angled at $\theta_{sg}$ to the freestream. No-slip isothermal wall conditions are enforced on the bottom wall, both sidewalls and on the upper surface between $L_{sg}$ and $L_{out}$.}\label{fig:Fig1}
  \end{centering}
\end{figure}
For span-periodic simulations of SBLI the standard method of generating an incident shock is to apply the inviscid Rankine–Hugoniot jump conditions on the upper or inlet boundary of the domain. For confined duct flows this is not valid as it creates a non-physical interface between the sidewall boundary layers and the shock jump conditions on the upper surface. In this work the oblique shock is generated by deflecting the flow with a no-slip ramp as shown in figure \ref{fig:Fig1}. Duct dimensions, aspect ratio and the length of the shock generating ramp are the primary considerations when selecting a computational domain. The domain must also be long enough in the streamwise direction to allow the central recirculation to fully develop. The baseline case is selected to have a one-to-one aspect ratio with non-dimensional dimensions of $\left(L_x \times L_y \times L_z\right) = \left(550, 175, 175\right)$ as in table \ref{tab:grid_sizes}. As these are laminar simulations, a modest flow deflection of $\theta_{sg} = 2.0^{\circ}$ is selected for all simulations to follow \cite{Katzer1989} unless otherwise stated. On the upper surface in figure \ref{fig:Fig1}, $L_{in}$, $L_{sg}$ and $L_{out}$ refer to the distance between the inlet and the shock generator, the length of the shock generator and the remaining distance to the outlet. For the baseline case the shock generator starts at $x=45$, with $L_{sg} = 300$ and $L_{out} = 205$. For this $L_{sg}$ the trailing edge expansion fan generated at $x=345$ leaves through the outlet of the domain without impinging on the bottom wall. The effect of $L_{sg}$ on the central recirculation bubble is given in section \ref{sec:ramp_length}. The other cases in table \ref{tab:grid_sizes} correspond to the aspect ratio study in section \ref{sec:aspect_ratio} for aspect ratios between one-quarter and four. 


\begin{figure}
\begin{centering}
	\includegraphics[width=0.45\textwidth]{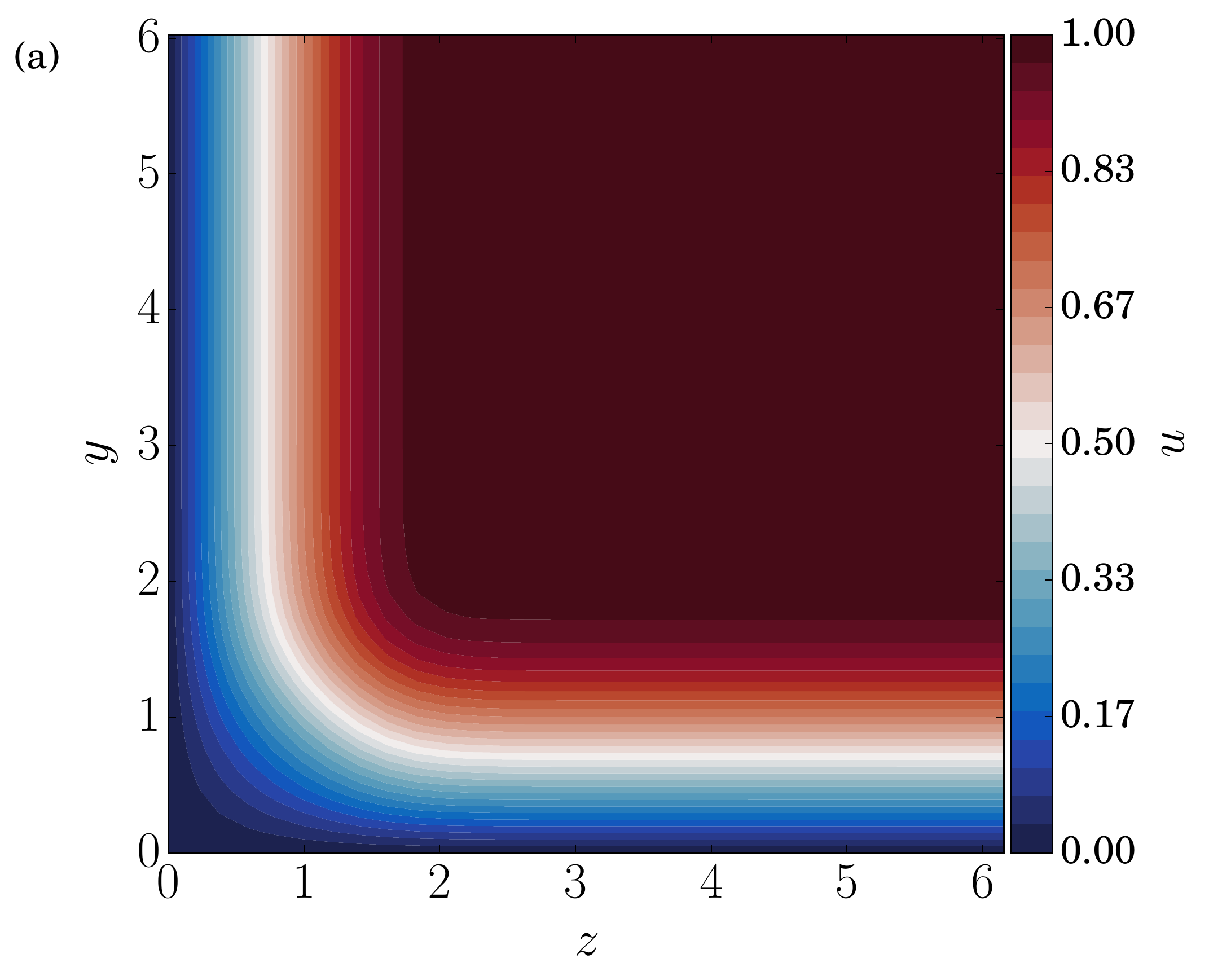}
  \includegraphics[width=0.49\textwidth]{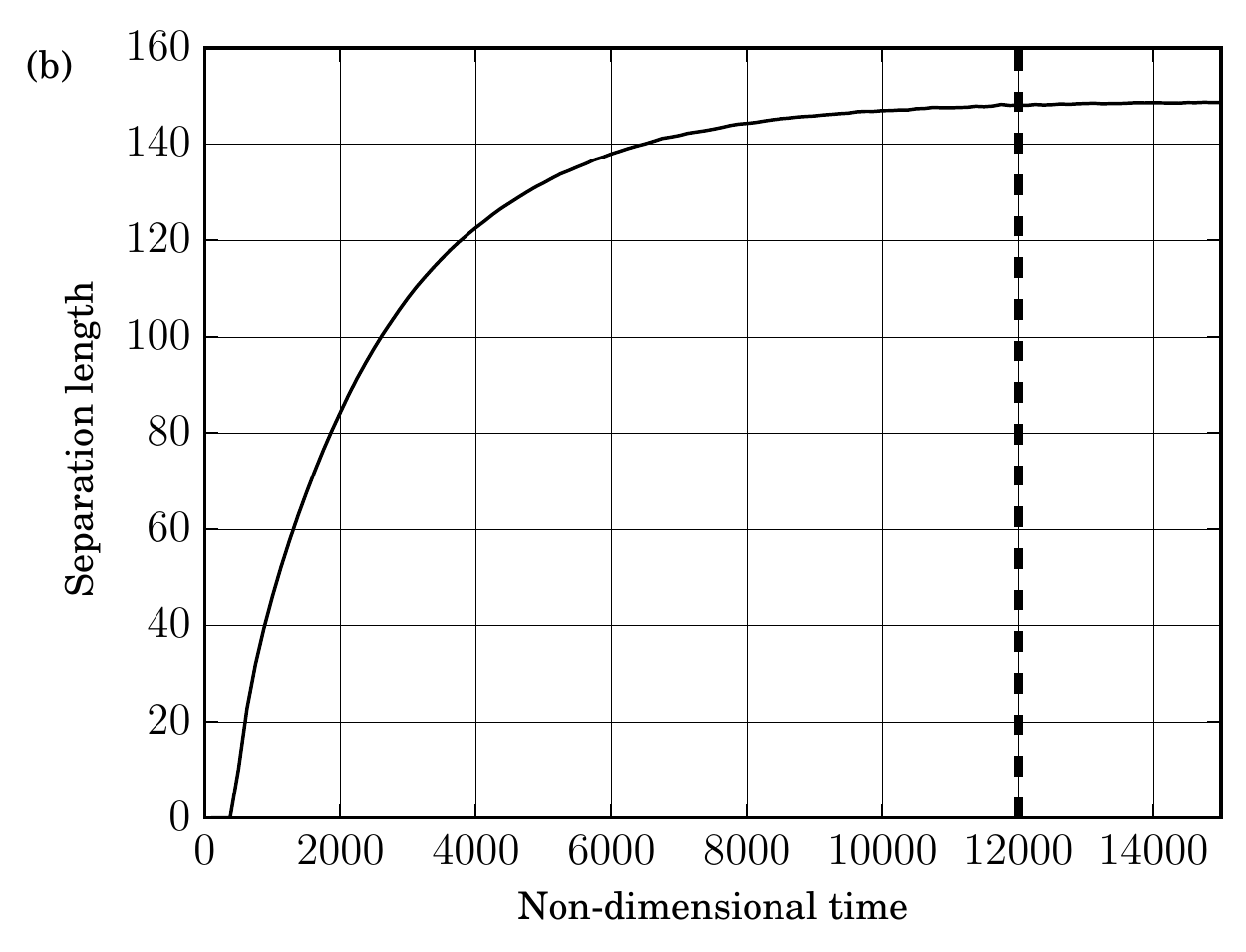}
  \caption{(a) Streamwise velocity contours of the inlet laminar boundary layer profile at the intersecting corner between two no-slip walls. (b) Convergence of the centreline separation bubble length in time. One flow-through time of the freestream is equal to $t=550$ time units.}\label{fig:Fig2}
  \end{centering}
\end{figure}

All simulations are performed at Mach 2 with a laminar boundary layer profile imposed at the inlet of the domain. The profile is obtained via the similarity solution of the compressible boundary layer equations \citep{white2006viscous}. The Reynolds number based on the displacement thickness at the start of the computational domain is $Re_{\delta *} = 750$. For the baseline configuration a flow deflection of $\theta_{sg} = 2^{\circ}$ is applied by the shock generator located at $x_{sg} = 45$, giving an inviscid impingement point of $x=328$ for the incident shock. Reynolds number based on the distance from the leading edge of the plate to the impingement point is $Re_{x} = 3\times 10^{5}$ as in one of the cases from \cite{Katzer1989}. For the variation of incident shock strength  in section \ref{sec:shock_strength}, the location of the shock generator is shifted to maintain the same $Re_{x}$ at impingement. On the bottom and both sidewalls of the domain a no-slip isothermal condition is applied with a constant non-dimensional temperature of $T_{w} = 1.676$ (4 s.f.), corresponding to the adiabatic wall temperature from the similarity solution. A zero gradient condition is applied on the upper boundary over $L_{in}$ in figure \ref{fig:Fig1} to maintain the freestream and sidewall boundary layers upstream of the shock generator. At $x_{sg}$ the upper surface becomes a no-slip wall with the same isothermal condition as on the bottom and sidewalls of the domain. The no-slip condition on the upper surface is maintained until the outlet. At the inlet and outlet a pressure extrapolation and low order extrapolation method are applied, respectively, to improve stability. No boundary layer is initialised on the shock generator; it is left to develop naturally during the initial stages of the simulation. An open condition upstream of the shock generator was selected to mimic experimental configurations where the freestream is incident on a shock generator plate. In the corner region boundary layer profiles of equal thickness from two adjacent walls are blended together as follows. The streamwise velocity profile for each wall is multiplied by the wall normal velocity component of the adjacent wall to create a combined profile that smoothly tends to zero in the corner. The similarity solution temperature profiles $T_{1}(y), T_{2}(z)$ for two intersecting walls are blended with a constant wall temperature $T_w$ as 
\begin{equation}
T(y,z) = T_w + T_{1}(y) T_{2}(z) \left(1 - T_w\right),
\end{equation}
giving a smooth profile that varies from $T=1$ in the freestream to $T=T_w$ at the wall. The wall normal velocity component from each of the sidewalls is of equal magnitude but opposite direction, requiring it to be damped with the $z$ coordinate in both directions to create a zero $w$ component of velocity on the centreline. Figure \ref{fig:Fig2} (a) shows the resulting profile that is imposed on the inlet; the normalised laminar flow is seen to vary smoothly from zero at the walls to one in the freestream.

As highlighted by \citet{Sansica2013}, laminar separation bubbles require long time integration to fully develop and previous studies have often reported shorter lengths from non-converged simulations. To verify the simulations were sufficiently converged for this work the evolution of centreline separation length is presented in figure \ref{fig:Fig2} (b). After impinging on the bottom wall boundary layer the incident shock rapidly creates a region of flow-reversal during the early stages of the simulation. With increasing time the separation length converges and a stable separation bubble length is observed. For all simulations in this work the convergence time is taken to be $t=12000$ ($\approx 22$ flow-through times of the domain), denoted by the vertical dashed line in figure \ref{fig:Fig2}(b). Integrating the simulation for a further $\approx 5.5$ flow-through times up to $t=15000$ only resulted in a 0.3\% change in separation length. Having defined the domain and physical parameters for the simulations, the next section demonstrates the grid independence of the solution.

\begin{table}
  \begin{center}
  \begin{tabular}{cccc}
      Simulation case & Domain dimensions $\left(L_x \times L_y \times L_z\right)$& Grid distribution $\left(N_x \times N_y \times N_z\right)$ \vspace{0.1cm}\\ 
      0.25AR &  $550 \times 175 \times 43.75$ &$ 750 \times 455 \times 87 $ \\
      0.5AR  &  $550 \times 175 \times 87.5 $& $750 \times 455 \times 177 $ \\
      Baseline 1.0AR  &  $550 \times 175 \times 175$ &$ 750 \times 455 \times 355$  \\
      2.0AR  &  $550 \times 175 \times 350$ &$ 750 \times 455 \times 715$  \\
      4.0AR  &  $550 \times 175 \times 700$ &$ 750 \times 455 \times 1435$ \\
  \end{tabular}
  \caption{Domain specification and grid distributions. Aspect ratio (AR) is defined as the ratio of duct width to height $\left(L_{z} / L_{y} \right)$. A one-to-one aspect ratio is taken as the baseline configuration.}
  \label{tab:grid_sizes}
  \end{center}
\end{table}
\subsection{Sensitivity to grid refinement}\label{sec:laminar_grid_study}
\begin{figure}
  \includegraphics[width=0.50\textwidth]{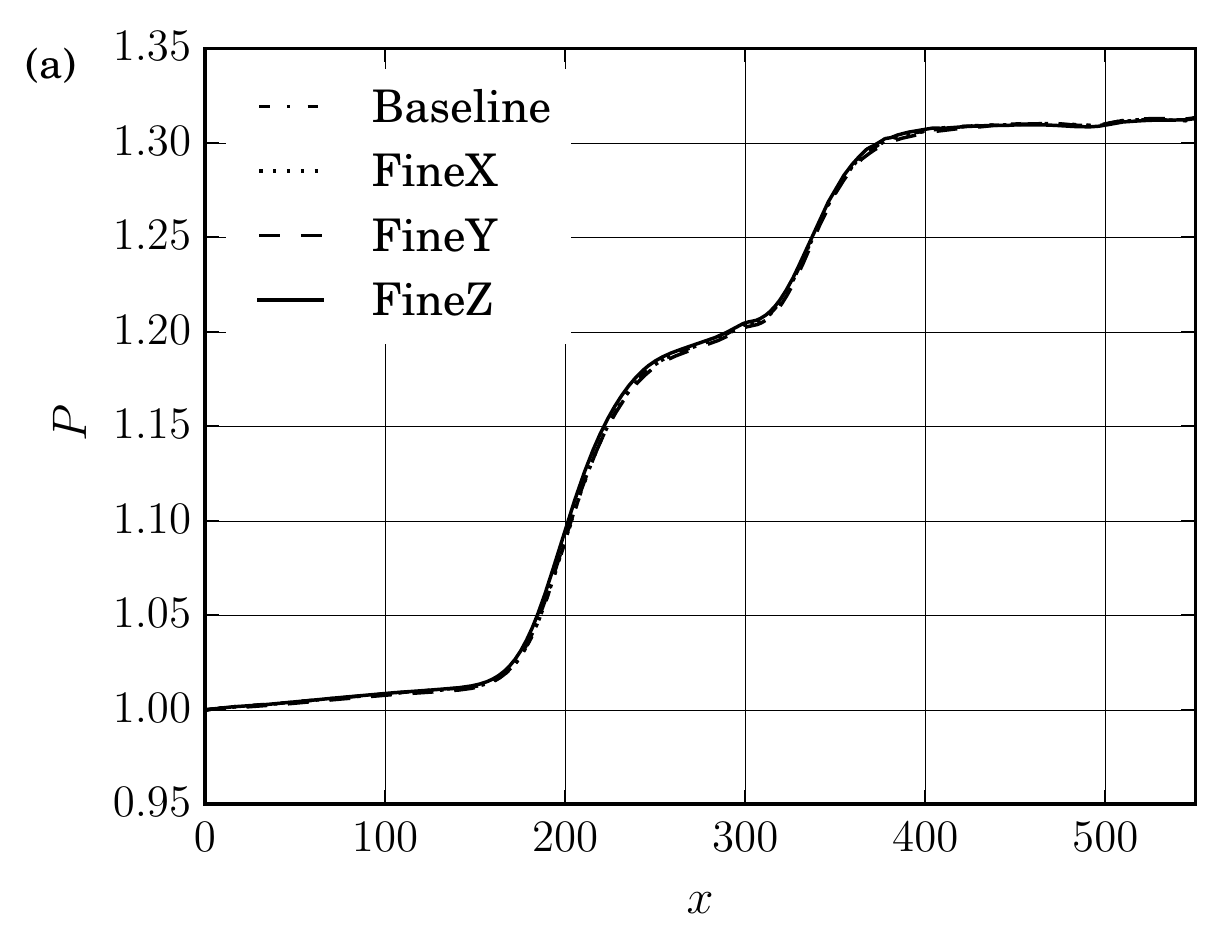}
  \includegraphics[width=0.50\textwidth]{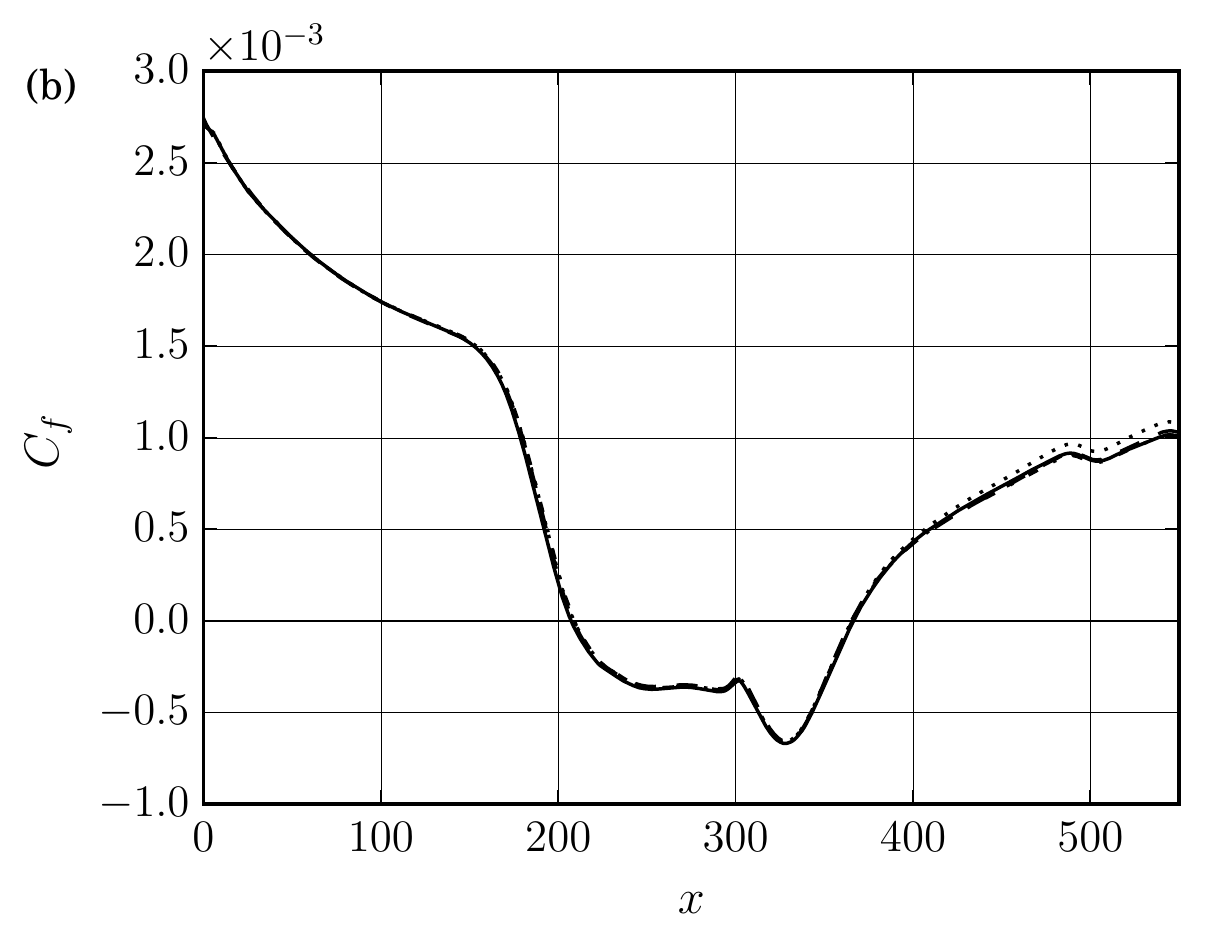}
  \caption{Sensitivity of the centreline (a) wall pressure and (b) skin friction to grid refinement for $AR=1$. In each direction 50\% additional grid points are added independently.}\label{fig:Fig3}
\end{figure}
Based on initial exploratory simulations, a starting grid resolution of $\left(N_x, N_y, N_{z}\right) = \left(700, 295, 295\right)$ was selected to perform the grid refinement study and investigate the effect of shock generator length. Insensitivity to grid refinement was assessed by increasing the number of grid points by 50\% in each spatial direction independently. Grid stretching is performed symmetrically in the $y$ and $z$ directions to cluster points in the boundary layers of each wall, with a uniform distribution in $x$. Grid points in $y$ and $z$ are distributed with a stretch factor $s=1.3$ as

\begin{equation}\label{grid_stretching}
y = \frac{1}{2}L_y \frac{1-\tanh\left(s\left(1-2\xi\right)\right)}{\tanh\left(s\right)}, \quad \quad z = \frac{1}{2}L_z \frac{1-\tanh\left(s\left(1-2\xi\right)\right)}{\tanh\left(s\right)},
\end{equation}
for uniformly distributed points $\xi = \left[0,1\right]$.
Figure \ref{fig:Fig3} (a) and (b) show the effect of increased grid resolution for the centreline wall pressure and skin friction respectively. For the baseline $\theta_{sg} = 2^{\circ}$ case with one-to-one aspect ratio the shock induced pressure rise normalised by the inlet is $p_3/p_1 = 1.31$. There is minimal discrepancy between each of the simulations and the centreline pressure is insensitive to further grid refinement. A similar picture is seen for the skin friction in figure \ref{fig:Fig3} (b), all grids produce the expected asymmetric twin trough shape of a laminar separation bubble. A small deviation is seen downstream of the reattachment point in the case of streamwise grid refinement. The separation bubble length is the streamwise extent of flow-reversal, defined as the distance between the two zero crossings of the skin friction curve in figure \ref{fig:Fig3} (b). The separation length is insensitive to grid refinement; the largest variation occurred for the `FineZ' case which was 1\% larger than the coarse grid. There is also a slight discrepancy at the outlet in the `FineX' case. Based on these results and to improve resolution on the shock generator a refined grid of $\left(N_x, N_y, N_{z}\right) = \left(750, 455, 355\right)$ was selected for the default one-to-one aspect ratio cases in this paper. Parametric studies of aspect ratio in section \ref{sec:aspect_ratio} use the grids outlined in table \ref{tab:grid_sizes}.


\subsection{The role of shock generator length and the trailing expansion fan}\label{sec:ramp_length}
During the selection of the computational domain it became apparent that in addition to the aspect ratio and dimensions of the domain, the length of the shock generator is an important factor. This influence is most significant when considering laminar SBLI as the separation regions are considerably larger than in the presence of turbulence and are therefore more likely to be crossed by expansion fans emitted from the trailing edge of the shock generator. To quantify this effect, a selection of shock generator lengths are reported in this section for 2D and 3D simulations, using the baseline grid from the previous section. The comparison of 2D to 3D is useful because it illustrates the role of the sidewalls when considering the shock generator length.
\begin{table}
  \begin{center}
  \begin{tabular}{cccc}
      Shock generator length $\left(L_{sg}\right)$& Interaction region $\left(x_{start}, x_{end}\right)$ & $L_{sep}$& Increase in $L_{sep}$ (\%) \vspace{0.1cm}\\ 
      200 &  (251.4, 369.5) & 118.09  &  -- \\
      250  &  (251.1, 370.4) & 119.32 & 1\% \\
      300  &  (251.0, 370.5) & 119.52 &  1\% \\
      350  &  (251.0, 370.5) & 119.55 &  1\% \\
  \end{tabular}
  \caption{Sensitivity of the centreline separation to increasing $L_{sg}$ for two-dimensional SBLI without sidewalls. Increasing $L_{sg}$ causes the trailing edge expansion fan to impinge further downstream on the bottom wall. Percentage increase is relative to the shortest $L_{sg} = 200$ case.}\label{tab:ramp_length_2d}
  \end{center}
\end{table}

Four shock generator lengths in the range $L_{sg} = \left[200, 350\right]$ are considered, which correspond to $(36-64)\%$ of the streamwise domain length. The lengths were chosen to ensure that the trailing expansion fan did not impinge directly on the separation bubble, but were close enough to ascertain the downstream influence on the main interaction region. As these are all laminar interactions, 2D simulations are equivalent to a 3D simulation with span-periodic boundary conditions. The 3D simulations include the effect of sidewalls and so a deviation from the quasi-2D results in this section can only be attributed to 3D effects resulting from physical flow confinement in the duct. 

\begin{figure}
  \includegraphics[width=0.50\textwidth]{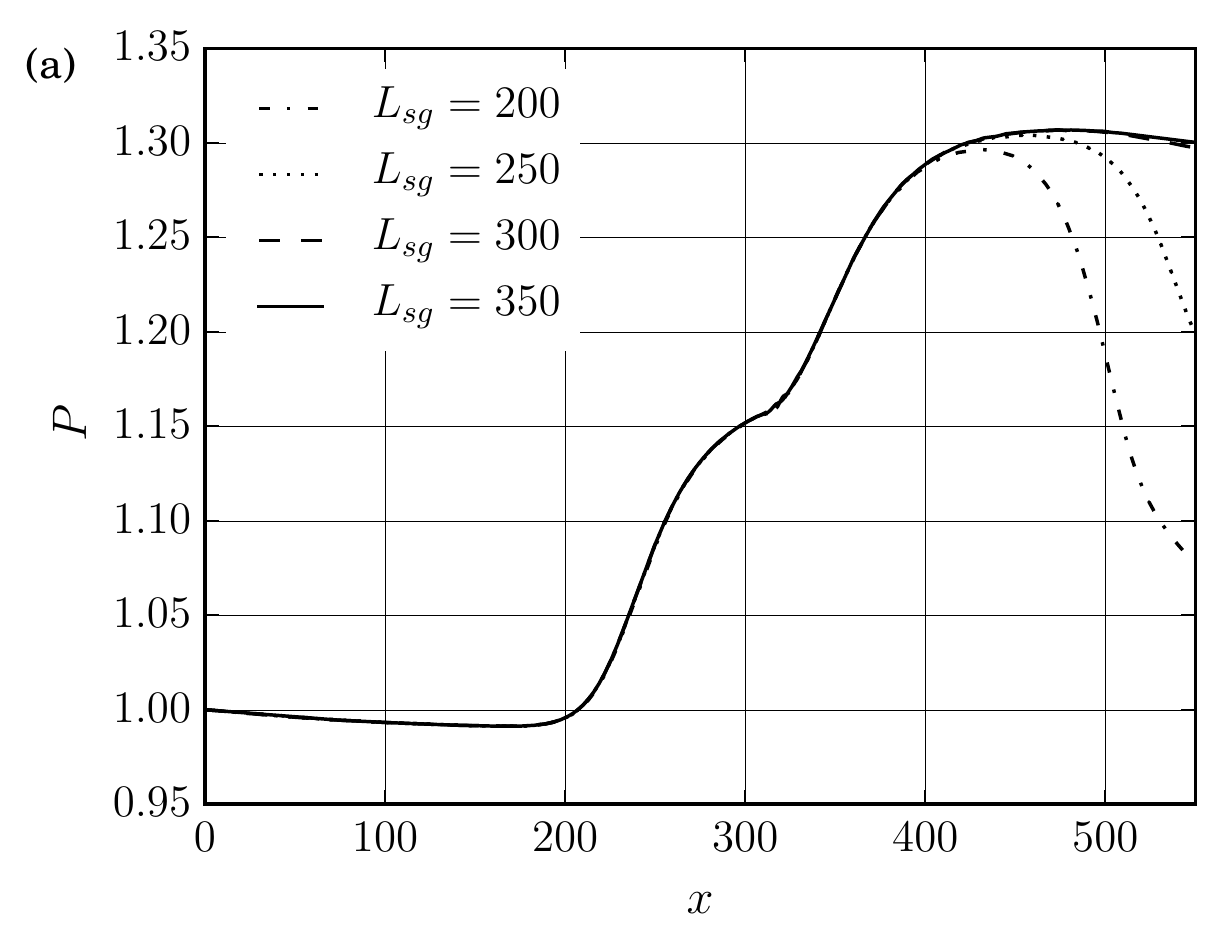}
  \includegraphics[width=0.50\textwidth]{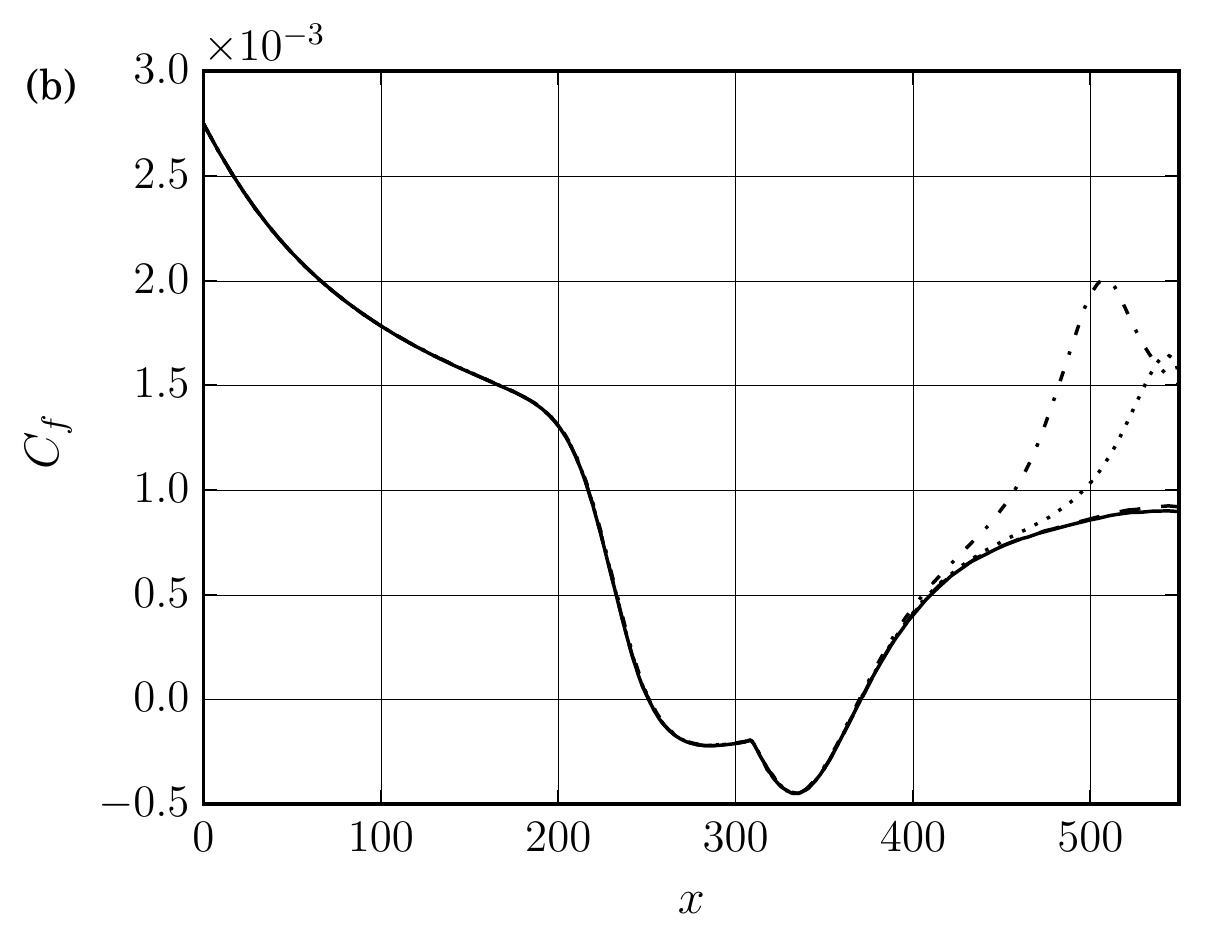}
  \caption{Sensitivity of the 2D simulation (a) wall pressure and (b) skin friction to shock generator length.}
\label{fig:Fig4}
\end{figure}

Figures \ref{fig:Fig4} (a) and (b) show the centreline wall pressure and skin friction for the 2D simulations as the shock generator length is varied. For the shortest two shock generator lengths an expansion fan impinges on the bottom wall of the domain downstream of the reattachment point. Although there is a significant decrease/increase in pressure/skin-friction near the outlet, the separation bubble is largely unchanged by this downstream influence. Table \ref{tab:ramp_length_2d} quantifies the effect the shock generator length has on separation for quasi-2D interactions. The shortest two shock generators agree to within 1\% of each other and further increases in shock generator length have no significant influence on the separation bubble.

\begin{table}
  \begin{center}
  \begin{tabular}{cccc}
      Shock generator length $\left(L_{sg}\right)$& Interaction region $\left(x_{start}, x_{end}\right)$ & $L_{sep}$& Increase in $L_{sep}$ (\%) \vspace{0.1cm}\\ 
      200 &  (220.0, 351.9) & 131.89  &  -- \\
      250  &  (213.5, 359.8) & 146.35 & 11\% \\
      300  &  (209.0, 365.8) & 156.77 &  19\% \\
      350  &  (205.2, 371.2) & 166.05 &  26\% \\
  \end{tabular}
  \caption{Sensitivity of the centreline separation to increasing $L_{sg}$ for three-dimensional SBLI at $AR=1$ with sidewall effects. Increasing $L_{sg}$ causes the trailing edge expansion fan to impinge further downstream on the bottom wall and also modifies the pressure distribution downstream of the interaction. Percentage increase is given relative to the shortest $L_{sg} = 200$ case.}
  \label{tab:ramp_length_3d}
  \end{center}
\end{table}

\begin{figure}
  \includegraphics[width=0.50\textwidth]{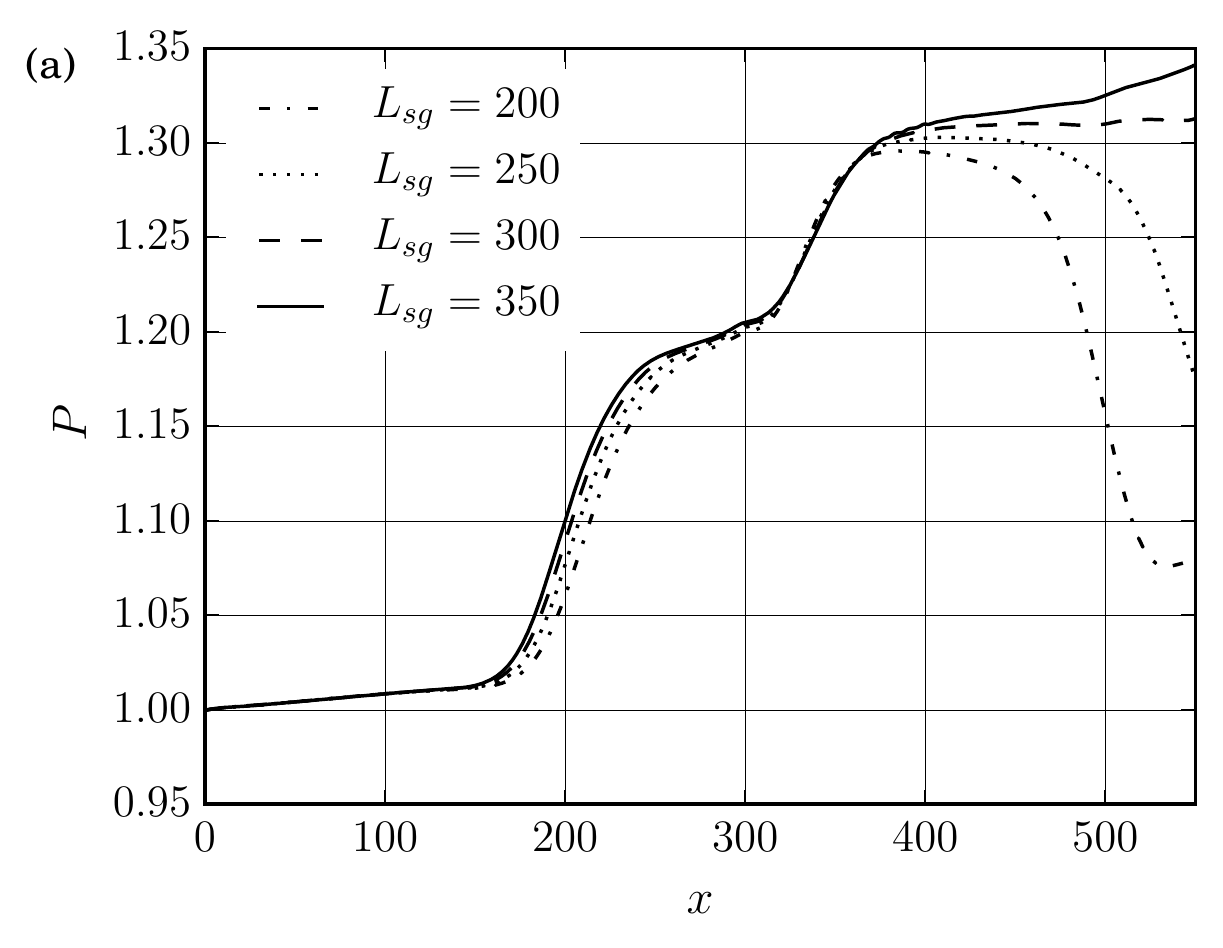}
  \includegraphics[width=0.50\textwidth]{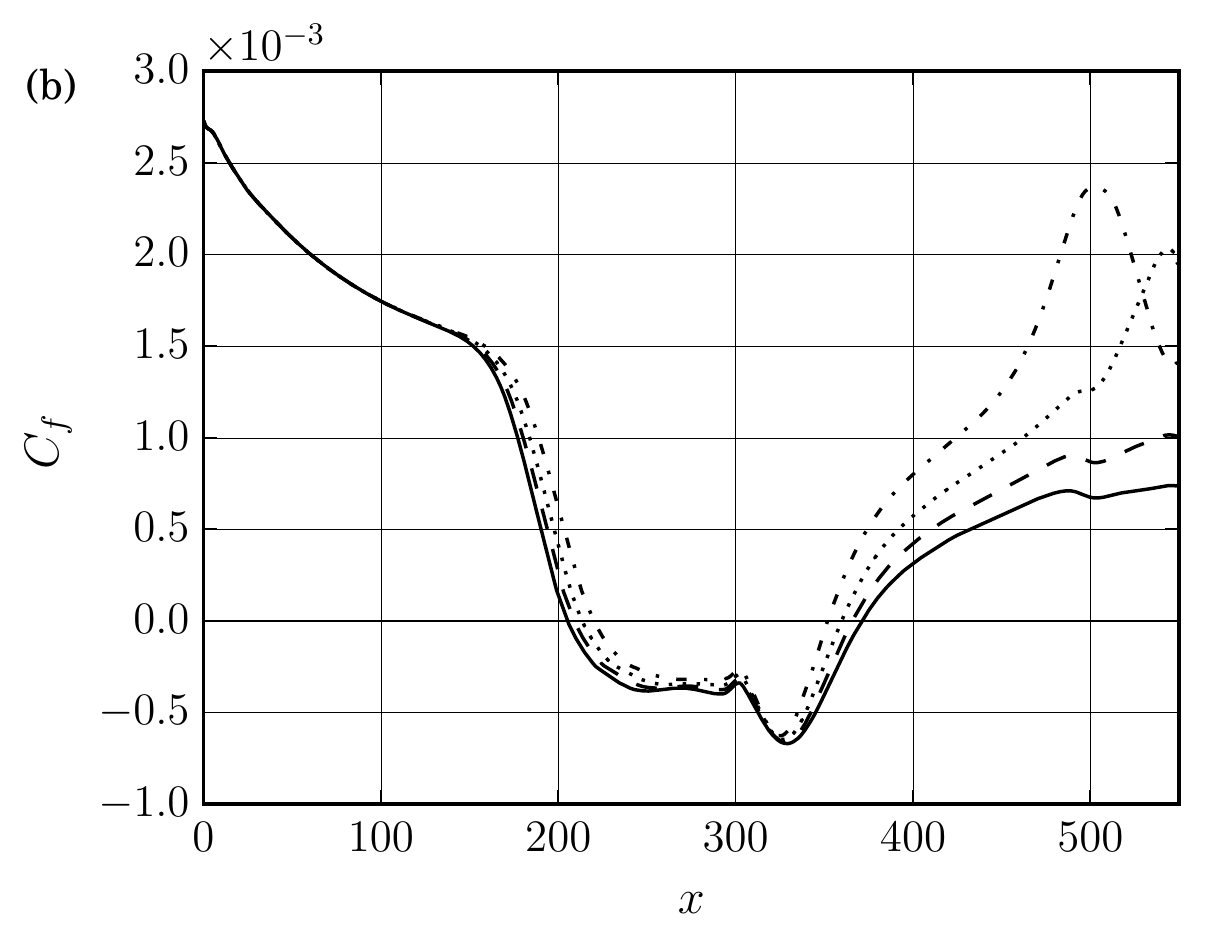}
  \caption{Sensitivity of the 3D simulation with sidewalls at $AR=1$ for the (a) wall pressure and (b) skin friction to shock generator length.}
\label{fig:Fig5}
\end{figure}

3D results with sidewall effects are shown in Figures \ref{fig:Fig5} (a) and (b) for the same range of shock generator lengths as in the 2D cases but with $AR=1$. In contrast to figure \ref{fig:Fig4} (b) the skin friction distribution of figure \ref{fig:Fig5} (b) shows a clear influence of the trailing expansion fan on the main interaction. In addition to the previously seen skin friction rise at the outlet, the central separation bubble has been shortened significantly in the 3D case for the shorter shock generator lengths. When the sidewall influence is included, the separation and reattachment locations of the separation bubble are both modified. A similar pattern is seen in figure \ref{fig:Fig5} (a), where the initial pressure rise at the point of separation is delayed downstream for shorter shock generators. 

Table \ref{tab:ramp_length_3d} gives the size of the interaction region and increases in separation length for the 3D cases. As the length of the shock generator is increased from $L_{sg}=200$ the separation and reattachment locations shift upstream and downstream respectively. This leads to $(11-26)\%$ increases in overall separation length compared to the shortest shock generator. Importantly we see there is an increase in separation length even between $L_{sg}=300$ and $L_{sg} = 350$, where in both cases the trailing expansion fan is leaving the computational domain before impinging on the bottom wall. As the largest two shock generators disagree with each other despite the expansion fans not directly hitting the bottom wall, the discrepancy can only be attributed to 3D effects of the trailing expansion fan on the sidewall flow and its subsequent influence on the central separation. Experimentally this effect has been observed for a turbulent case by \cite{Grossman2017}, in which the physical thickness of the shock generator was varied to move the location of the expansion fan. The authors noted that as the expansion fan is moved downstream, there is an increase in the strength and streamwise length of the central separation accommodated by an upstream shift in the separation point. Despite the differences in incident shock strength and boundary layer state to the present work, their findings are consistent with those of figure \ref{fig:Fig5} (b). The main difference to this work is that in the laminar case a downstream shift of the reattachment location is observed while remaining largely independent of the expansion fan location in \cite{Grossman2017}.

Having quantified the role of the shock generator length for the 3D simulations, we select a domain with $L_{sg} = 300$ as the default configuration for all following simulations unless otherwise stated. It must be emphasised that this is a design choice of the duct and differences in the separation length would occur for different configurations. Including three-dimensional flow confinement into the problem increases the complexity of the flow field and naturally adds a dependence of the duct aspect ratio, domain dimensions, and shock generator length to any reported results. This is in contrast to quasi-2D simulations where the SBLI depends only on the incident shock strength and incoming boundary layer state. The use of the trailing expansion fan to control the central interaction is investigated further in section \ref{sec:long_domain_control} for a longer domain with a considerably longer shock generator.

\section{3D laminar duct SBLI with sidewall effects}
\begin{figure}
\begin{centering}
  \includegraphics[width=0.5\textwidth]{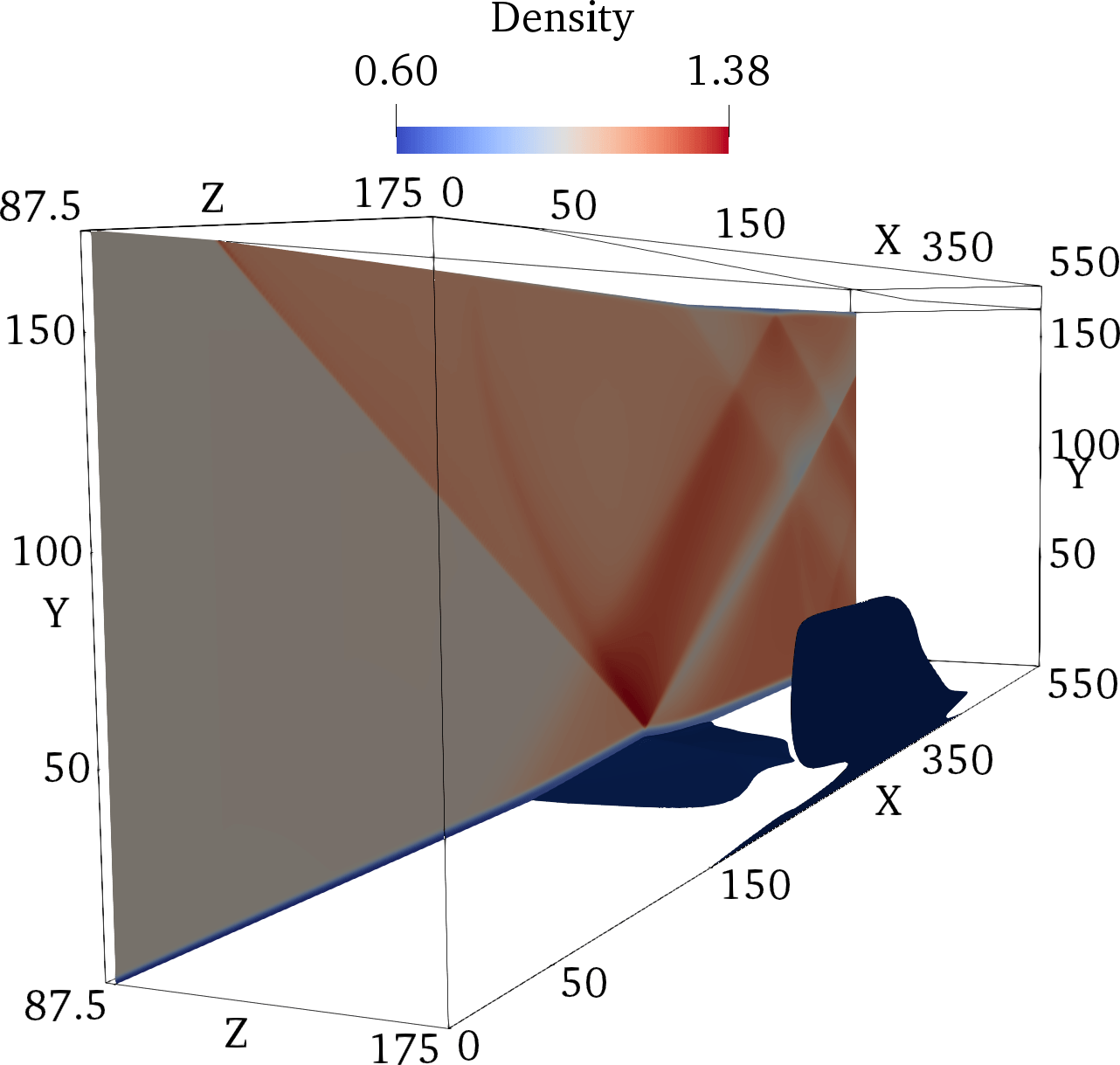}
  \caption{Baseline duct SBLI density contours ($AR=1$). Displaying a centreline density slice $\left(z=87.5\right)$ with regions of reverse flow $\left(u \leq 0\right)$ on the bottom and sidewall highlighted in dark blue.}\label{fig:Fig6}
  \end{centering}
\end{figure}

\subsection{Baseline duct configuration}\label{sec:duct_flow}
Figure \ref{fig:Fig6} shows density contours for the laminar base flow obtained for the default configuration of aspect ratio one and $\theta_{sg} = 2^{\circ}$. The regions of flow-reversal are highlighted in dark blue on the sidewall and in the centre. Despite the relatively weak initial shock, large regions of reverse flow develop in the corners and on the bottom and side walls of the domain. This is in contrast to turbulent SBLI such as \citet{wang_sandham_hu_liu_2015}, where the greatly enhanced mixing rates in the boundary layer help prevent flow separation on the sidewalls. A slice of density along the centreline shows that the separation bubble extends far upstream of the impingement point, with a series of compression waves emitted from the start of the bubble due to a thickening of the boundary layer. For the laminar base flow the features are symmetric about the centreline $\left(z=87.5\right)$, with each sidewall containing a large region of reverse flow. A long thin corner separation is seen that extends further upstream than both the central and sidewall separations. Between the sidewall and central separations is a distinct region of attached flow where the initial shock has been weakened by the sidewall influence. At the trailing edge of the shock generator an expansion fan can be seen crossing the reflected shock and leaving the computational domain. The reflected shock creates a secondary separation bubble on the upper wall of the domain before passing through the outlet.

\begin{figure}
  \includegraphics[width=0.50\textwidth]{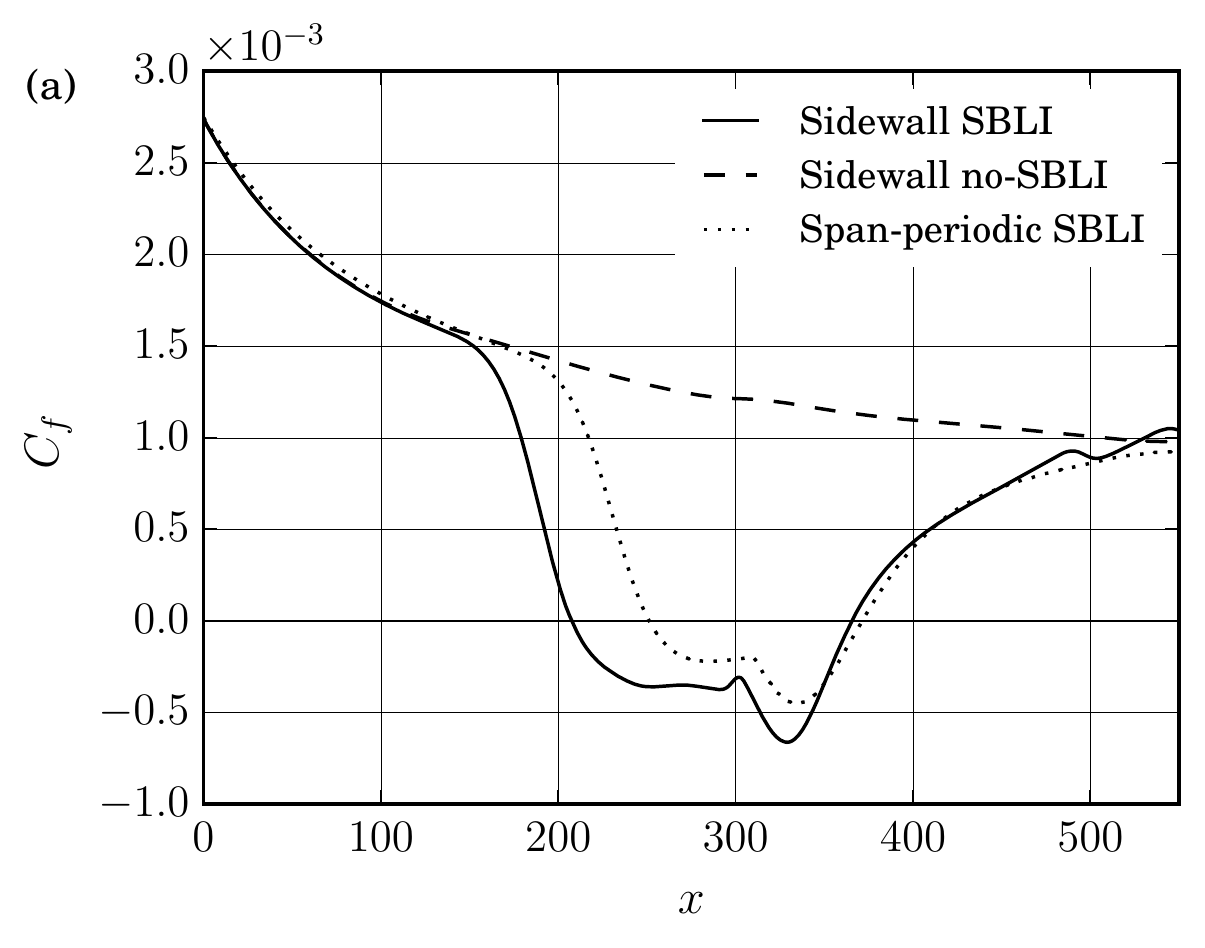}
  \includegraphics[width=0.50\textwidth]{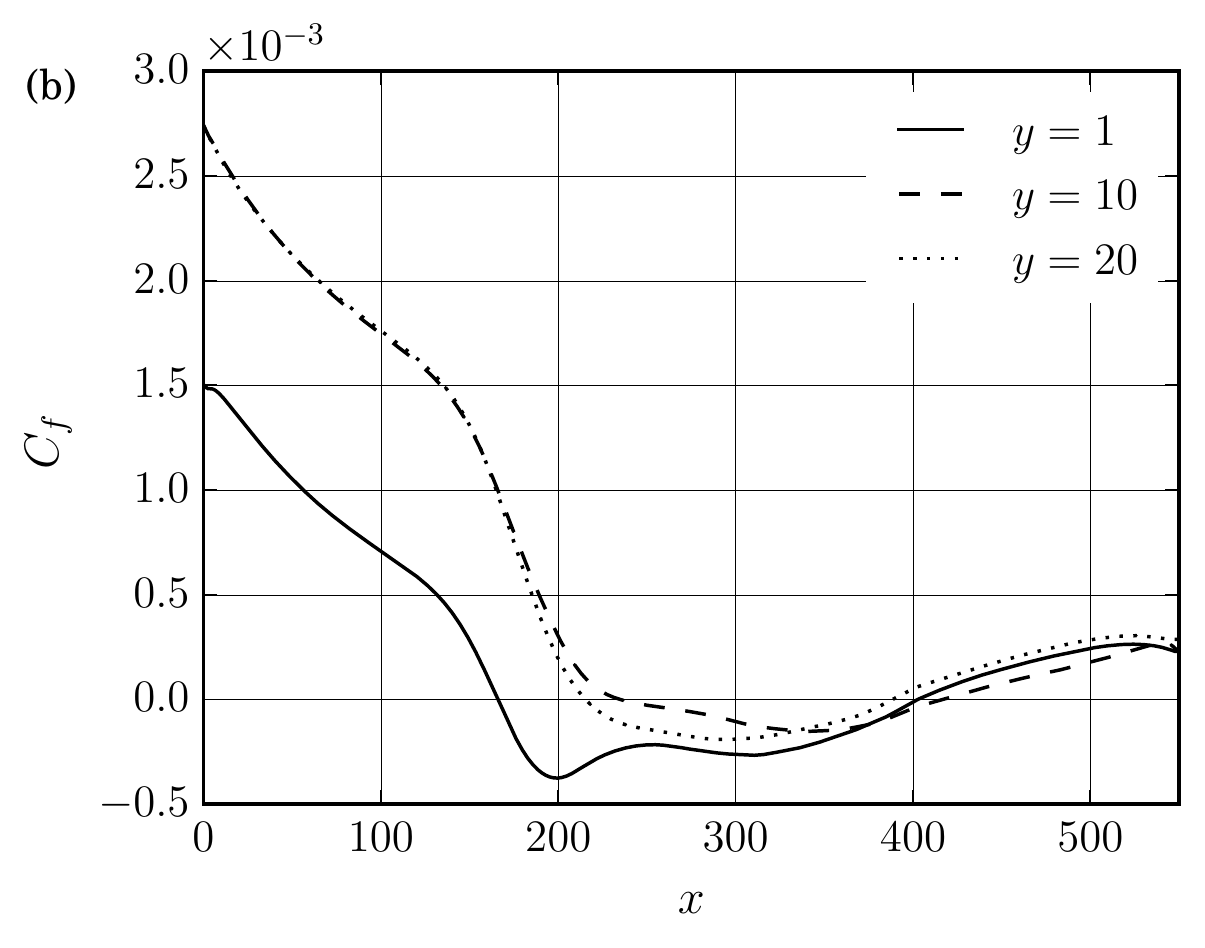}
  \caption{(a) Centreline skin friction on the bottom wall $\left(y=0\right)$ for a duct with and without SBLI at $AR=1$, compared to a case without sidewalls. (b) Skin friction relative to the sidewall $\left(z=0\right)$ for the duct SBLI at various $y$ heights, showing the early streamwise onset of the corner separation.}\label{fig:Fig7}
\end{figure}

Figure \ref{fig:Fig7} (a) compares the centreline skin friction at $y=0$ for the duct with and without a shock generator. It can be seen that the reattached flow downstream of the SBLI recovers to match the laminar boundary layer near the outlet. A further comparison is made to a span-periodic case to demonstrate the effect that sidewall confinement has on the central flow. The strengthening of the incident shock from the sidewalls leads to an increase in central separation length of 31.5\%. It is again emphasised that as in Table \ref{tab:ramp_length_3d}, this percentage increase is highly dependent on the shock generator length and subsequent position of the trailing expansion fan. Separation and reattachment locations $\left(x_{start}, x_{end}\right)$ are found at $x = \left(251.2, 371.4\right)$ and $x = \left(207.7, 365.6\right)$ for the span-periodic and duct SBLI respectively. Although the reattachment locations are similar, the separation point has moved upstream substantially due to the sidewall influence. The early onset of the corner separation relative to the sidewall separation can be seen in figure \ref{fig:Fig7} (b). Skin-friction relative to the sidewall (\ref{cf_definition}) is shown at three different $y$ locations on the $z=0$ side of the domain. Within the corner boundary layer at $y=1$ the flow first detaches at $x=166.1$, at which point the centreline and sidewall boundary layers are still attached. The skin friction distributions at $y=10$ and $y=20$ agree well up until the point of separation, occurring at $x=215.3$ and $x=236.5$ respectively. The strongest flow-reversal on the sidewall occurs early within the corner region, as seen in the trough around $x=200$. From this we conclude that the corner regions of the duct are most susceptible to shock-induced separation, as there is a build up of low-momentum fluid being simultaneously retarded by no-slip walls in two directions.

\begin{figure}
  \includegraphics[width=0.435\textwidth]{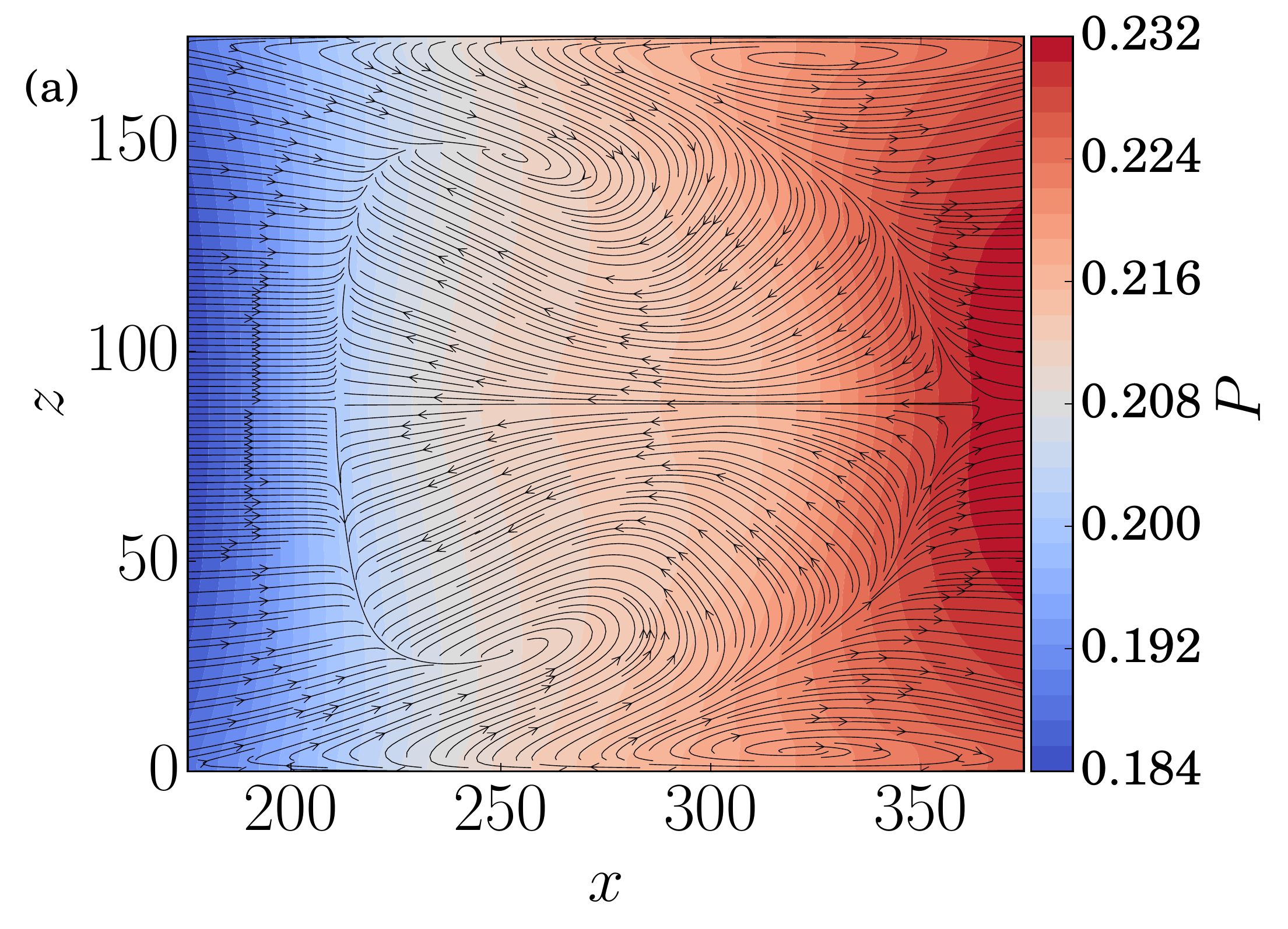}
  \includegraphics[width=0.565\textwidth]{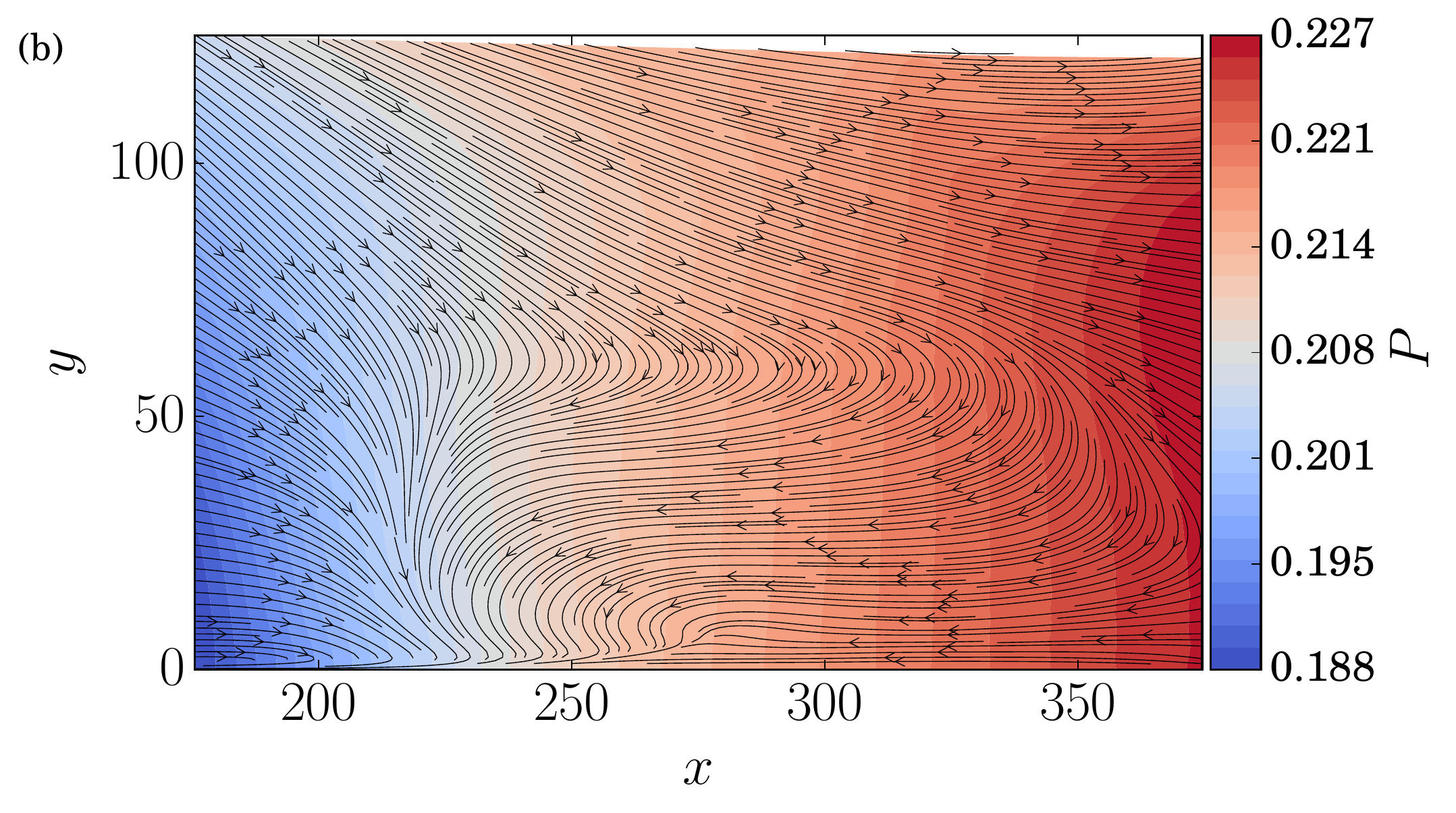}
  \caption{Streamline patterns coloured by the shock-induced pressure jump of the main interaction for $AR=1$. Displaying (a) $u\textrm{-}w$ streamlines above the bottom wall at $y=0.14$ and (b) $u\textrm{-}v$ streamlines above the sidewall at $z=0.14$.}\label{fig:Fig8}
\end{figure}

\begin{figure}
  \includegraphics[width=1.0\textwidth]{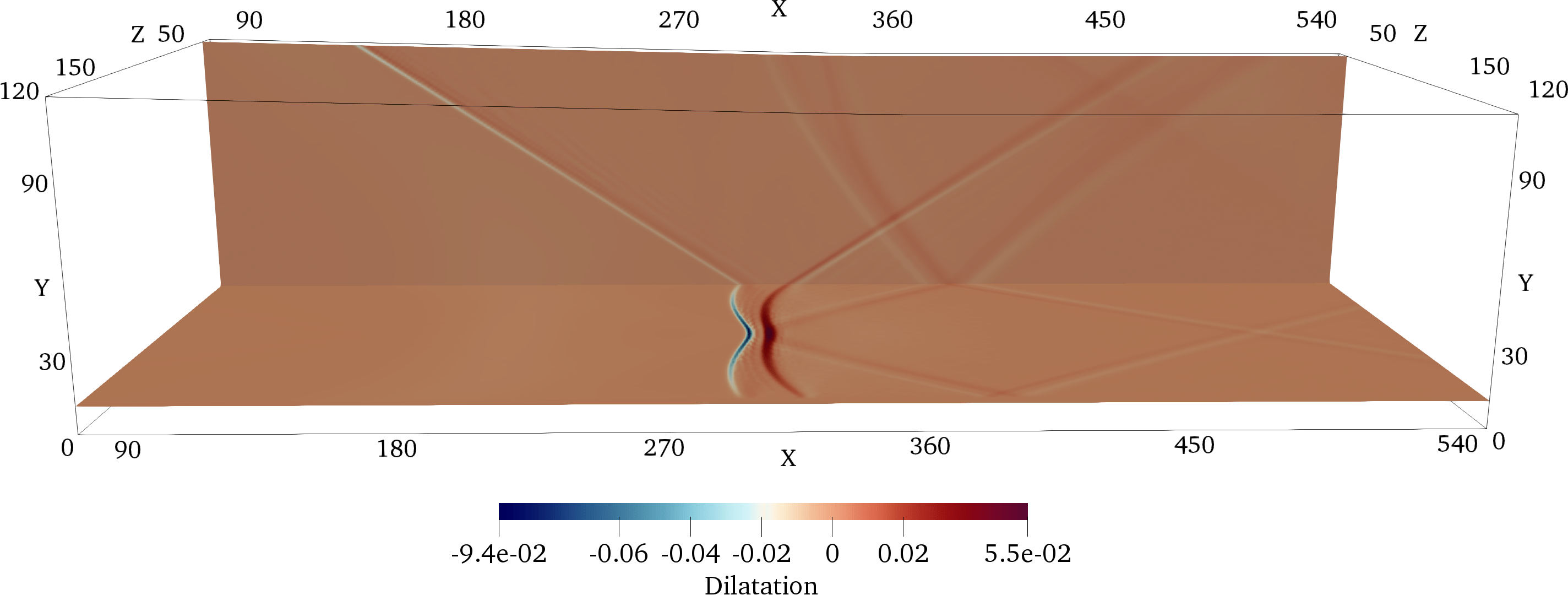}
  \caption{2D slices of dilatation $\left(\nabla \cdot \bar{u}\right)$ plotted at $y=10$ and $z=10$ for the $AR=1$ baseline case. The influence of the sidewalls leads to a curving of the incident shock. Negative dilatation shown in blue corresponds to the strongest regions of the incident shock. A pair of transverse shockwaves are seen to emanate from the interaction region and reflect off the sidewalls.}
\label{fig:Fig9}
\end{figure}

To further elucidate the regions of flow-reversal in figure \ref{fig:Fig6}, velocity streamline patterns are shown in the near-wall region in figure \ref{fig:Fig8} for (a) the bottom wall and (b) the sidewall of the domain. A more formal analysis of flow topology is given in the next section after the main recirculation zones are highlighted here. The structure of the central recirculation is clear to see by noting the direction of the streamlines; flow is ejected from each corner and towards the centreline where it travels upstream. Streamlines diverge at the separation (blue) and reattachment (red) regions of the interaction and recirculation is also visible in each of the corners. In between the central and corner separations the attached flow region in figure \ref{fig:Fig6} is seen as the region where velocity streamlines diverge away from the attachment line and continue downstream. Figure \ref{fig:Fig8} (b) shows the down-wash of fluid near the sidewall as a result of the swept SBLI. Streamlines from all directions are directed into a nodal point at $x=220$ with an accompanying focal point similar to the type 1 separation of \cite{Tobak1982}. Flow-reversal dominates a large portion of the sidewall and extends to almost 50\% of the duct height. Finally we draw attention to the curved shape of the incident and reflected shocks shown by the dilatation plot $\left(\nabla \cdot \bar{u}\right)$ in figure \ref{fig:Fig9}. Two orthogonal intersecting slices are plotted at $y=10$ and $z=10$, with negative (blue) and positive (red) dilatation representing the incident shock and expansion fan above the bubble respectively. There is a curving of the shock, that is consistent with figure 10 (a) of \cite{wang_sandham_hu_liu_2015}, despite the differences in flow conditions and boundary layer state. The shock deforms downstream at the centreline, which is also seen to be the strongest part of the incident shock. Away from the centre the incident shock decreases in strength, consistent with the conditions needed to produce regions of attached flow bordering the central recirculation. 

\subsection{Topology of the interaction}\label{sec:laminar_topology}
\begin{figure}
\begin{centering}
  \includegraphics[width=0.7\textwidth]{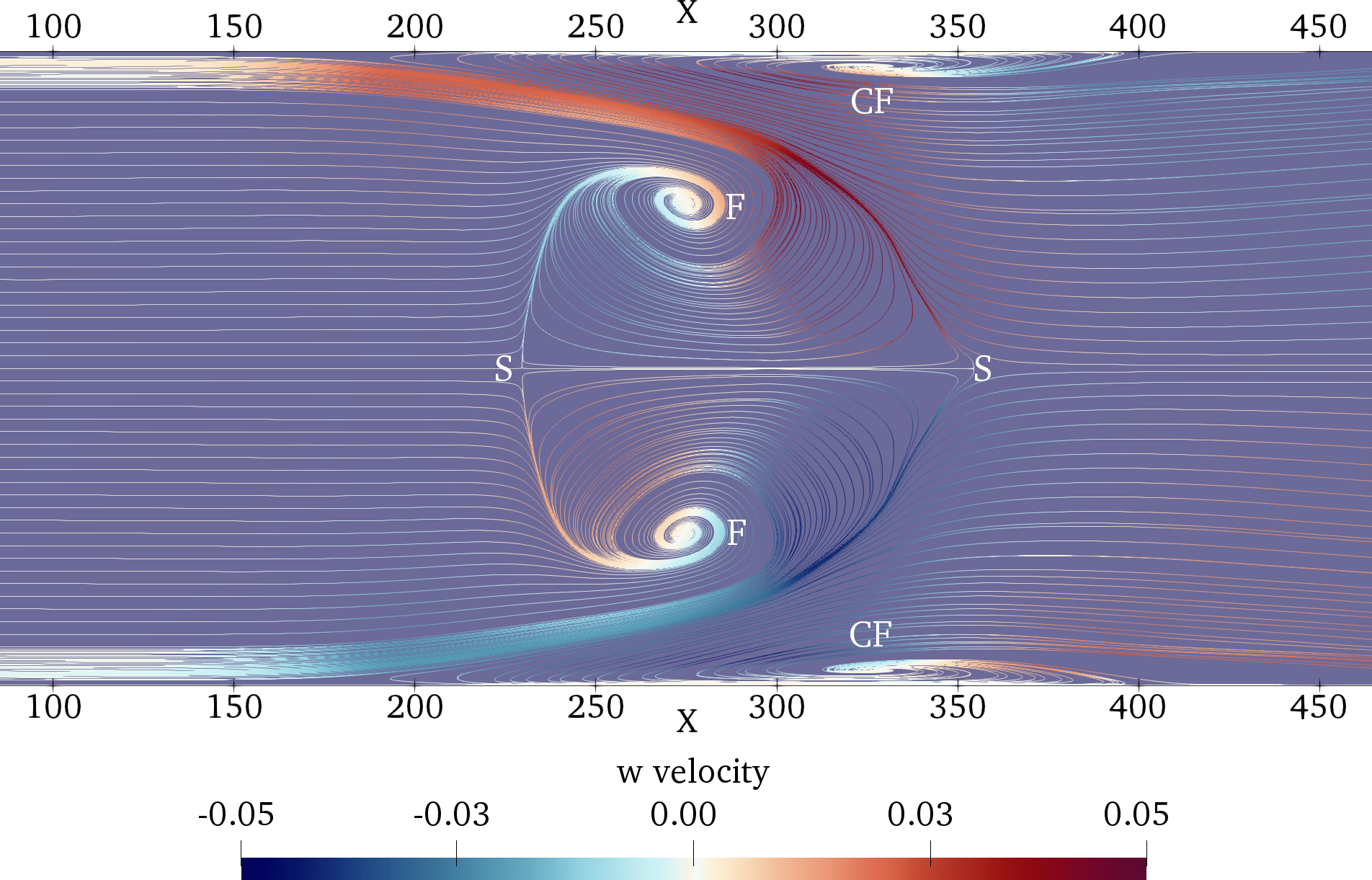}
  \caption{Streamlines evaluated in the $x$-$z$ plane at $y=1$ for the baseline $\theta_{sg} = 2.0^{\circ}$ case. Streamlines are coloured by the transverse velocity component $w$ with a constant colour background. The flow diverges at saddle points (S) at the front and back of the main recirculation region. The SBLI generates strong transverse velocity gradients that cause an ejection of the corner flow towards the centreline. Streamlines within the separation bubble are directed into two foci (F) that are symmetric relative to the centreline. Two additional foci (CF) required for topological consistency are labelled in each corner region.}\label{fig:2deg_owl_pattern}
\end{centering}
\end{figure}
To understand three-dimensional SBLI, it is important to look at both global shock structures and the topological features visible in near-wall streamline traces. Experimental streamline patterns are typically obtained via an injection of an oil mixture upstream of the interaction, which gives an imprint of the mean flow on the walls of the test chamber. Examples of oil injection include figure 11 of \cite{Grossman2018}, where the oil injection points are clearly visible upstream of the central SBLI. In addition to the potential for oil injection to cause undesirable modification of the flow, care must also be taken to avoid imprints of the transient behaviour during wind tunnel start-up/shutdown. Modern experimental techniques such as stereo-PIV can capture velocity data in three dimensions, allowing for the construction of limiting streamline patterns \citep{Eagle2014}. Among the benefits of numerical work is the access to full three-dimensional time-dependent flow data which can supplement observations made experimentally. 

Critical point analysis is a useful tool for identifying three-dimensional separations from streamline patterns. Critical points occur where skin friction lines terminate on a surface or, equally, where the magnitude of two-dimensional skin friction vectors becomes zero. Points are classified into either nodes or saddle points, with further subdivisions of nodes into nodal points and foci of either attachment or separation depending on the direction of the streamlines \citep{Tobak1982}. While usually described in the context of skin friction, the same analysis can be performed on streamlines obtained from velocity fields \citep{babinsky_harvey_2011}. Attachment nodes (N) are classified as the source of streamlines emerging from an object and separation nodes are found where they terminate. A focus (F) is a point about which streamlines spiral around and ultimately terminate. Saddle points (S) are defined as singular locations at which only two streamlines enter, one inwards and one outwards. All other streamlines diverge away from a saddle point hyperbolically, separating the streamlines that emerge from adjacent nodes. A two-dimensional separation bubble is characterised by a streamline that lifts off a surface at a separation point and reattaches at a point downstream of the bubble. Within the bubble closed streamlines circulate around a single common point and do not escape to the outer flow. This description is incompatible with three-dimensional separations where streamlines instead have a decaying orbit around a focus point that terminates them. In three dimensions the criteria for identifying flow separation can be defined as streamline patterns that contain at least one saddle point \citep{Delery2001}. At a focus fluid escapes laterally and signals the presence of a tornado-like vortex \citep{babinsky_harvey_2011}. A vortex above a surface acts to lift fluid entering the focus upwards and transfer it downstream to the outer flow. In this sense three-dimensional separations are denoted `open' separations as flow attaching downstream of the interaction is distinct from that which separated previously \citep{Eagle2014}.

\begin{figure}
\begin{centering}
  \includegraphics[width=0.6\textwidth]{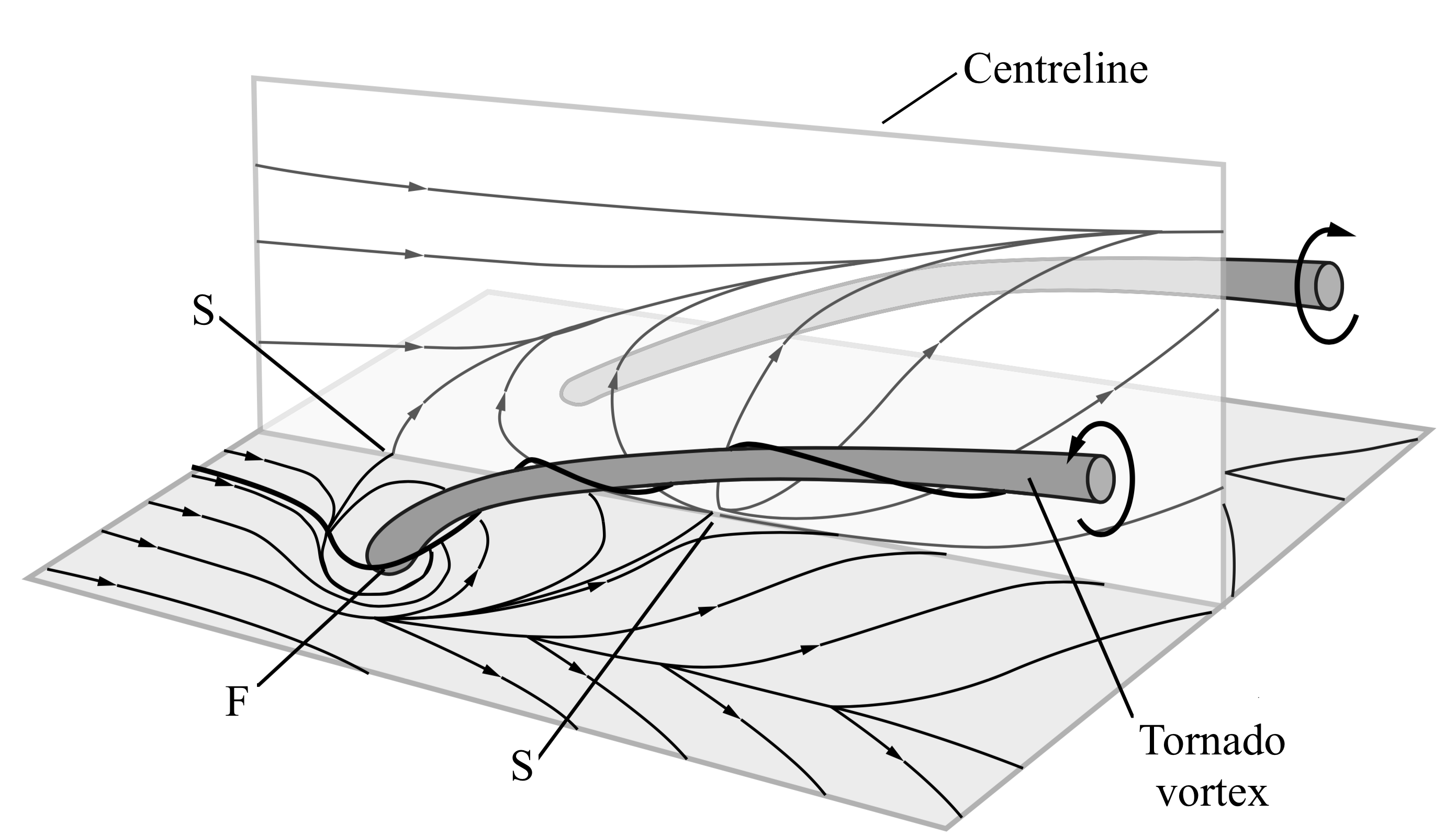}
  \caption{Schematic of the `owl-like' separation of the first kind adapted from \cite{Colliss2016}, based on the work of \cite{1984ZFl}. The front saddle point (S) acts as a separating line at the start of the recirculation bubble. A focus (F) either side of the centreline signifies a tornado-like vortex that lifts fluid away from the surface. The description is consistent with the results presented in figure \ref{fig:2deg_owl_pattern} for the baseline case.}\label{fig:owl_schematic}
\end{centering}
\end{figure}

Figure \ref{fig:2deg_owl_pattern} shows $u$-$w$ velocity streamlines evaluated near the bottom wall at $y=1$. Streamlines are coloured by the transverse velocity component $w$ over a constant colour background. Additional streamlines are added in the corner regions to demonstrate the ejection of the corner flow into the central separation. Streamlines are deflected at $x=150$ as the corner profile thickens, with strong transverse velocity directing the flow towards the centreline on either side. At $x=300$ the streamlines diverge between the saddle point (S) on the reattachment line and the focus (F) within the separation. Flow ejected from the corner spirals into the tornado vortex at each focus, is lifted up from the surface and transported downstream. At the front of the bubble a well defined saddle point (S) is observed; a single streamline is seen entering the saddle point laterally along the separation line from both sides indicating the presence of a surface lifting off the wall. Streamlines adjacent to the separating line are deflected hyperbolically into one of the foci. The pattern agrees well with `owl-like' separations of the first kind introduced by \cite{1984ZFl} as shown in figure \ref{fig:owl_schematic}. There is a noticeable bulge in the reattachment line as the saddle point is shifted downstream at the centre of the span. The shift of the saddle point was less pronounced for the weaker interactions simulated in section \ref{sec:shock_strength}, which were observed to have a reattachment line approximately perpendicular to the downstream flow. We also note the presence of two additional foci located in the near-wall corner region denoted as CF in figure \ref{fig:2deg_owl_pattern}. These satisfy the topological rule that, for a given surface, the number of nodes (nodal points or foci) must exceed the number of saddle points by two \citep{Tobak1982}. Downstream of the central circulation the attached flow follows a smooth laminar profile with streamlines remaining mostly parallel to each other.

\begin{figure}
\begin{centering}
  \includegraphics[width=0.7\textwidth]{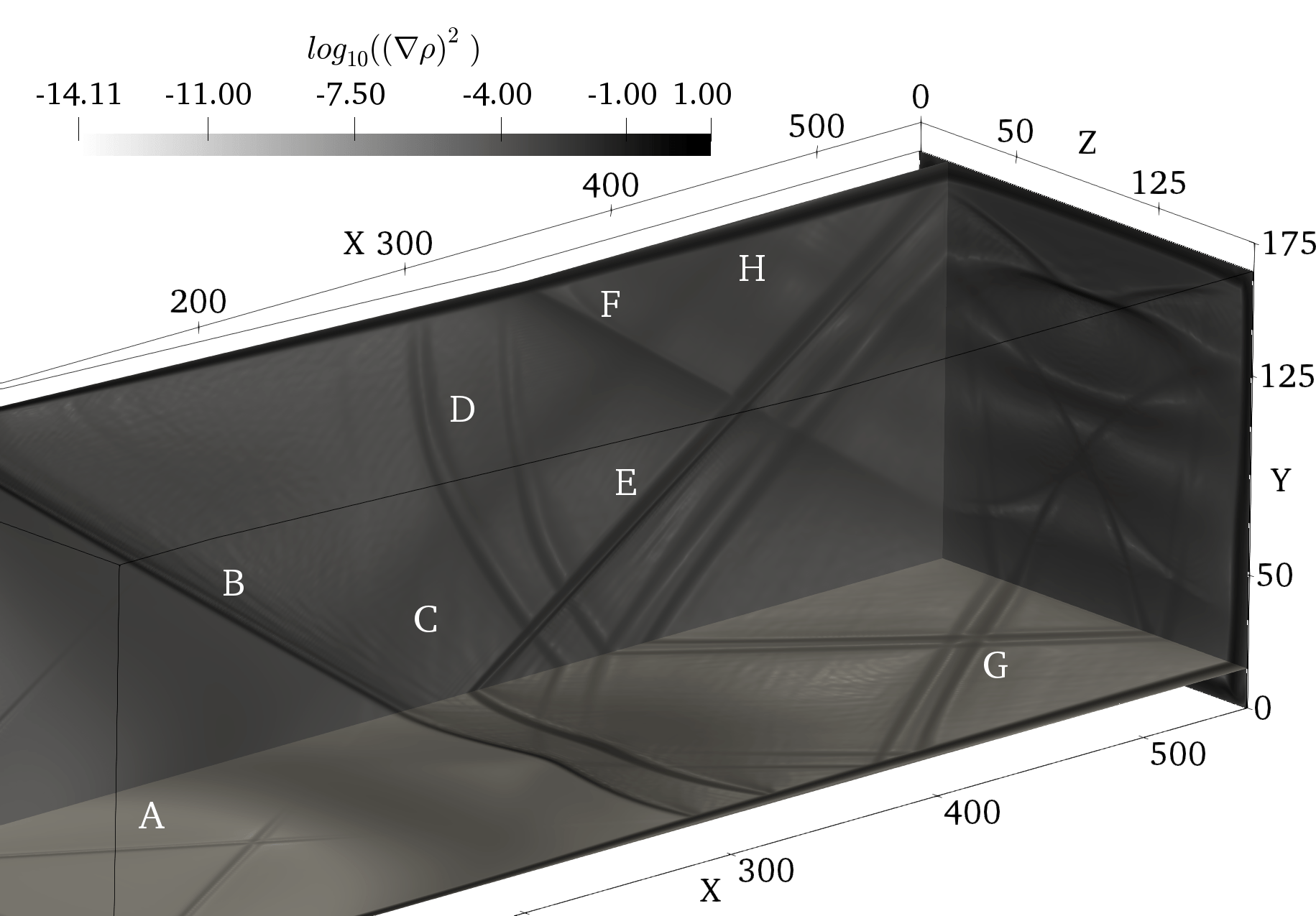}
  \caption{Numerical schlieren of density gradient $log_{10}(\left(\nabla \rho\right)^2)$ showing the complex shock structure downstream of the three-dimensional SBLI at $AR=1$. Three intersecting slices are shown at $x=550$, $y=15$ and $z=15$. Notable features include: (A) Compression waves from the initial sidewall boundary layer development. (B) Main incident shock. (C) Compression waves from the start of the central separation. (D) Two conical shocks from the corner of the shock generator. (E) Expansion fan formed from the reflection of the incident shock. (F) Trailing edge expansion fan. (G) First crossing point of the reflected conical shocks. (H) Secondary reflection of the central compression waves.}\label{fig:shock_patterns3d}
\end{centering}
\end{figure}

Computing the logarithm of density gradient magnitudes $log_{10}(\left(\nabla \rho\right)^2)$ is an effective way of numerically detecting shock structures, providing a more sensitive version of the schlieren photography techniques found in experiments. Figure \ref{fig:shock_patterns3d} shows a numerical schlieren of three intersecting orthogonal slices evaluated at $x=550$, $y=15$ and $z=15$. The main shock structures are identified as follows: (A) Weak compression waves from the initial development of the imposed boundary layer profile, coalescing into weak intersecting shocks. (B) The trace of the incident swept shock through the $z=15$ plane with curvature visible at $y=15$. (C) Reflected compression wave caused by the thickened boundary layer at the base of the SBLI. (D) Reflections of the conical shocks generated at the corner of the sidewalls and shock generator ramp, discussed in more detail later in this section. (E) Expansion fan developing as the flow turns over the apex of the recirculation bubble. (F) The trailing expansion fan generated at $x = x_{sg} + L_{sg}$. (G) Crossing of the conical shocks after reflection from the sidewalls. The crossing point is the visible kink at $x=500$ in the $C_f$ plot of figure \ref{fig:Fig7} (a) (solid line). (H) A secondary reflection of the compression wave (C) as it reaches the upper boundary layer. It is clear that the numerical schlieren is a more sensitive metric for detecting shock structures than the dilatation of figure \ref{fig:Fig9}. Although there is a subtle shading upstream of the incident shock in the $y$ plane, we do not identify significant corner shocks as suggested by \cite{Xiang2019} for a stronger turbulent interaction. This may be due to the turbulent boundary layer in the experiments, but we note that such corner shocks were also not clearly visible in \cite{wang_sandham_hu_liu_2015}. We highlight that even the compression waves resulting from the streamline curvature caused by boundary layer development at (A) are more prominent than the corner compression. By far the strongest structures seen downstream of the SBLI are the reflecting conical shocks of (D) and (G), consistent with the turbulent case of \cite{wang_sandham_hu_liu_2015}.

\begin{figure}
\begin{centering}
  \includegraphics[width=0.45\textwidth]{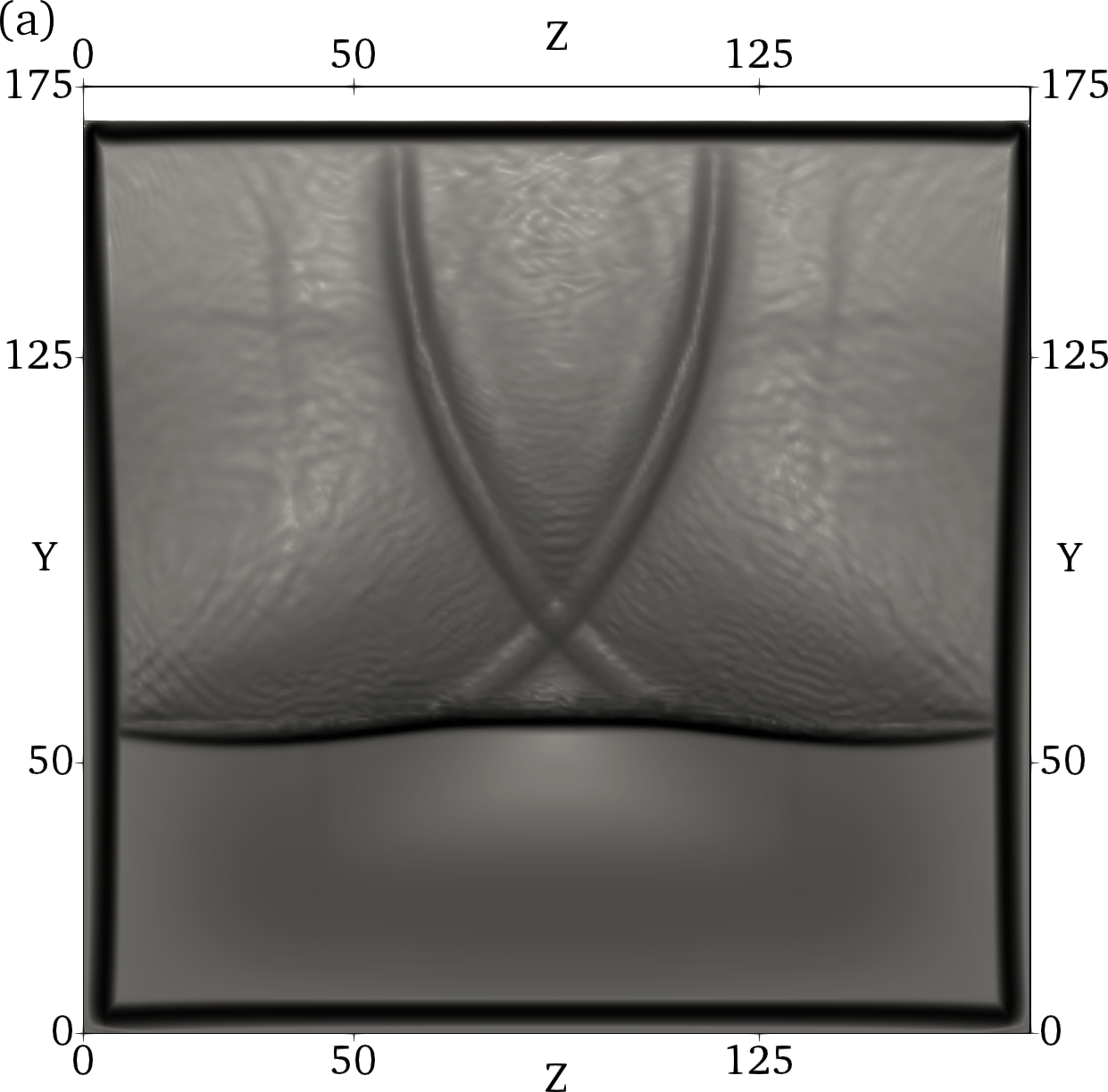}
\includegraphics[width=0.45\textwidth]{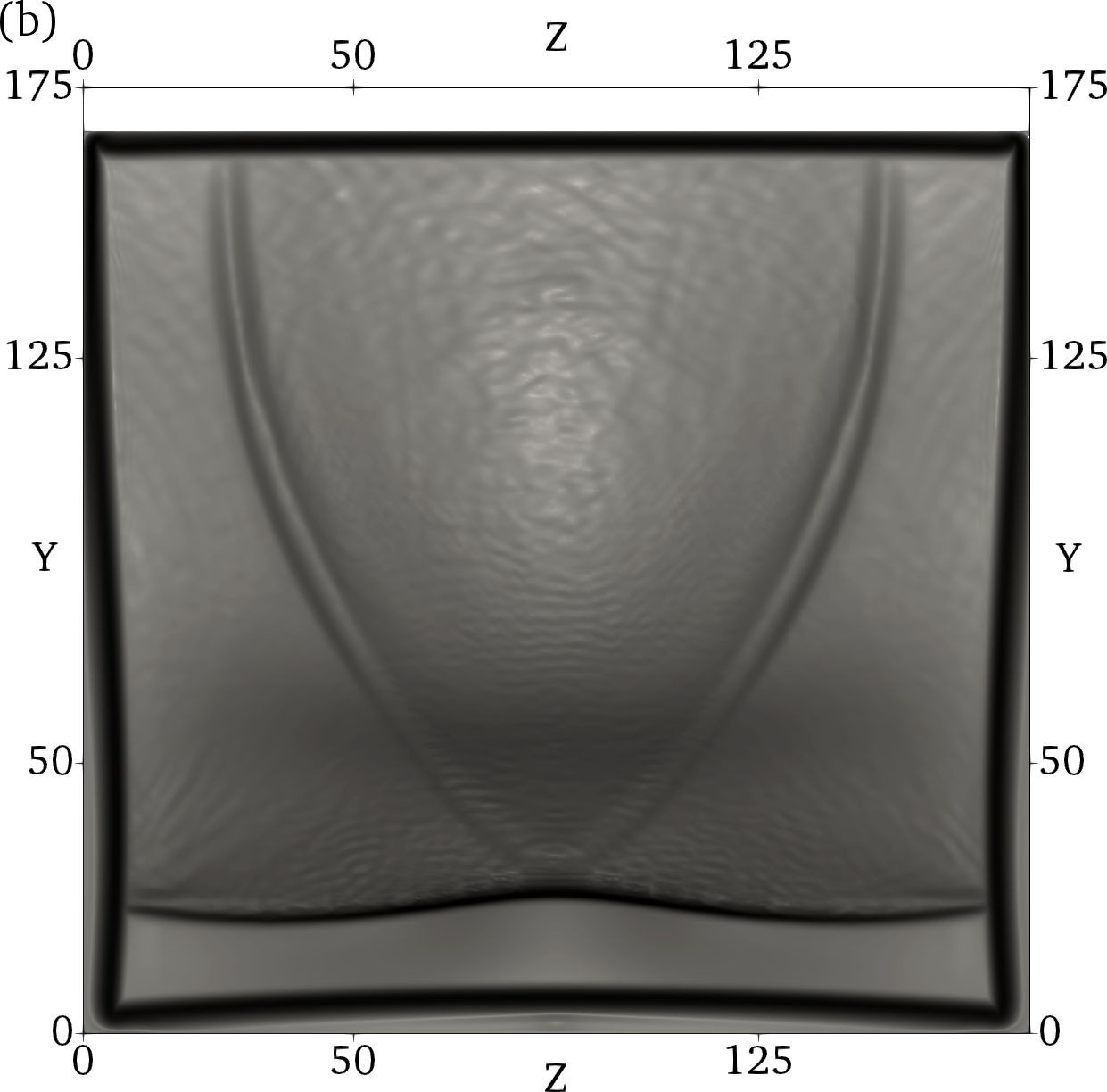}\\
\includegraphics[width=0.45\textwidth]{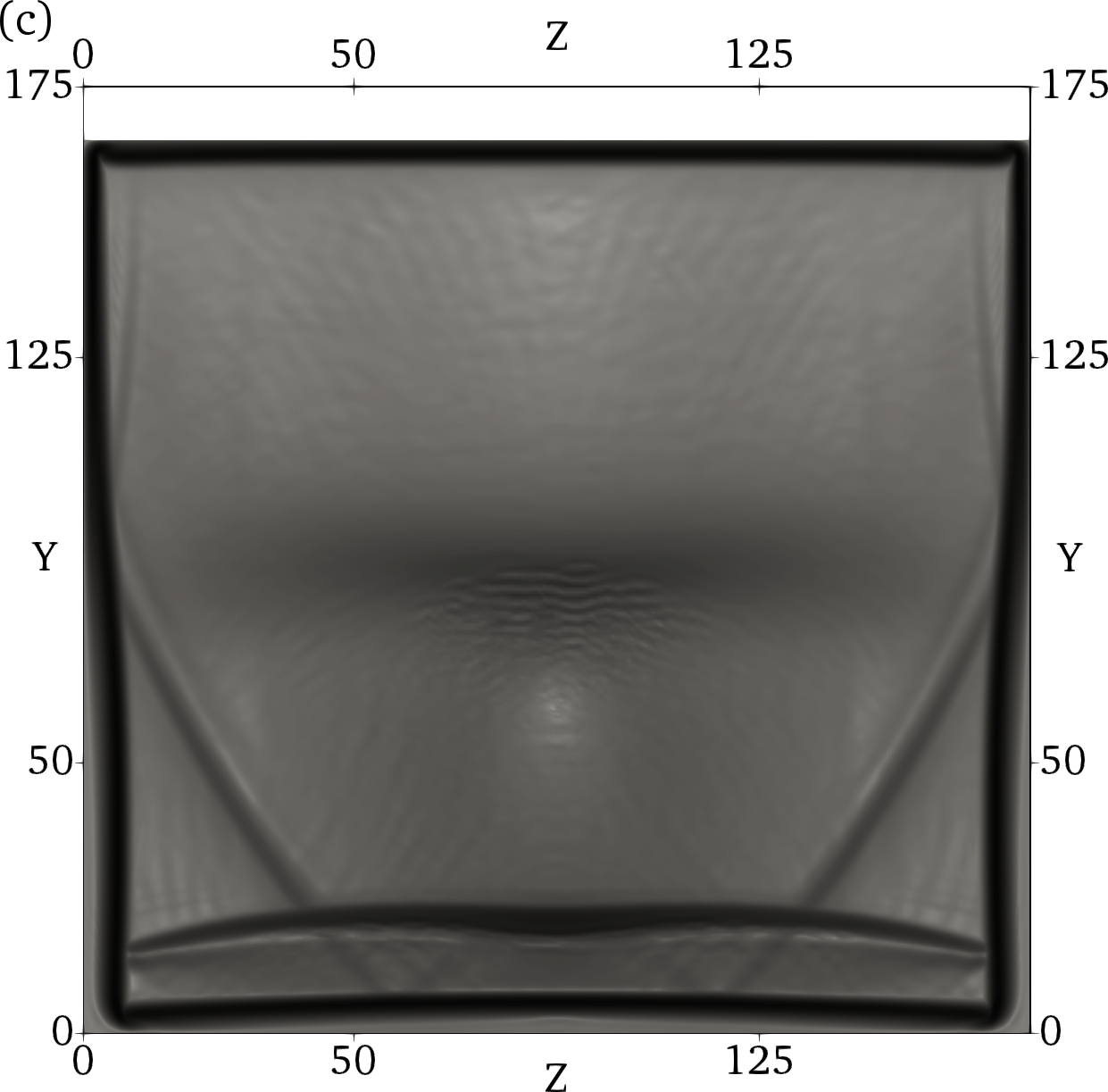}
\includegraphics[width=0.45\textwidth]{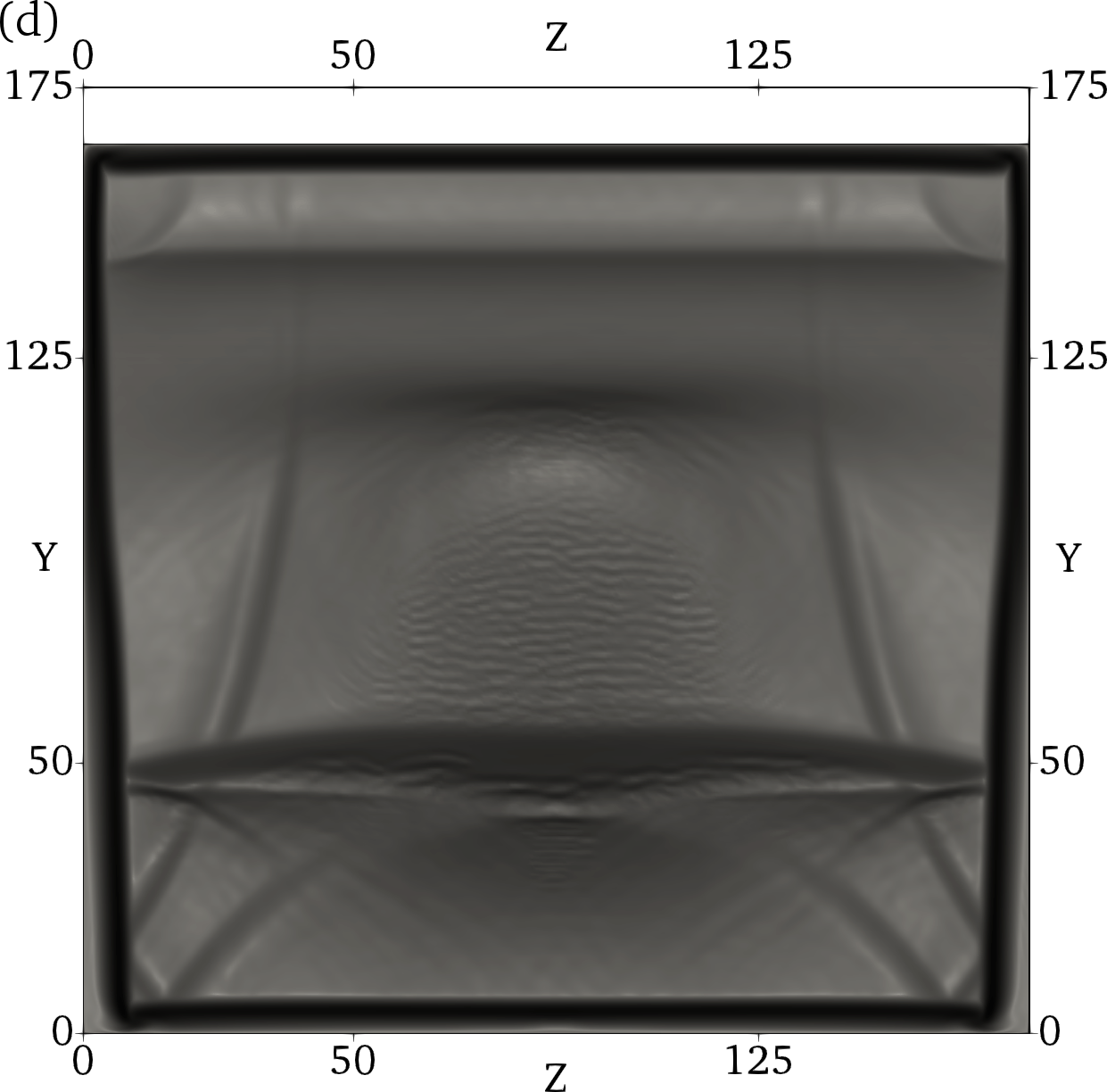}
  \caption{Numerical schlieren density gradient $log_{10}(\left(\nabla \rho\right)^2)$ showing the streamwise development of the conical swept SBLI. $y-z$ slices are displayed at streamwise locations (a) $x=225$ (b) $x=275$ (c) $x=325$ and (d) $x=375$. Two conical shocks generated by the initial swept SBLI in the upper left and right corner cross through each other in (a) and (b), before reflecting off the bottom wall and opposite sidewall in (c) and (d). The dark horizontal line in (a) and (b) is the main incident shock, while in (c) and (d) it is the expansion after the interaction. The start of the trailing edge expansion fan can also be seen in the upper region of (d).}\label{fig:2d_conical_shock}
\end{centering}
\end{figure}

The role of the reflected conical shocks (D) is clearer to see when looking at the $y$-$z$ slices at different $x$ locations in figure \ref{fig:2d_conical_shock}. The slices show the same numerical schlieren as before, this time located at $x=\left(225, 275, 325, 375\right)$ in (a)-(d). The conical shocks are generated in the upper left and right corners of the plane at the intersections between the shock generator ramp and the two sidewalls. Stepping forward in $x$ along the duct, the conical shocks expand outwards from their starting location. In figure \ref{fig:2d_conical_shock} (a) and (b) the conical shocks are seen to intersect each other and continue towards the opposite sidewall. The thin dark line is the $y$-$z$ projection of the incident shock which is attached to the conical shock fronts. At the edges of (a) and (b) we see that the boundary layer has thickened as a result of the swept SBLI. Figures \ref{fig:2d_conical_shock} (c) and (d) show the impact of the conical shocks on the opposite wall to that which they originated as well as a dark horizontal line marking the expansion above the apex of the main separation bubble. The upper and lower portions of the shock front reflect first in (c) and can be seen to be propagate back towards their starting sidewall in (d). Comparing these slices again to the $x=550$ end slice of figure \ref{fig:shock_patterns3d} we can see that the conical shocks trace a path across the span and reach their original sidewall near the outlet. The slice in figure \ref{fig:2d_conical_shock} (d) also shows the start of the trailing edge expansion fan generated on the upper surface at $x_{sg}+L_{sg} = 345$. An extensive search was performed using various forms of density gradients, pressure, and dilatation, but no obvious corner compressions were observed near the bottom of the duct. It is possible that there are fundamental differences between simulation and experiment causing the discrepancy, or that the laminar interaction is simply too weak to generate strong corner compressions. Many experimental configurations also feature gaps between the sidewall and the shock generator, which would lead to a weakened swept SBLI effect compared to enclosed ducts. For the present work the dominant structures crossing the centreline originate from the conical shocks generated between the upper ramp and sidewalls.

\section{Parametric sensitivity}

\subsection{The effect of duct aspect ratio}\label{sec:aspect_ratio}

\begin{table}
  \begin{center}
  \begin{tabular}{cccccc}
      Aspect ratio $\left(W/H\right)$  & $p_3/p_1$ & Interaction $\left(x_{sep}, x_{reattach}\right)$ & $L_{sep}$& \% of baseline $L_{sep}$ & $L_{f}$ \vspace{0.1cm}\\ 
      1/4  & 1.439 &  (167.5, 199.9) & 32.43  &  21\% & -\\
      1/2  & 1.325  & (185.0, 301.8) & 116.80 & 74\% & 28.0 \\
      1  & 1.313  & (207.7, 365.6) & 157.93 &  - & 42.1 \\
      2  & 1.306  & (231.8, 388.2) & 156.42 &  99\% & 45.3 \\
      4  & 1.321 & (243.6, 375.8) & 132.15 & 84\% & 49.0 \\
      Span-periodic & 1.300 & (251.2, 371.4) & 120.12 & 76\% & - \\
  \end{tabular}
  \caption{The effect of aspect ratio on the baseline $\theta_{sg} = 2^{\circ}$ shock generator case. Comparison is also made to a span-periodic simulation without sidewalls demonstrating the strengthened three-dimensional interaction. Separation length is shown as a percentage of the one-to-one aspect ratio sidewall case. $L_{sep}$ is the separation length along the centreline and $L_{f}$ is the distance between the foci and the sidewalls.}
  \label{tab:aspect_ratios}
  \end{center}
\end{table}

The aim of this section is to determine how laminar SBLI are affected by a varying degree of flow confinement for the selection of narrow and wide duct configurations given in table \ref{tab:grid_sizes}. For each case the upstream flow conditions and shock strength are held constant to the $\theta_{sg} = 2^{\circ}$ one-to-one aspect ratio baseline case in the previous section. Aspect ratios ranging from one-quarter to four are considered, with the largest aspect ratio expected to show strong two-dimensional behaviour on the centreline. By comparing the span-periodic result to the larger aspect ratios, an estimation can be made of how wide a duct must be before span-periodicity is a valid assumption. The narrow aspect ratio cases will assess whether the observed strengthening of the SBLI due to sidewalls continues in the presence of even stronger flow confinement.

\begin{figure}
  \includegraphics[width=0.50\textwidth]{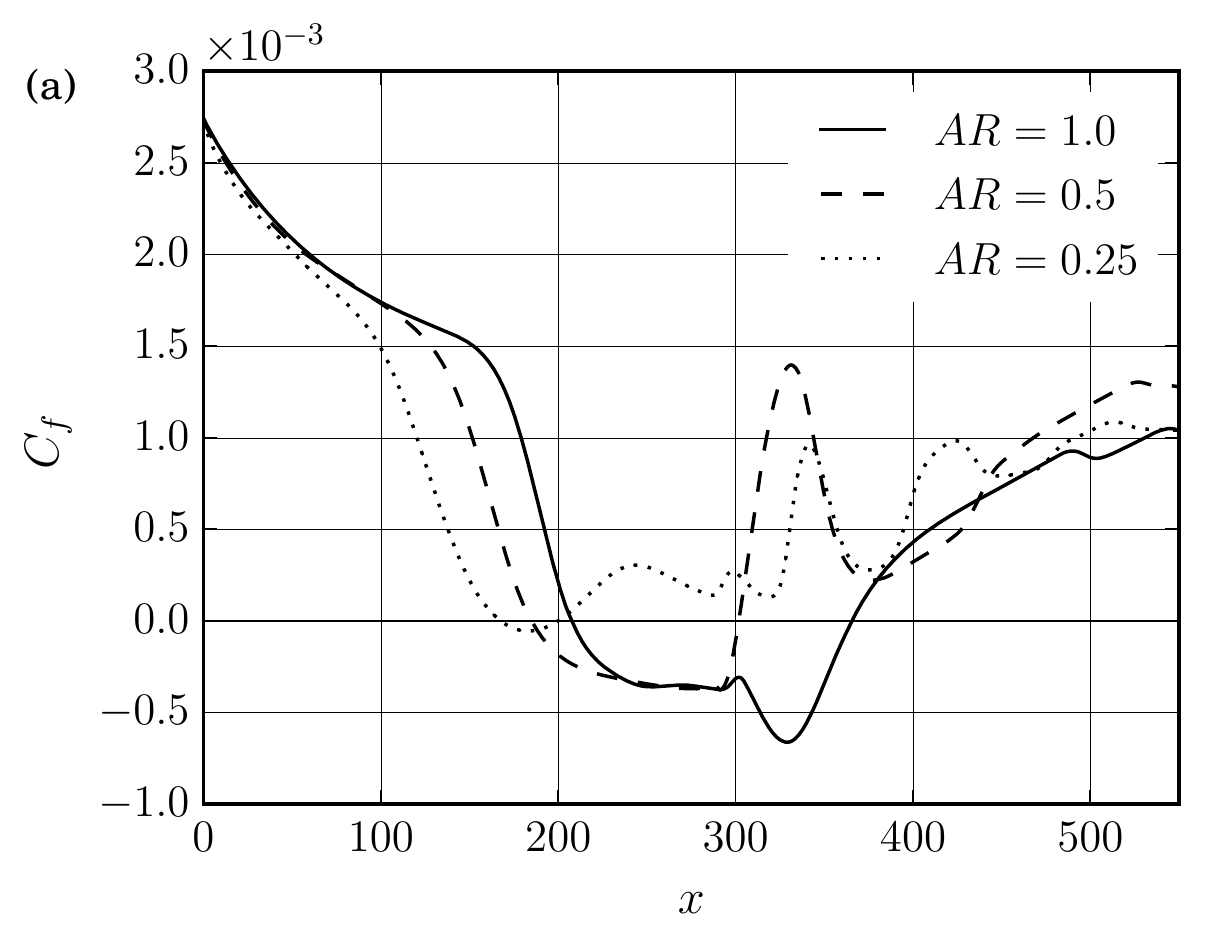}
  \includegraphics[width=0.50\textwidth]{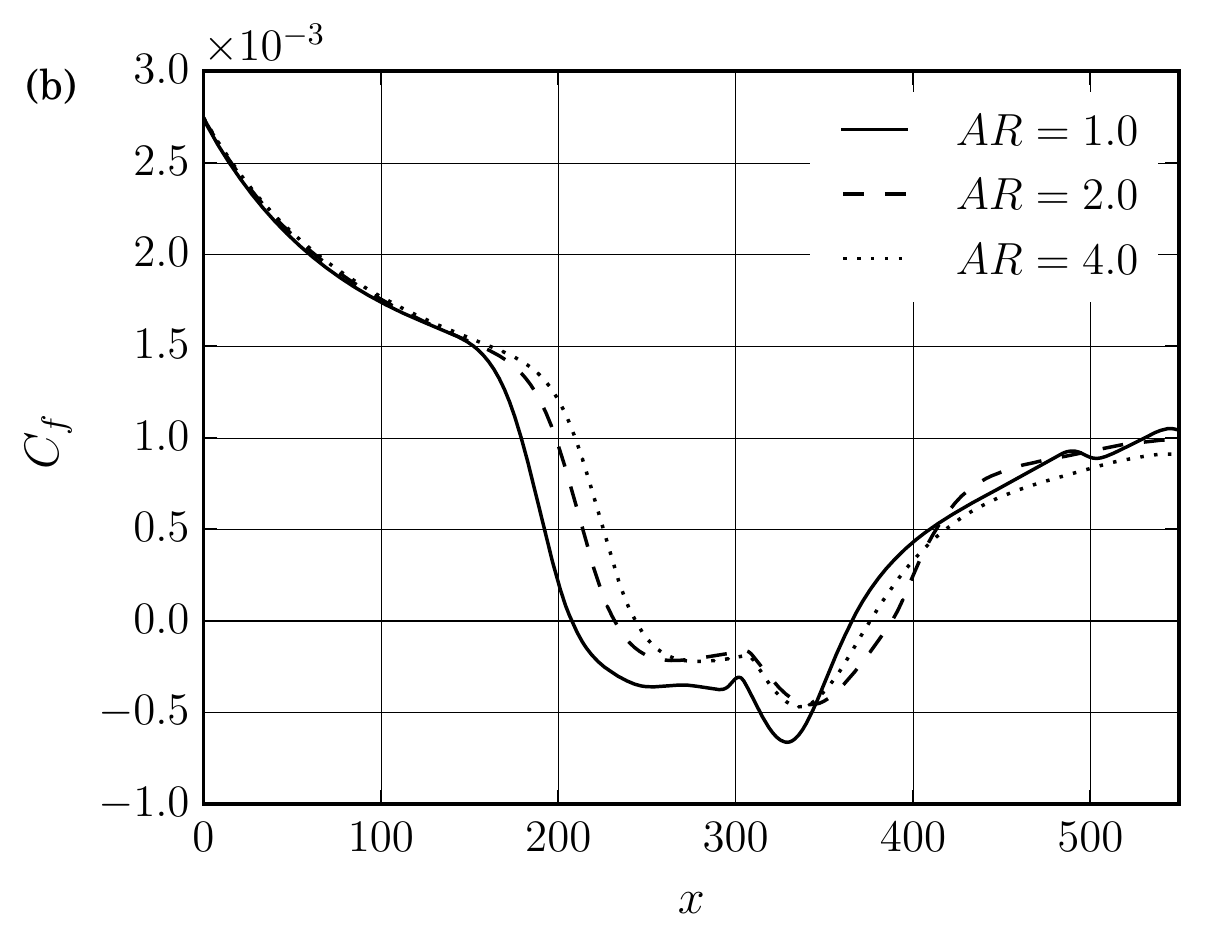}
  \caption{The effect of varying aspect ratio on the centreline bottom wall skin friction in the case of (a) narrowing and (b) widening aspect ratios. In each case the skin friction is compared to the one-to-one aspect ratio baseline duct (solid line).}
\label{fig:Fig10}
\end{figure}

Figure \ref{fig:Fig10} shows the centreline skin friction along the bottom wall for (a) narrow and (b) wide aspect ratios. In each of the two plots the solid line represents the one-to-one aspect ratio case from figure \ref{fig:Fig7} (a). For the narrow ducts in figure \ref{fig:Fig10} (a) a severe reduction in central separation length is seen and the separation point has noticeably been shifted upstream. For an aspect ratio of $AR=0.25$ the flow is almost entirely attached; the separation bubble from the initial SBLI covers only $5\%$ of the streamwise duct length. At $AR=0.5$ the expected asymmetric double-trough profile of a laminar separation bubble \citep{Katzer1989} is also not seen since the flow abruptly reattaches at the back of the separation bubble. The most notable features of the narrower aspect ratios are the multiple peaks in skin friction downstream of the initial interaction. These correspond to the successive crossings of the incident swept-interaction that was highlighted in figure \ref{fig:shock_patterns3d}. As the width of the duct has been reduced, the waves reflecting between the sidewalls have less distance to travel and cross the centreline multiple times before reaching the outlet. Rather than a further strengthening of the interaction, aspect ratios below unity are seen to suppress the interaction and exhibit shock-trains that traverse the span of the domain. 

More regular behaviour is found for the larger aspect ratios in figure \ref{fig:Fig10} (b), in which the three-dimensionality of the sidewall flow has less of an impact on the centreline dynamics. Consistent with the narrower cases, an increase in aspect ratio causes a downstream shift of the separation point. The same is not true for the reattachment location however, with the widest $AR=4$ case reattaching before $AR=2$. The kink in $C_f$ at $x=500$ for $AR=1$, due to crossing of the sidewall reflections is not seen for larger aspect ratios, the wider span results in the crossing occurring downstream of the computational domain. Table \ref{tab:aspect_ratios} shows the strong dependence of aspect ratio on the centreline SBLI. The third column gives the separation and reattachment locations for each of the aspect ratios. As seen in figure \ref{fig:Fig10} the downstream shift of the separation location is a consistent trend each time the aspect ratio is widened, with the largest $AR=4$ case still farther downstream than for an infinite span. 
\begin{figure}
  \includegraphics[width=0.49\textwidth]{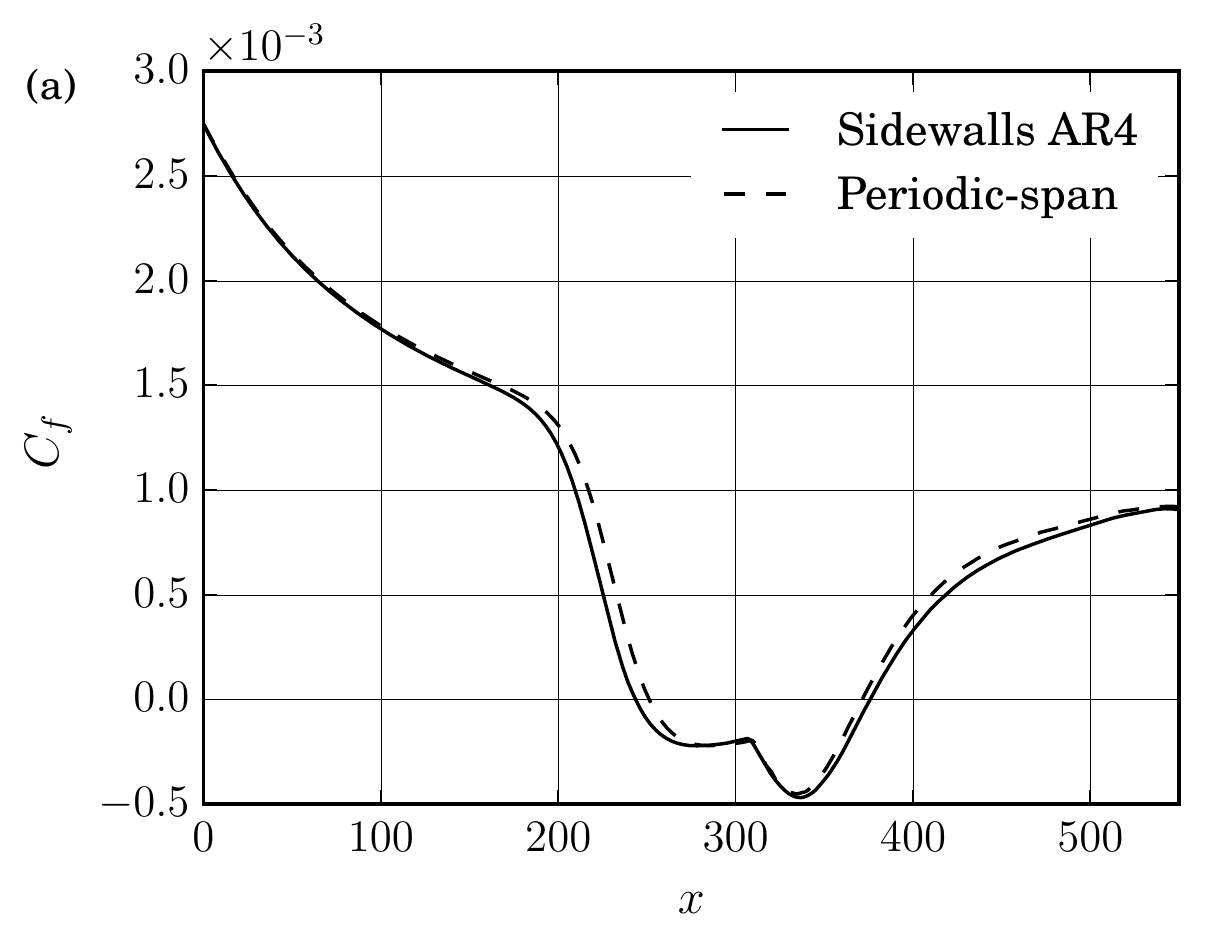}
  \includegraphics[width=0.50\textwidth]{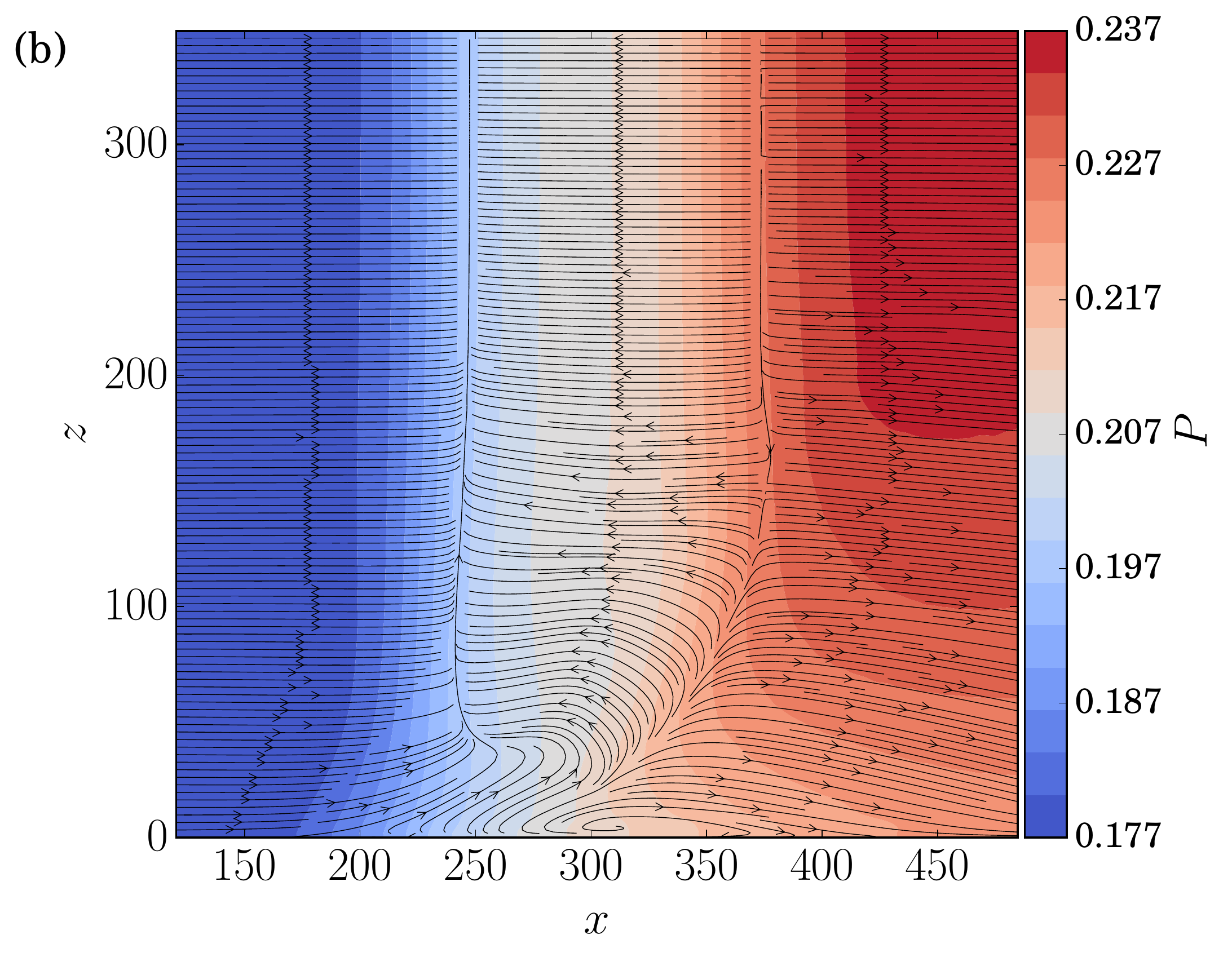}
  \caption{(a) Centreline skin friction comparison of the $AR=4$ wide duct with sidewalls to an infinite span. (b) Velocity streamline pattern above the bottom wall for the $AR=4$ duct, coloured by pressure. Half of the span is shown $z = \left[0, 350\right]$ due to the centreline symmetry.}
\label{fig:Fig11}
\end{figure}
For the $AR=1$ and $AR=2$ cases the centreline $L_{sep}$ differs by only 1\%, despite the downstream shift of the interaction at the higher aspect ratio. At the two smallest aspect ratios the separation length was reduced to 21\% and 74\% of the $AR=1$ result. Above $AR=2$ the bubble decreases in size as three-dimensional effects become less important to the central flow. Figure \ref{fig:Fig11} (a) shows a comparison of the widest duct with sidewalls to an infinite span. The shape of the skin friction distribution agrees well at an aspect ratio of four, although $L_{sep}$ is still 9\% longer than for the idealised infinite span. The extent to which the corner interaction affects the centreline flow is of interest when assessing the viability of the infinite span assumption for modelling SBLI.

Figure \ref{fig:Fig11} (b) shows velocity streamlines coloured by pressure above the bottom wall. Half of the span is displayed, as the flow is symmetric about the centreline. The near wall structures are similar to those seen for the narrower aspect ratio in figure \ref{fig:Fig8} (a); an attached region of flow is turned inwards by the swept shock and feeds into the central separation bubble. Foci are seen in the corner and at the edge of the central separation, but they are not as pronounced as in the $AR=1$ case. The bubble is longest in the streamwise direction at roughly 20\% of the span away from the sidewall. Separation length then reduces to a constant value at $z\approx 200$, which is maintained as the flow becomes two-dimensional towards the centreline. Beyond this point the streamlines are seen to be anti-parallel to the oncoming flow. The sidewall influence causes visible deflection of the streamlines over almost 30\% of the span, which would explain the strong dependence of aspect ratio on $L_{sep}$ for $AR=2$ and narrower. For example at $AR=2$ the centreline is located at $z=175$, within the range of influence seen here. These results are consistent with the experimental laminar SBLI of \cite{Degrez1987}, in which an aspect ratio of at least 2.5 was required to see two-dimensional behaviour of the interaction.

\begin{figure}
  \includegraphics[width=0.49\textwidth]{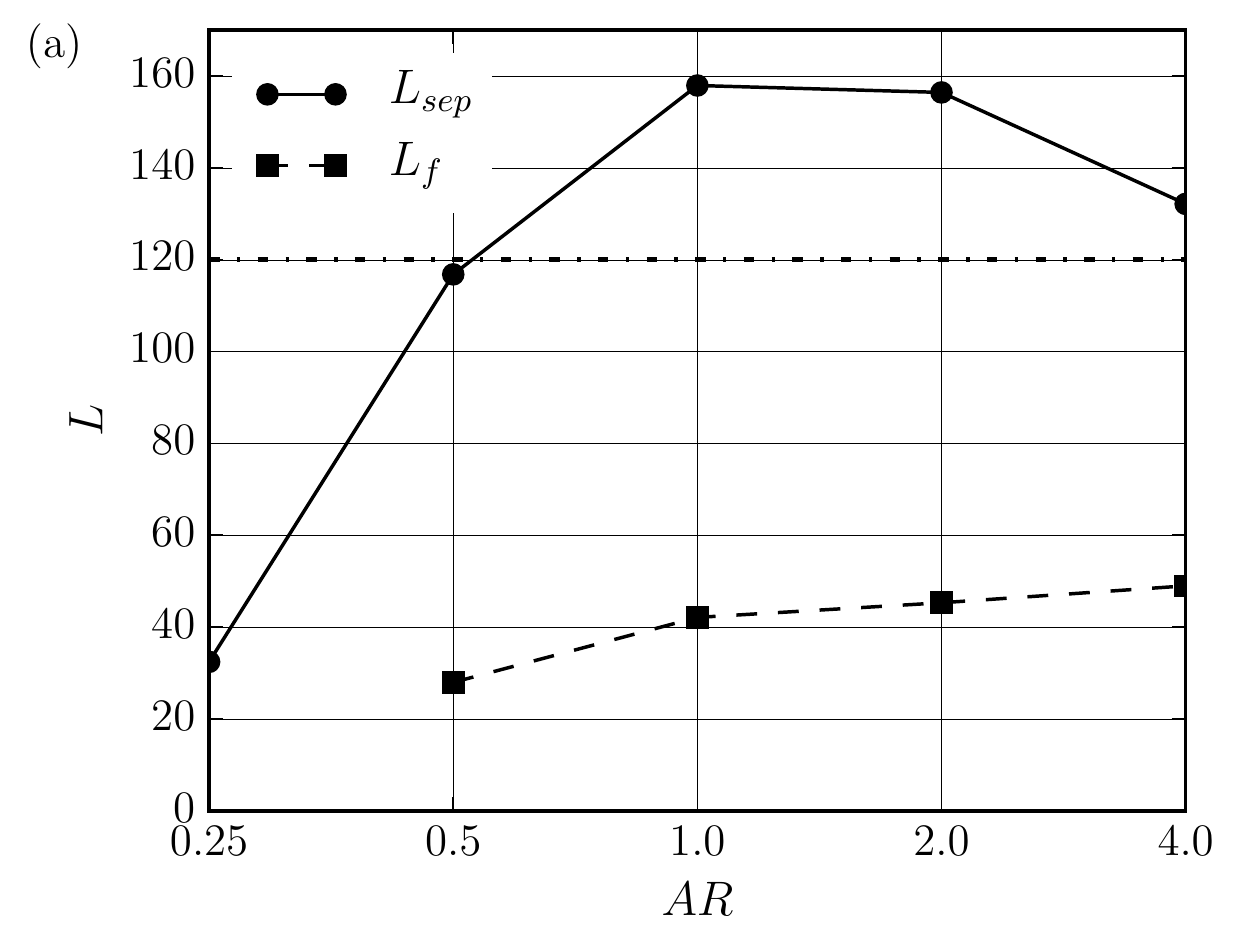}
  \includegraphics[width=0.50\textwidth]{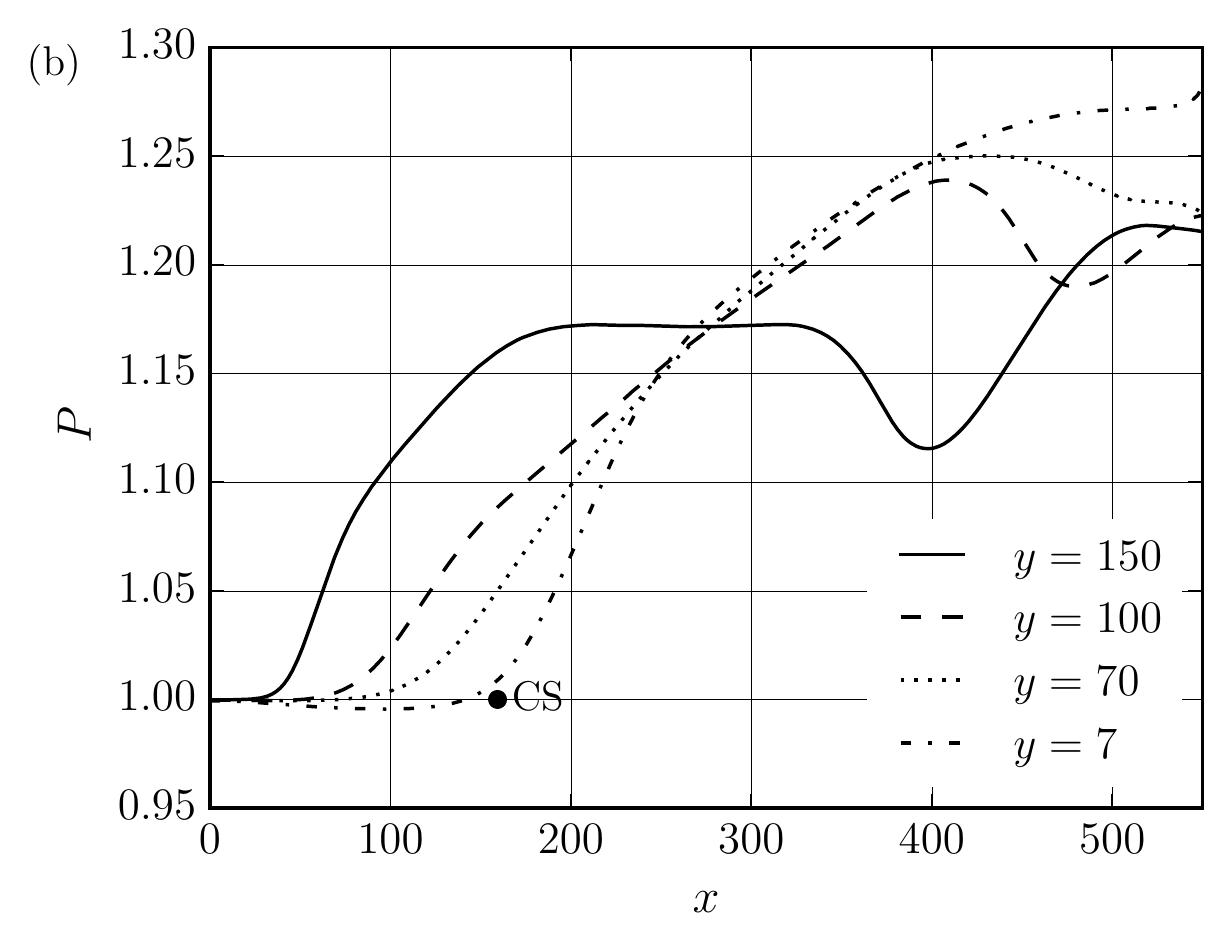}
  \caption{(a) The effect of duct aspect ratio on centreline streamwise separation length $L_{sep}$ and the distance between the foci and sidewalls $L_{f}$. The dashed horizontal line denotes $L_{sep}$ for the span-periodic case. (b) Normalized pressure on the sidewall ($z=0$) at increasing $y$ heights above the interaction. The `CS' marker is the start of corner separation near the bottom wall.}
\label{fig:sidewall_pressure_lines}
\end{figure}

Figure \ref{fig:sidewall_pressure_lines} (a) compares the effect of aspect ratio on the streamwise separation length $L_{sep}$ and the lateral distance in $z$ between the foci and the sidewall $L_{f}$. Similar trends are found to those in the experimental literature (\cite{Babinsky2013}, \cite{Xiang2019}); smaller aspect ratios lead to suppression of the central separation compared to the quasi-2D result denoted by the dashed line. At medium aspect ratios a peak occurs that is in good agreement to figure 11 of \cite{Babinsky2013} and asymptotes towards the quasi-2D result at the largest aspect ratio. The distance of the foci from the sidewall $L_{f}$ increases with aspect ratio, noting that at the smallest aspect ratio of $AR=0.25$ a clear focus could not be identified. The distance the foci shift is small relative to the width of the duct. Increasing the aspect ratio from one to four led to an increase of only 16.5\% in $L_{f}$, suggesting that the ejection of the corner flow remains mostly localised to the sides of the duct at large aspect ratios.

\begin{figure}
\begin{centering}
  \includegraphics[width=0.49\textwidth]{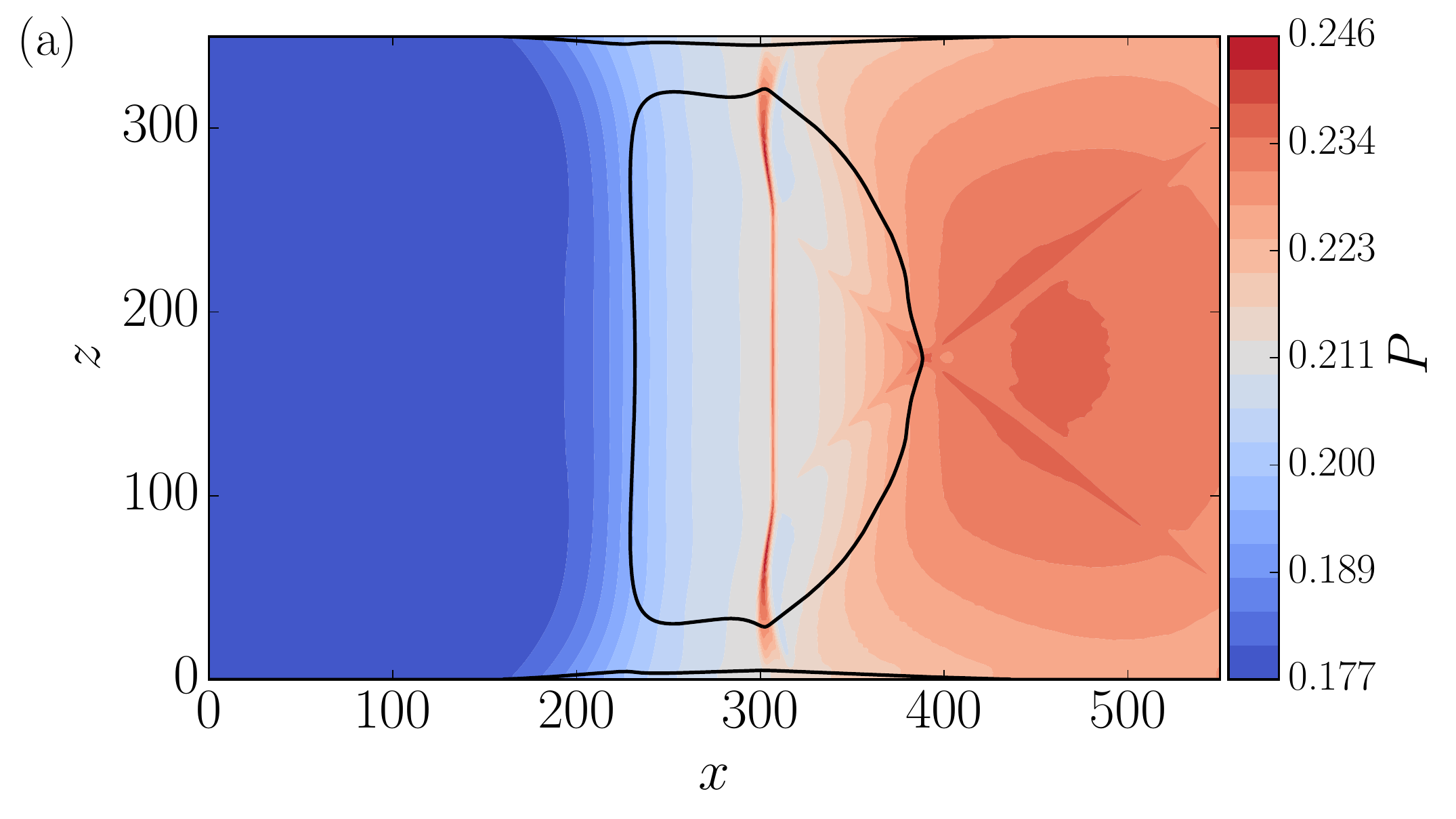}
  \includegraphics[width=0.50\textwidth]{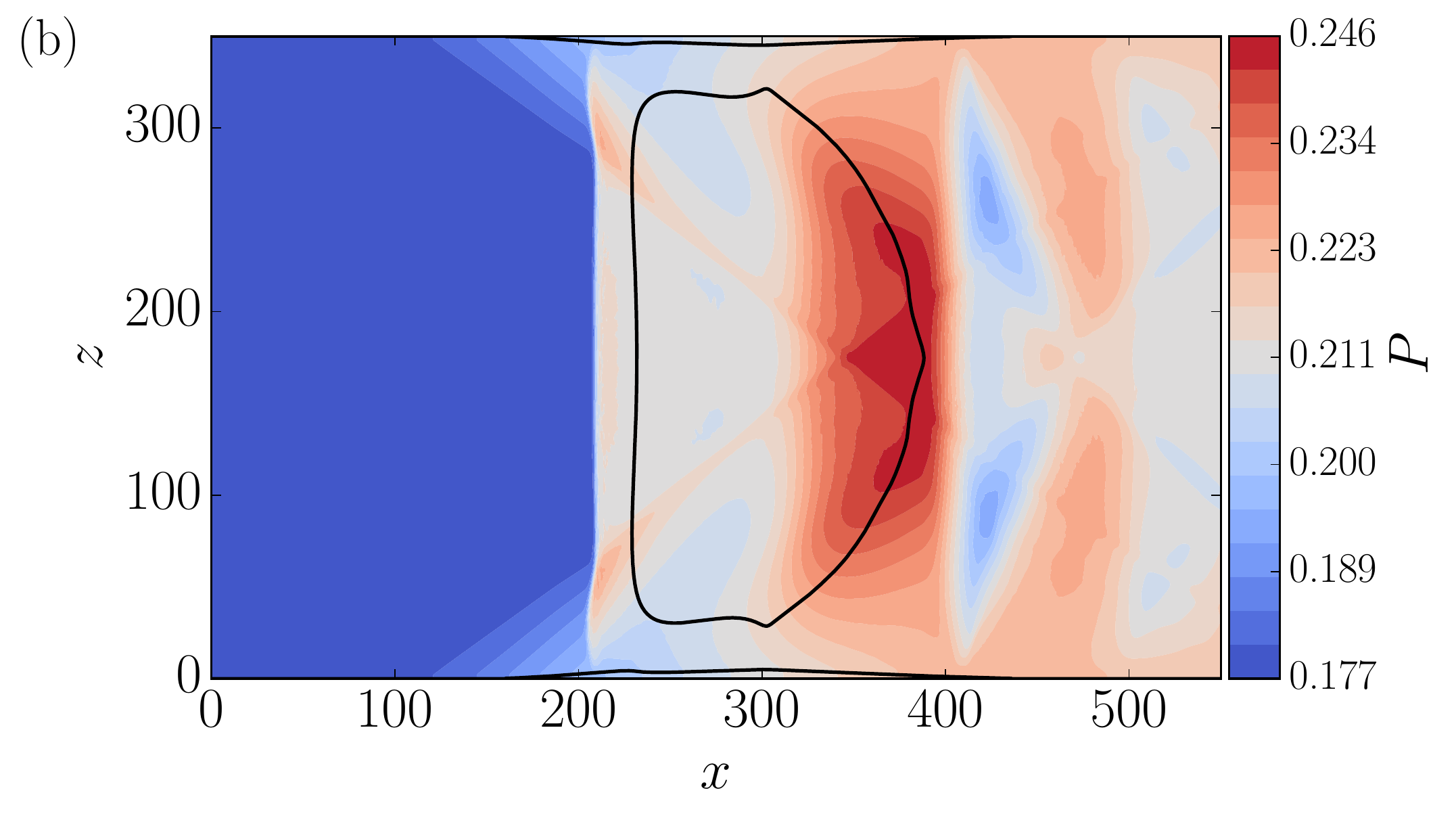}
  \caption{Slices of $x$-$z$ pressure for the $AR=2$ case at (a) $y=7$ above the separation bubble and at (b) $y=70$, corresponding to $40\%$ of the duct height above the bottom wall. In both cases the black line is the zero crossing of skin friction on the bottom wall ($y=0$). The crossing shocks are observed to be generated by sidewall compressions from the initial swept SBLI, independent of the onset of corner separation near the bottom wall.}\label{fig:shock_crossing_contours}
\end{centering}
\end{figure}

A method for predicting the central separation size in duct SBLI was proposed by the experimental turbulent studies of \cite{Babinsky2013}, \cite{Xiang2019}. The key criteria was proposed to be the streamwise crossing location of shockwaves generated at the onset of corner separation. Corner shocks crossing upstream of the central interaction led to a weakened SBLI compared to quasi-2D predictions. For shocks crossing within the interaction a strengthened SBLI and larger separation bubble were observed. At larger aspect ratios the flow was two-dimensional on the centreline, as the shocks crossed far downstream of the interaction. Figure \ref{fig:shock_crossing_contours} (a) shows pressure contours for the $AR=2$ case in the present study at a height of $y=7$. This height corresponds to the apex of the central separation bubble, with the imprint of the incident shock seen at $x=300$ in this plane. The solid black line represents the $C_f=0$ crossing on the bottom wall of the domain ($y=0$), to serve as a reference point for the separation bubble. Two oblique structures are seen to traverse the span and cross at the back of the central interaction. As the $AR=2$ case has the largest recorded $L_{sep}$, this crossing is consistent with the trends noted in the experimental work. The same experimental trends were observed for the other aspect ratios in this work (not shown), with the smaller aspect ratio cases having structures that crossed upstream of the interaction. 

Experimental work has attributed these crossing shockwaves to a bottom wall corner compression effect. However, in the present simulations, observing pressure contours higher up in the duct at $y=70$ as in figure \ref{fig:shock_crossing_contours} (b), they are seen to originate from the swept SBLI of the incident shock with the sidewalls. It is important to note that the sidewall compressions leading to the crossing shockwaves occur at an earlier upstream location than the onset of bottom wall corner separation. At $x=205$ the two conical shocks interact with the incident shock and cause it to strengthen, consistent in shape with figure 9 of \cite{wang_sandham_hu_liu_2015}. The streamwise development of pressure on the sidewall is given at four $y$ heights in figure \ref{fig:sidewall_pressure_lines} (b), relative to the start of the corner separation (CS). While there is a pressure rise close to the onset of corner separation at $y=7$, this effect is also present all the way up the height of the duct due to the swept SBLI and occurs upstream of the corner separation.

This section has reaffirmed that aspect ratio is a crucial parameter for confined SBLI, highlighting that span-periodicity is not a suitable assumption for internally confined flows even for relatively wide ($AR=2$) ducts. The flow at $AR=4$ still showed small differences compared to span-periodic predictions. The general impact of aspect ratio on central separation length agreed well with experimental findings. We were, however, unable to attribute this effect to the crossing of shocks from bottom wall corner compressions. It was shown that for the present study the origin of the crossing shockwaves was the initial swept conical SBLI as previously identified in figure \ref{fig:2d_conical_shock}. This effect was shown to occur at a height above the influence of bottom wall corner compressions, and substantially further upstream than any corner separations. As a result, the crossing location is dependent on the height $y$. There are several differences in this study to the experimental work, which could be causing the difficulty in identifying strong corner shocks. The present study is laminar and far weaker ($\theta_{sg} = 2^{\circ}$ vs $\theta_{sg} = 8^{\circ}$), plus there is no gap between the shock generator and the sidewalls which would emphasise the swept conical SBLI relative to weak corner compressions. Nevertheless, the present results are consistent with previous LES \citep{wang_sandham_hu_liu_2015} of turbulent interactions with sidewalls. 

\subsection{Variation of incident shock-strength}\label{sec:shock_strength}
\begin{figure}
  \includegraphics[width=0.50\textwidth]{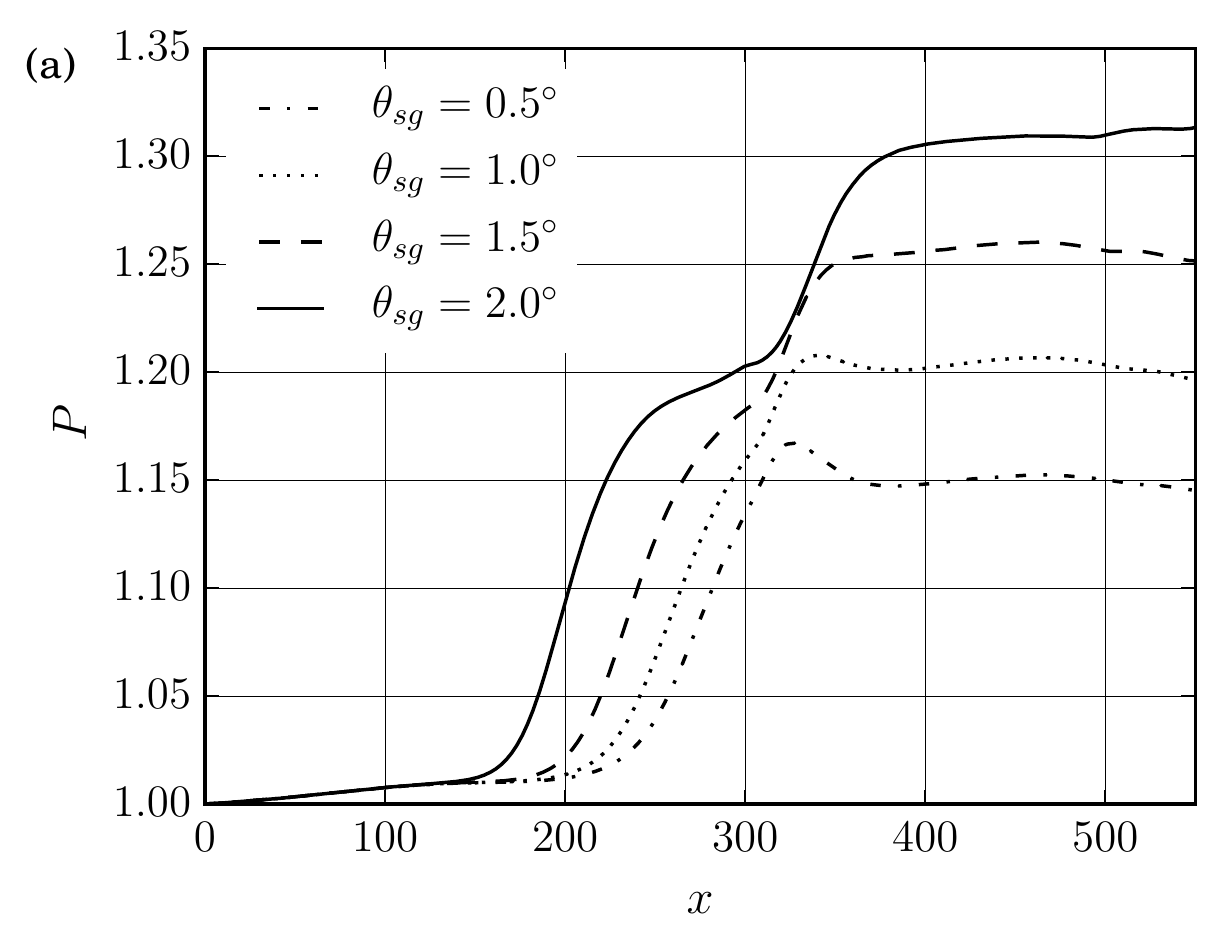}
  \includegraphics[width=0.50\textwidth]{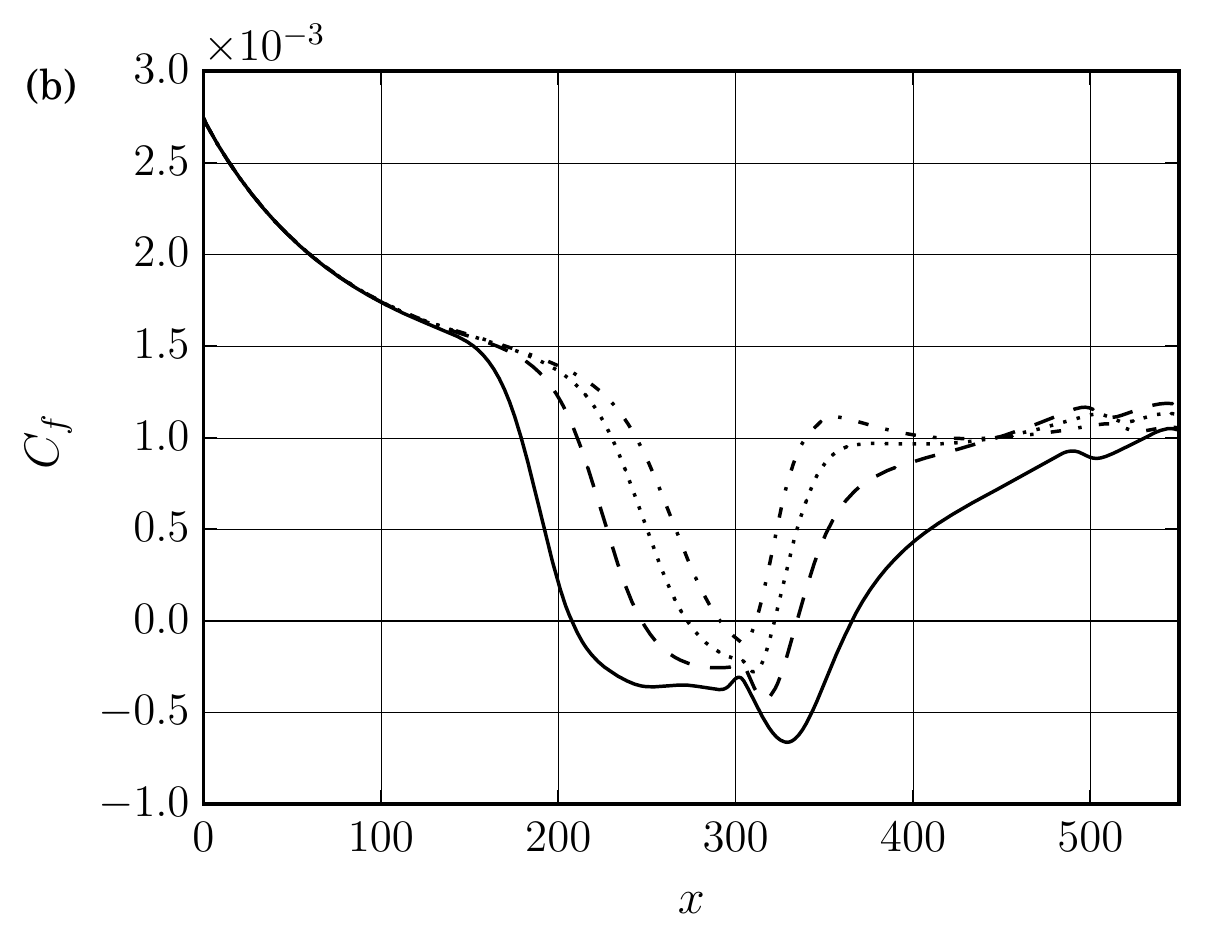}
  \caption{Sensitivity of the centreline (a) wall pressure and (b) skin friction to incident shock strength for the $AR=1$ duct. The solid line represents the baseline configuration.}
\label{fig:shockcompare_line}
\end{figure}

 In this section the initial flow deflection $\theta_{sg}$ is modified to determine the sensitivity of the SBLI to variations in incident shock strength. For each case the starting location of the shock generator $x_{sg}$ is modified to maintain the same value of $Re_{x}$ at the inviscid impingement location. One stronger and three weaker interactions are considered for this section, corresponding to flow deflections of $\theta_{sg} = \left[0.5, 1, 1.5, 2.5\right]$. Figure \ref{fig:shockcompare_line} (a) shows the normalised centreline pressure along the bottom wall for the baseline and three weaker interactions. As the interaction is weakened the pressure rise at the start of separation shifts downstream. The outlet pressure ratio $p_3/p_1$ for the three weaker interactions in given in table \ref{tab:shock_strength}. For each shock strength there is a lack of a pressure plateau in the middle interaction, often seen in quasi-2D laminar SBLI (\cite{Katzer1989}, \cite{Sansica2013}). Instead, the flow reattaches quickly and for the two weakest interactions there is actually a pressure reduction after the initial compression.

 \begin{table}
  \begin{center}
  \begin{tabular}{ccccc}
      Flow deflection $\left(\theta_{sg}^{\circ}\right)$ &  $p_3/p_1$ & Interaction region $\left(x_{sep}, x_{reattach}\right)$ & $L_{sep}$& \% of baseline $L_{sep}$ \vspace{0.1cm}\\
      0.5  & 1.145 &  (291.2, 311.5) & 20.28 & 13\% \\ 
      1.0  & 1.197 &   (272.7, 321.9) & 49.20 & 31\% \\ 
      1.5  & 1.252 &  (247.1, 334.6) & 87.53 & 55\% \\ 
      2.0  & 1.313 &  (207.7, 365.6) & 157.93 & -- \\ 
  \end{tabular}
  \caption{Reduction in streamwise separation length $L_{sep}$ of the main interaction with decreasing incident shock strength. Comparison is made to the centreline skin friction for the baseline $\theta_{sg} = 2.0^{\circ}$ case with $AR=1$.}\label{tab:shock_strength}
  \end{center}
\end{table}

 The effect of the weakened shock on the central separation is shown in figure \ref{fig:shockcompare_line} (b). Both the separation and reattachment locations shift down and upstream respectively for the weaker interactions, with the $\theta_{sg} = 0.5^{\circ}$ case being close to incipient separation on the centreline. It was observed during the simulations that the corner region was the first to separate, owing to the large regions of low-momentum fluid in the corners. The crossing at $x=500$ of the reflected conical shocks identified in section \ref{sec:laminar_topology} is less obvious for the weaker interactions. Aside from the baseline case only the $\theta_{sg}=1.5^{\circ}$ case shows the asymmetric double trough $C_f$ distribution; the weaker cases reattach abruptly in a manner similar to the narrow aspect ratio ducts in figure \ref{fig:Fig10} (a). Table \ref{tab:shock_strength} shows the separation and reattachment points for each of the shock strengths and gives the size of $L_{sep}$ relative to the baseline. Although the separation and reattachment are both sensitive to incident shock strength, the effect on the separation point is more severe. A reduction of $1.5^{\circ}$ in the initial flow deflection leads to an overall 87\% reduction in centreline separation. Each of the weaker interactions followed owl-like topologies of the first kind, with two distinct foci and saddle points. As the interaction was strengthened there was an elongation of the separated region in the streamwise direction.

\begin{figure}
\begin{centering}
  \includegraphics[width=0.6\textwidth]{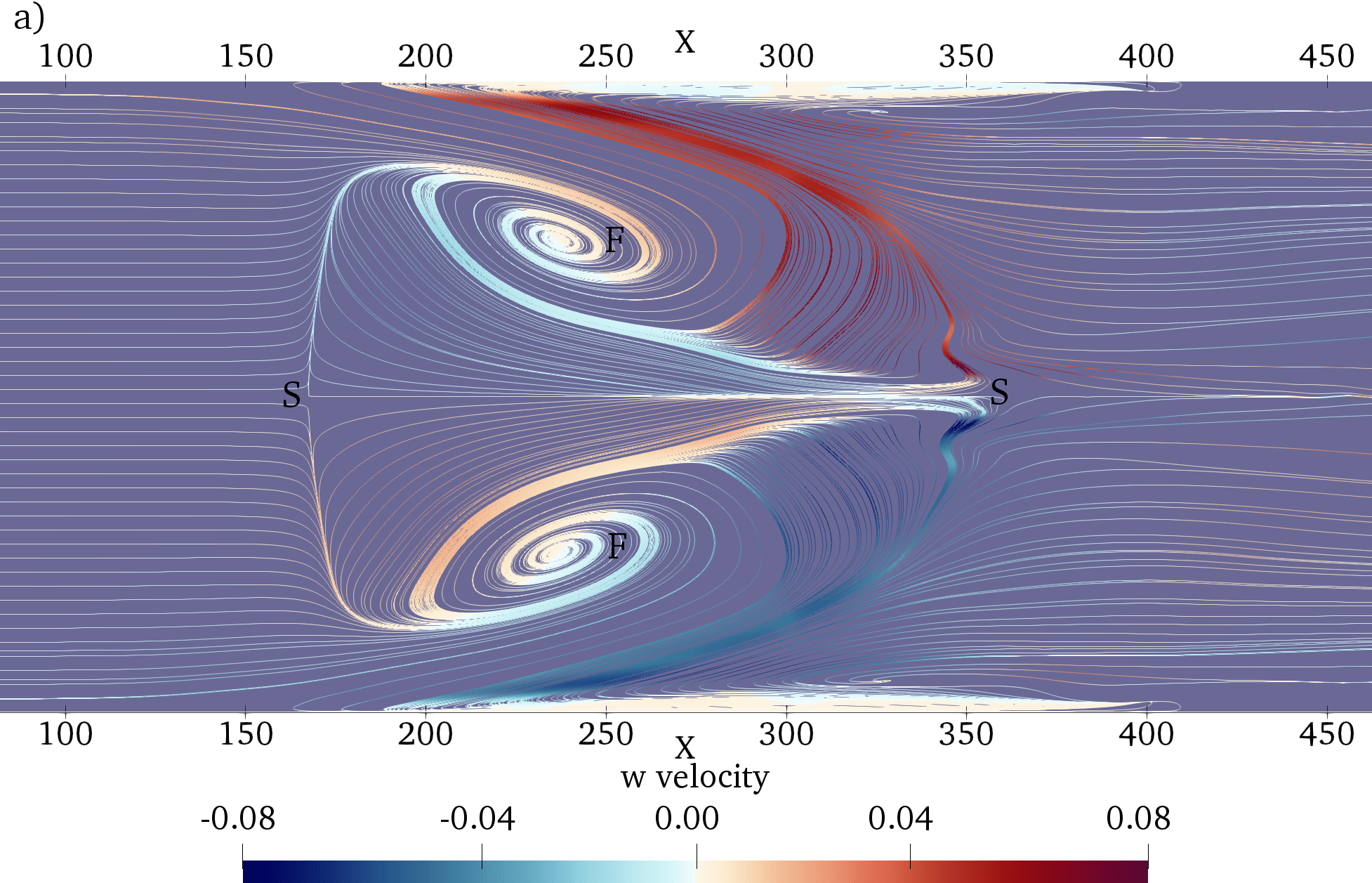}
  \includegraphics[width=0.6\textwidth]{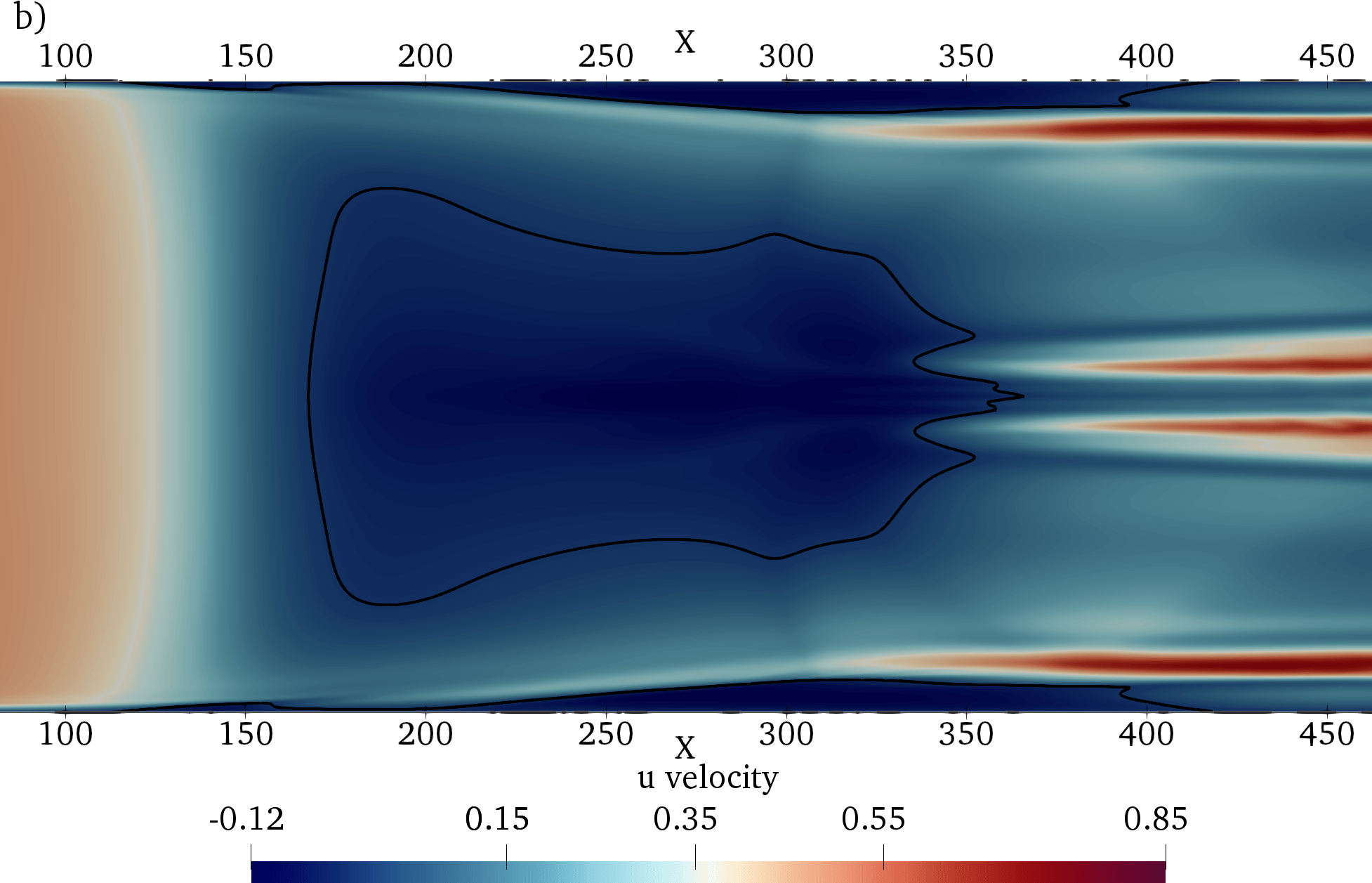}
  \caption{(a) $u$-$w$ velocity contours evaluated at $y=1$ above the bottom wall. Coloured by transverse velocity component $w$. Saddle (S) and foci (F) are highlighted as in figure \ref{fig:2deg_owl_pattern}. (b) Streamwise velocity at $y=1$ above the bottom wall. The solid black line shows the $u=0$ line to highlight regions of recirculation. Four high-speed streaks are observed downstream of the interaction in red. A darkened imprint of the conical swept shock is also visible.}\label{fig:2half_contours}
\end{centering}
\end{figure}

 A stronger interaction at $\theta_{sg} = 2.5^{\circ}$ was also performed and was found to be very close to the limits of stability downstream of the main interaction. The transition process for duct SBLI is beyond the scope of the present laminar study and will be the subject of future work. The results for the stronger interaction are included here to give insight into the limiting behaviour of large separations. Figure \ref{fig:2half_contours} (a) shows $u-w$ velocity streamlines at $y=1$ over the same range as in figure \ref{fig:2deg_owl_pattern}. The same combination of saddle points (S) and foci (F) are observed for the stronger interaction but the size and magnitude of the recirculation is increased. The separation line has shifted upstream and each of the foci have been elongated in the streamwise direction relative to the weaker interaction in figure \ref{fig:2deg_owl_pattern}. The deformation of the saddle point at the reattachment line is increased, there is no longer a well defined set of two streamlines entering the saddle point. Downstream of the interaction the streamlines no longer remain parallel to each other and a steady laminar flow was not obtained. A grid refinement study was performed with 100\% additional grid points in each direction independently. A laminar solution was not obtained for any of the refined grids, with intermittent regions of transition observed near the outlet. The streamwise velocity contours of figure \ref{fig:2half_contours} (b) show four high-speed streaks downstream of the main interaction. The flow was found to accelerate rapidly after the apex of each of the corner separations and on either side of the centreline saddle point. While the interaction is still roughly `owl-like' of the first kind, the deformation of the attachment line saddle point could indicate an intermediate state approaching the second owl-like state shown schematically in figure 4 (a) of \cite{Eagle2014}. For owl-like patterns of the second kind, the rear saddle point transitions into a node with multiple streamlines directed into it. Owl-like patterns of the second kind for turbulent SBLI have been shown to correspond to stronger interactions in experiments \citep{Xiang2019}. Based on the topological trends seen in this case, it is feasible that the transition mechanism for stronger three-dimensional laminar SBLI involves the bifurcation from topologies of the first kind to the second.

\subsection{Controlling the interaction with the trailing expansion fan}\label{sec:long_domain_control}

\begin{figure}
  \includegraphics[width=0.90\textwidth]{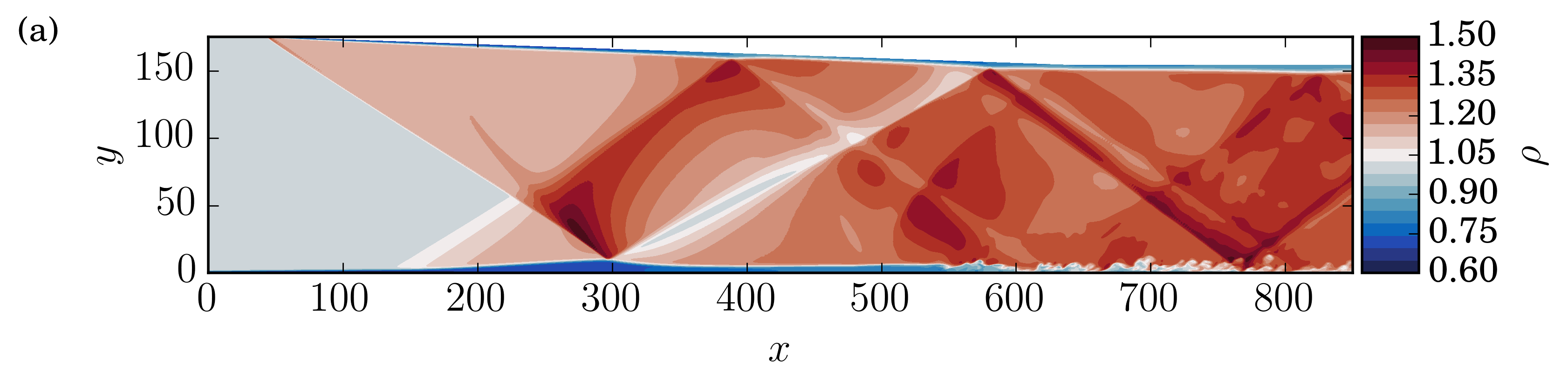}
  \includegraphics[width=0.90\textwidth]{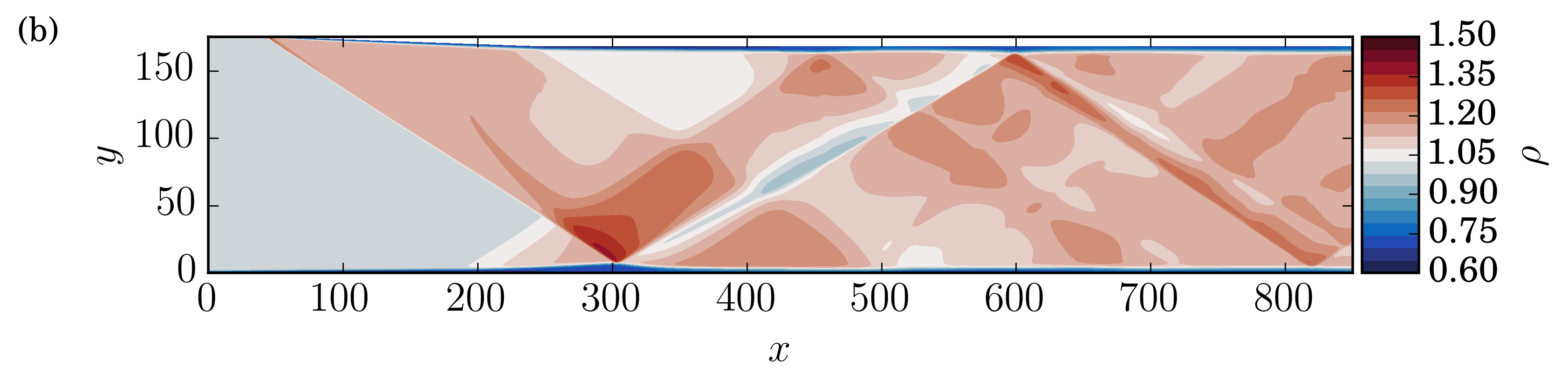}
  \includegraphics[width=0.90\textwidth]{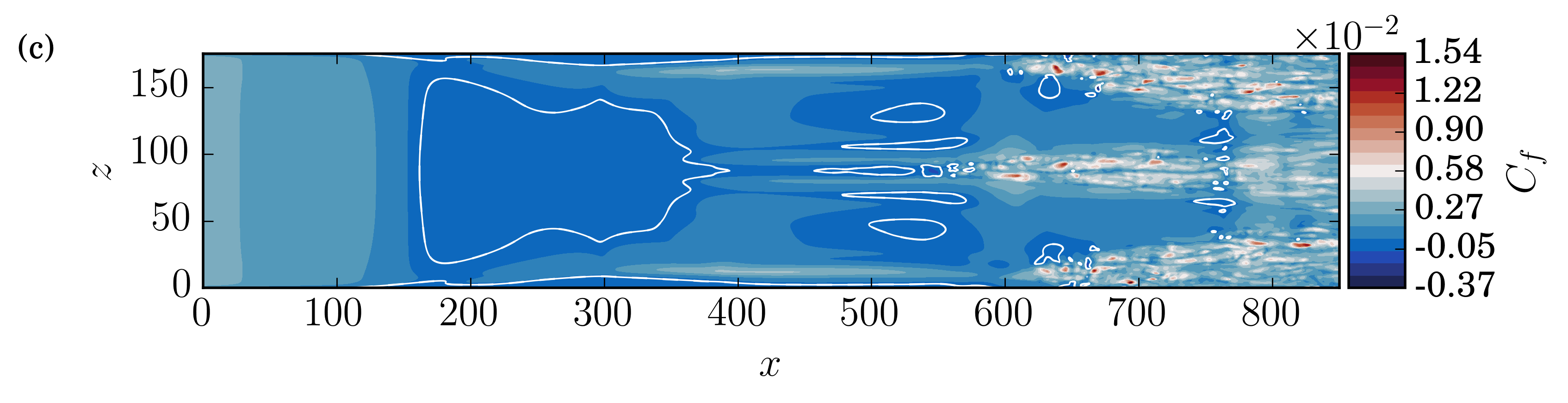}
   \includegraphics[width=0.90\textwidth]{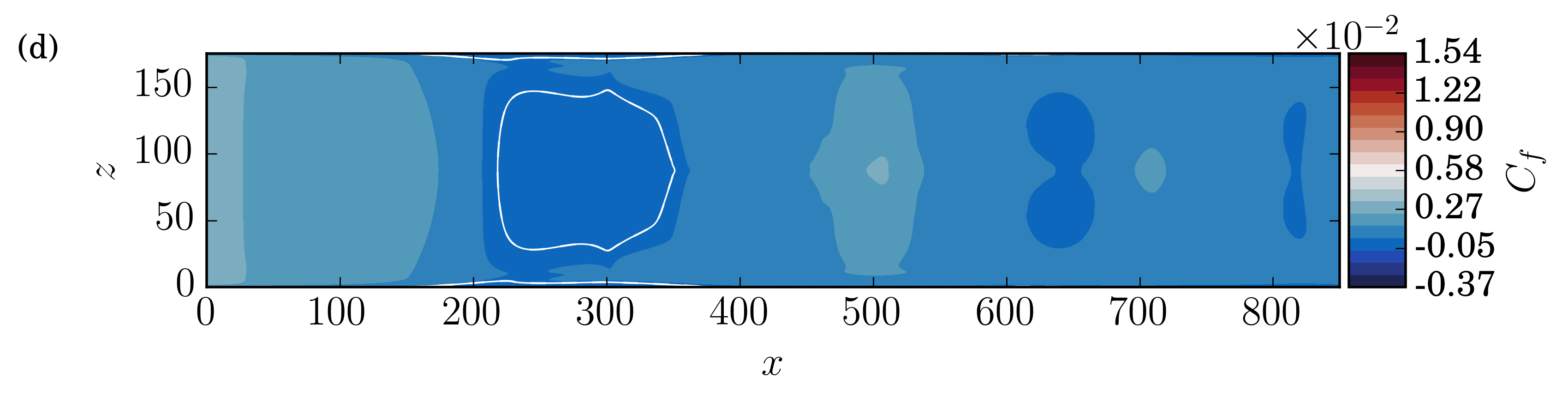}
  \includegraphics[width=0.90\textwidth]{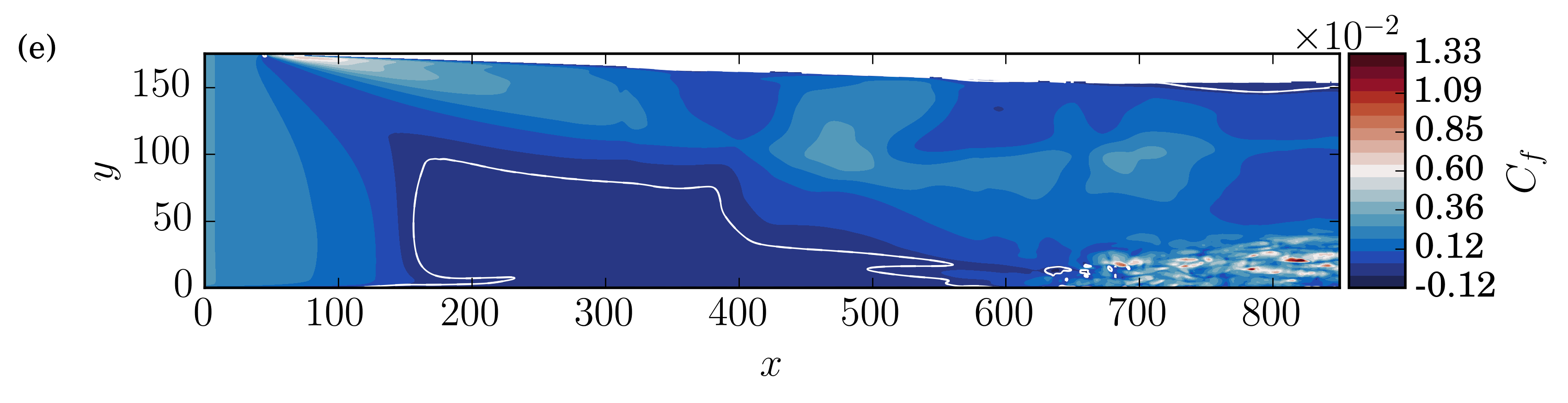}
  \includegraphics[width=0.90\textwidth]{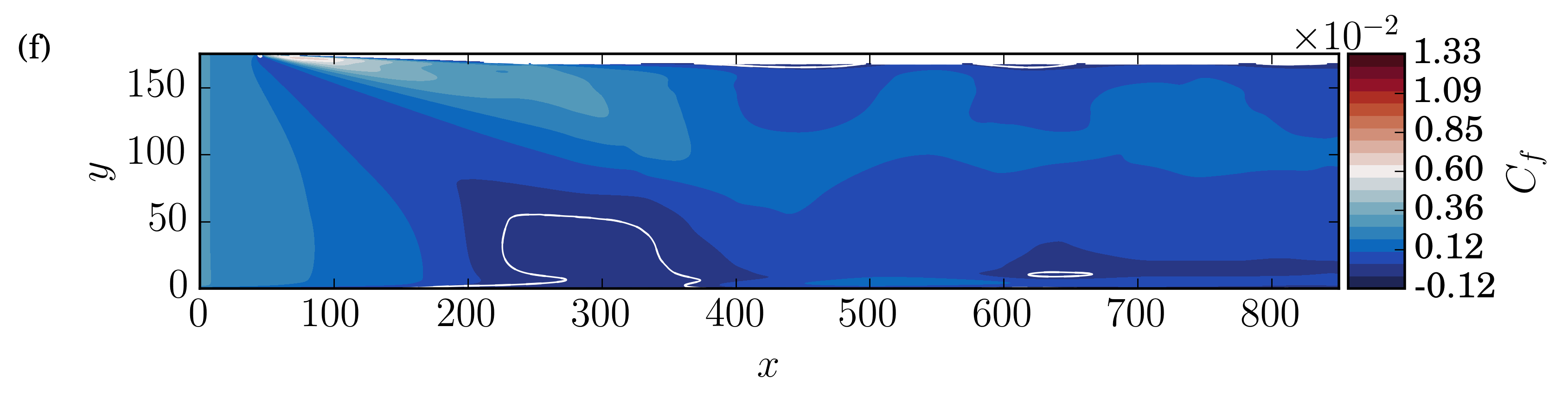}
  \caption{Long domain duct ($L_{x}=850$, $\theta_{sg} = 2^{\circ}$, $AR=1$) demonstrating the ability of the trailing expansion fan to control the interaction. The figures correspond to shock generators with $L_{sg} = 600$ in (a,c,e) and $L_{sg} = 200$ in (b,d,f). Showing centreline density ($z=87.5$) (a,b), skin friction on the bottom wall ($y=0$) (c,d) and skin friction on the sidewall ($z=0$) (e,f). The solid white line highlights the zero crossing of skin friction, enclosing regions of flow recirculation.}
\label{fig:Fig14}
\end{figure}

It was demonstrated in section \ref{sec:ramp_length} that for laminar duct SBLI the flow on the centreline is sensitive to the length of shock generator used. For longer shock generators the trailing expansion fan recovery occurs further downstream and leads to an increase in the central separation bubble length (figure \ref{fig:Fig5}(b)). A comparison to figure \ref{fig:Fig4}(b) showed this is a purely 3D effect that is not present in quasi-2D laminar SBLI. The only difference between the quasi-2D and 3D simulations is the lateral confinement imposed by the sidewalls. Table \ref{tab:ramp_length_3d} reported a 6\% increase in $L_{sep}$ between $L_{sg} = 300$ and $L_{sg} = 350$, a limiting case of maximum $L_{sep}$ was not found for the range of $L_{sg}$ tested. This section aims to rectify this by investigating the limiting behaviour of the interaction for a longer duct with both short ($L_{sg} = 200$) and very long ($L_{sg} = 600$) shock generators. In both cases the trailing expansion fan is located far downstream of the central interaction and does not impinge directly on the bottom wall separation bubble. The baseline domain from table \ref{tab:grid_sizes} is extended by $\sim 55\%$ in the streamwise direction to a length of $L_x = 850$. A refined grid resolution of $ \left(1150 \times 455 \times 455\right)$ is selected to maintain the same streamwise resolution. 

Figure \ref{fig:Fig14} shows the instantaneous flow features of the two long domain cases after $t=12000$ for shock generators of length $L_{sg} = 600$ (a,c,e) and $L_{sg} = 200$ (b,d,f). Instantaneous skin friction relative to the bottom ($\frac{\partial u}{\partial y}$) and sidewalls ($\frac{\partial u}{\partial z}$) are shown in (c,d) and (e,f) respectively, with centreline ($z=87.5$) density contours given in (a,b). For the skin friction plots the solid white line highlights the $C_f = 0$ crossing of the component relative to that wall. In both of the density plots of figure \ref{fig:Fig14} (a,b) a shock train is clearly visible; compression waves at the start of the central separation coalesce into a shock that impinges on the upper surface and causes a small region of secondary flow-reversal. A further reflection occurs and this wave impinges on the bottom wall. The reflected incident shock also reflects between the upper and lower walls of the domain, highlighting the complex secondary reflections present for internally confined flows. A trailing edge expansion fan can be seen originating from the upper surface at $x=245$ for the short shock generator, first hitting the bottom wall boundary layer around $x=500$. In the long case an expansion fan is generated at $x=645$ which exits through the outlet before hitting the bottom wall. For the long shock generator the flow transitions to turbulence while remaining laminar for the shorter $L_{sg} = 200$. This suggests that if geometry permits, the strength and size of the central separation can be controlled by the use of shorter shock generators.

The difference in the reattached boundary layer state is clearer to see in figures \ref{fig:Fig14}(c,d). For the long shock generator in figure \ref{fig:Fig14}(c) the flow transitions at $x=600$ on the centreline and in both corners of the duct. The corner separation is truncated as the boundary layer transitions and remains attached in the corner towards the outlet. The central separation is substantially larger for the longer shock generator and is characterized by a flat separation line across the span and a distortion of the reattachment line peaking on the centreline. The separated region covers more of the span and is also noticeably thicker in the corner regions. The shape of the interaction is very similar to the stronger $\theta_{sg} = 2.5^{\circ}$ interaction shown in figure \ref{fig:2half_contours} (b). This demonstrates that for two different methods of achieving a stronger interaction, the shape of the central recirculation has a topology common to both. Comparing to the short shock generator of figure \ref{fig:Fig14}(d) we note the downstream shift of the expansion fan has led to an upstream and downstream shift of the separation and reattachment points respectively. The downstream shift of the expansion fan in the longer case also leads to smaller secondary recirculation zones at $x=500$. Comparing their location to the shock pattern in figure \ref{fig:Fig14}(a) it is clear the secondary separations cannot be attributed to the first vertical secondary reflection that impinges at $x=600$ and instead are likely caused by the lateral reflections shown downstream of the interaction in figure \ref{fig:shock_patterns3d}. Finally the impact of shock generator length on the sidewall flow is illustrated in figures \ref{fig:Fig14}(e,f). While the shape of the sidewall separation is similar in both cases, the downstream shift of the trailing expansion fan leads to a significant enlargement of the sidewall recirculation. For the short case it can be seen that the expansion fan causes a recovery of the sidewall flow that wards off separation after $x=400$. Smaller separation bubbles are seen further downstream due to the secondary reflections present in figure \ref{fig:Fig14}(b). For the long shock generator case in figure \ref{fig:Fig14}(e) the sidewall recirculation covers a much larger portion of the sidewall and is only truncated as the flow begins to transition around $x=600$. As the expansion fan originates from the upper surface at $x=645$ and is directed towards the outlet, we conclude that the sidewall separation is limited by the onset of transition in the sidewall boundary layer. This is in contrast to the short shock generator case where the sidewall separation is prematurely terminated by the upstream shift of the trailing expansion fan. The importance of transition in limiting sidewall separation suggests that attempts such as \cite{Giepman2016} to suppress the size of laminar SBLI with transition could also benefit from placing trips on the sides of the duct. We have seen that the central separation and state of the reattached boundary layer can be controlled simply by shortening the shock generator. Shorter shock generators modify the sidewall flow considerably and limit the growth of both the corner and sidewall separations. The modified interaction is observed to be weaker and leads to reduced three-dimensionality and a suppressed central separation bubble.

\section{Conclusions}
Three-dimensional laminar SBLI at Mach 2 have been investigated numerically for enclosed rectangular ducts. Similar to previous turbulent cases (\cite{Bermejo-Moreno2014}, \cite{wang_sandham_hu_liu_2015}), a strong dependence of duct aspect ratio was observed in the laminar case, albeit with much larger recirculation regions. Lateral confinement of laminar SBLI from sidewalls leads to a strengthened and highly three-dimensional interaction. Span-periodic analysis is unable to predict centreline skin friction distributions except in the limit of very wide aspect ratios ($AR \geq 4$). The streamwise extent of the central separation was almost identical for ducts with an aspect ratio of one or two. Aspect ratios less than unity had a substantial decrease in separation length and showed multiple reflections of laterally travelling shockwaves. At an aspect ratio of four the three-dimensionality of the interaction was limited to a region 30\% of the width of the span away from the sidewall. The centreline separation length was observed to still be 9\% different from quasi-2D predictions. Due to the dependency of the interaction on duct geometry and shock generator length, it is not possible to predict centreline separations without prior knowledge of these parameters. The baseline one-to-one aspect ratio configuration had a 30\% stronger interaction compared to quasi-2D predictions, but this was shown to be dependent on the length of shock generator used. 

In addition to aspect ratio, the interaction was found to be strongly influenced by the expansion fan generated from the trailing edge of the shock generator. For expansion fan impingement points further upstream, a decrease in the size and magnitude of central recirculation was observed. The expansion fan effect was purely a three-dimensional effect; for quasi-2D simulations there was no dependence on shock generator length over the same range. Critical point analysis showed that the confined SBLI had `owl-like' topologies of the first kind as introduced by \cite{1984ZFl}. For a stronger interaction the central recirculation was elongated in the streamwise direction and a distortion of the attachment line was observed. Preliminary results highlighted four high-speed streaks downstream of the interaction near the centreline and in the corner. Therefore, streak instability is one potential transition mechanism of confined SBLI. Shock structures identified downstream of the interaction were shown to result from reflections of the conical swept SBLI generated between the shock generator and sidewalls. Significant corner compressions from the bottom of the domain could not be identified; instead the primary mechanism behind the strengthened interaction was the swept conical SBLI. The swept interaction was shown to begin substantially further upstream than the onset of bottom wall corner separation. A 55\% longer domain case was simulated to demonstrate the ability to control the central separation with shorter shock generators. The recovery from the trailing expansion fan suppresses sidewall recirculation and modifies the strength of the interaction as a whole. Future work will focus on the transition mechanism for confined SBLI. Differences in the origin of crossing shocks identified in numerical work and experiment, which have been attributed to both the swept SBLI and bottom wall corner compressions respectively  should also be investigated.

\newpage
\noindent David J. Lusher is funded by an EPSRC Centre for Doctoral Training grant (EP/L015382/1). Compute resources used in this work were provided by the `Cambridge Service for Data Driven Discovery' (CSD3) system operated by the University of Cambridge Research Computing Service (\url{http://www.hpc.cam.ac.uk}) funded by EPSRC Tier-2 capital grant EP/P020259/1, and the IRIDIS5 High Performance Computing Facility, and associated support services at the University of Southampton. The OpenSBLI code is available at \url{https://opensbli.github.io}. Data from this report will be available from the University of Southampton institutional repository.

\bibliographystyle{jfm}
\bibliography{laminar_paper}

\begin{thebibliography}{42}
\expandafter\ifx\csname natexlab\endcsname\relax\def\natexlab#1{#1}\fi
\def\au#1{#1} \def\ed#1{#1} \def\yr#1{#1}\def\at#1{#1}\def\jt#1{\textit{#1}}
  \def\bt#1{#1}\def\bvol#1{\textbf{#1}} \def\vol#1{#1} \def\pg#1{#1}
  \def\publ#1{#1}\def\arxiv#1{#1}\def\org#1{#1}\def\st#1{\textit{#1}}

\bibitem[Adamson \& Messiter(1980)]{Adamson1980}
{\sc \au{Adamson, T~C} \& \au{Messiter, A~F}} \yr{1980}  \at{{Analysis of
  Two-Dimensional Interactions Between Shock Waves and Boundary Layers}}.
  \jt{Annual Review of Fluid Mechanics}  \bvol{12}~(1),  \pg{103--138}.

\bibitem[Babinsky \& Harvey(2011)]{babinsky_harvey_2011}
{\sc \au{Babinsky, Holger} \& \au{Harvey, John~K.}} \yr{2011} {\em Shock
  Wave-Boundary-Layer Interactions\/}.  \publ{Cambridge University Press}.

\bibitem[Babinsky {\em et~al.\/}(2013)Babinsky, Oorebeek \&
  Cottingham]{Babinsky2013}
{\sc \au{Babinsky, Holger}, \au{Oorebeek, Joseph} \& \au{Cottingham, Trey}}
  \yr{2013}  \at{{Corner effects in reflecting oblique
  shock-wave/boundary-layer interactions}}.  \bt{In {\em 51st AIAA Aerospace
  Sciences Meeting including the New Horizons Forum and Aerospace
  Exposition\/}}.  \publ{American Institute of Aeronautics and Astronautics}.

\bibitem[Bermejo-Moreno {\em et~al.\/}(2014)Bermejo-Moreno, Campo, Larsson,
  Bodart, Helmer \& Eaton]{Bermejo-Moreno2014}
{\sc \au{Bermejo-Moreno, Iv{\'{a}}n}, \au{Campo, Laura}, \au{Larsson, Johan},
  \au{Bodart, Julien}, \au{Helmer, David} \& \au{Eaton, John~K.}} \yr{2014}
  \at{{Confinement effects in shock wave/turbulent boundary layer interactions
  through wall-modelled large-eddy simulations}}.  \jt{Journal of Fluid
  Mechanics}  \bvol{758},  \pg{5--62}.

\bibitem[Borges {\em et~al.\/}(2008)Borges, Carmona, Costa \& Don]{Borges2008}
{\sc \au{Borges, Rafael}, \au{Carmona, Monique}, \au{Costa, Bruno} \& \au{Don,
  Wai~Sun}} \yr{2008}  \at{{An improved weighted essentially non-oscillatory
  scheme for hyperbolic conservation laws}}.  \jt{Journal of Computational
  Physics}  \bvol{227}~(6),  \pg{3191--3211}.

\bibitem[Bruce {\em et~al.\/}(2011)Bruce, Burton, Titchener \&
  Babinsky]{Bruce2011}
{\sc \au{Bruce, P.~J.K.}, \au{Burton, D.~M.F.}, \au{Titchener, N.~A.} \&
  \au{Babinsky, H.}} \yr{2011}  \at{{Corner effect and separation in transonic
  channel flows}}.  \jt{Journal of Fluid Mechanics}  \bvol{679},
  \pg{247--262}.

\bibitem[Burton \& Babinsky(2012)]{Burton2012}
{\sc \au{Burton, D.~M.F.} \& \au{Babinsky, H.}} \yr{2012}  \at{{Corner
  separation effects for normal shock wave/turbulent boundary layer
  interactions in rectangular channels}}.  \jt{Journal of Fluid Mechanics}
  \bvol{707},  \pg{287--306}.

\bibitem[Carpenter \& Kennedy(1994)]{carpenter_kennedy_1994}
{\sc \au{Carpenter, Mark~H} \& \au{Kennedy, Christopher~A}} \yr{1994}
  \at{Fourth-order 2n-storage runge-kutta schemes}.  \jt{NASA Langley Research
  Center} .

\bibitem[Carpenter {\em et~al.\/}(1998)Carpenter, Nordstr{\"{o}}m \&
  Gottlieb]{Carpenter1998}
{\sc \au{Carpenter, Mark~H}, \au{Nordstr{\"{o}}m, Jan} \& \au{Gottlieb, David}}
  \yr{1998}  \at{{A Stable and Conservative Interface Treatment of Arbitrary
  Spatial Accuracy}}.  \jt{Journal of Computational Physics}  \bvol{365}~(98),
  \pg{341--365}.

\bibitem[Clemens \& Narayanaswamy(2014)]{Clemens2014}
{\sc \au{Clemens, Noel~T} \& \au{Narayanaswamy, Venkateswaran}} \yr{2014}
  \at{{Low-Frequency Unsteadiness of Shock Wave/Turbulent Boundary Layer
  Interactions}}.  \jt{Annual Review of Fluid Mechanics}  \bvol{46}~(1),
  \pg{469--492}.

\bibitem[Colliss {\em et~al.\/}(2016)Colliss, Babinsky, N{\"{u}}bler \&
  Lutz]{Colliss2016}
{\sc \au{Colliss, S~P}, \au{Babinsky, H}, \au{N{\"{u}}bler, K} \& \au{Lutz, T}}
  \yr{2016}  \at{{Vortical Structures on Three-Dimensional Shock Control
  Bumps}}.  \jt{AIAA Journal}  \bvol{54}~(8),  \pg{2338--2350}.

\bibitem[Degrez {\em et~al.\/}(1987)Degrez, Boccadoro \& Wendt]{Degrez1987}
{\sc \au{Degrez, G}, \au{Boccadoro, C~H} \& \au{Wendt, J~F}} \yr{1987}
  \at{{The interaction of an oblique shock wave with a laminar boundary layer
  revisited. An experimental and numerical study}}.  \jt{Journal of Fluid
  Mechanics}  \bvol{177},  \pg{247--263}.

\bibitem[D{\'{e}}lery(2001)]{Delery2001}
{\sc \au{D{\'{e}}lery, Jean~M}} \yr{2001}  \at{{Robert Legendre and Henri
  Werl{\'{e}}: Toward the Elucidation of Three-Dimensional Separation}}.
  \jt{Annual Review of Fluid Mechanics}  \bvol{33}~(1),  \pg{129--154}.

\bibitem[Dolling(2001)]{Dolling2001}
{\sc \au{Dolling, David~S}} \yr{2001}  \at{{Fifty Years of
  Shock-Wave/Boundary-Layer Interaction Research: What Next?}}  \jt{AIAA
  Journal}  \bvol{39}~(8),  \pg{1517--1531}.

\bibitem[Dwivedi {\em et~al.\/}(2017)Dwivedi, Nichols, Jovanovic \&
  Candler]{Dwivedi2017}
{\sc \au{Dwivedi, Anubhav}, \au{Nichols, Joseph~W}, \au{Jovanovic, Mihailo~R}
  \& \au{Candler, Graham~V}} \yr{2017}  \at{{Optimal spatial growth of streaks
  in oblique shock/boundary layer interaction}}.  \bt{In {\em 8th AIAA
  Theoretical Fluid Mechanics Conference\/}}.  \publ{American Institute of
  Aeronautics and Astronautics}.

\bibitem[Eagle \& Driscoll(2014)]{Eagle2014}
{\sc \au{Eagle, W~Ethan} \& \au{Driscoll, James~F}} \yr{2014}  \at{{Shock
  wave–boundary layer interactions in rectangular inlets: three-dimensional
  separation topology and critical points}}.  \jt{Journal of Fluid Mechanics}
  \bvol{756},  \pg{328--353}.

\bibitem[Eagle {\em et~al.\/}(2011)Eagle, Driscoll \& Benek]{Eagle2011}
{\sc \au{Eagle, W~Ethan}, \au{Driscoll, James~F} \& \au{Benek, John~a}}
  \yr{2011}  \at{{Experimental Investigation of Corner Flows in Rectangular
  Supersonic Inlets with 3D Shock-Boundary Layer Effects}}.  \jt{49th AIAA
  Aerospace Sciences Meeting including the New Horizons Forum and Aerospace
  Exposition} ~(January),  \pg{1--11}.

\bibitem[Gaitonde(2015)]{Gaitonde2015}
{\sc \au{Gaitonde, Datta~V}} \yr{2015}  \at{{Progress in shock wave/boundary
  layer interactions}}.  \jt{Progress in Aerospace Sciences}  \bvol{72},
  \pg{80--99}.

\bibitem[Garnier(2009)]{Garnier2009}
{\sc \au{Garnier, Eric}} \yr{2009}  \at{{Stimulated Detached Eddy Simulation of
  three-dimensional shock/boundary layer interaction}}.  \jt{Shock Waves}
  \bvol{19}~(6),  \pg{479--486}.

\bibitem[Giepman {\em et~al.\/}(2016)Giepman, Louman, Schrijer \& van
  Oudheusden]{Giepman2016}
{\sc \au{Giepman, Rogier H~M}, \au{Louman, Renee}, \au{Schrijer, Ferry F~J} \&
  \au{van Oudheusden, Bas~W}} \yr{2016}  \at{{Experimental Study into the
  Effects of Forced Transition on a Shock-Wave/Boundary-Layer Interaction}}.
  \jt{AIAA Journal}  \bvol{54}~(4),  \pg{1313--1325}.

\bibitem[Giepman {\em et~al.\/}(2015)Giepman, Schrijer \& van
  Oudheusden]{Giepman2015}
{\sc \au{Giepman, R H~M}, \au{Schrijer, F F~J} \& \au{van Oudheusden, B~W}}
  \yr{2015}  \at{{High-resolution PIV measurements of a transitional shock
  wave–boundary layer interaction}}.  \jt{Experiments in Fluids}
  \bvol{56}~(6),  \pg{113}.

\bibitem[Giepman {\em et~al.\/}(2018)Giepman, Schrijer \& van
  Oudheusden]{Giepman2018}
{\sc \au{Giepman, R H~M}, \au{Schrijer, F F~J} \& \au{van Oudheusden, B~W}}
  \yr{2018}  \at{{A parametric study of laminar and transitional oblique shock
  wave reflections}}.  \jt{Journal of Fluid Mechanics}  \bvol{844},
  \pg{187--215}.

\bibitem[Gross \& Fasel(2016)]{Gross2016}
{\sc \au{Gross, Andreas} \& \au{Fasel, Hermann~F}} \yr{2016}  \at{{Numerical
  Investigation of Shock Boundary-Layer Interactions}}.  \bt{In {\em 54th AIAA
  Aerospace Sciences Meeting\/}}.  \publ{American Institute of Aeronautics and
  Astronautics}.

\bibitem[Grossman \& Bruce(2017)]{Grossman2017}
{\sc \au{Grossman, Ilan~J} \& \au{Bruce, Paul~J}} \yr{2017}  \at{{Effect of
  Test Article Geometry on Shock Wave-Boundary Layer Interactions in
  Rectangular Intakes}}.  \bt{In {\em 55th AIAA Aerospace Sciences Meeting\/}}.
   \publ{American Institute of Aeronautics and Astronautics}.

\bibitem[Grossman \& Bruce(2018)]{Grossman2018}
{\sc \au{Grossman, Ilan~J} \& \au{Bruce, Paul J~K}} \yr{2018}  \at{{Confinement
  effects on regular–irregular transition in
  shock-wave–boundary-layer interactions}}.  \jt{Journal of Fluid Mechanics}
   \bvol{853},  \pg{171--204}.

\bibitem[Hakkinen {\em et~al.\/}(1959)Hakkinen, Greber, Trilling \&
  Abarbanel]{hakkinen1959}
{\sc \au{Hakkinen, R.~J.}, \au{Greber, I.}, \au{Trilling, L.} \& \au{Abarbanel,
  S.S.}} \yr{1959}  \at{The interaction of an oblique shock wave with a laminar
  boundary layer}.  \jt{NASA Memorandum 2-18-59W} .

\bibitem[Hildebrand {\em et~al.\/}(2018{\natexlab{{\em a\/}}})Hildebrand,
  Dwivedi, Nichols, Jovanovi{\'{c}} \& Candler]{Hildebrand2018_1}
{\sc \au{Hildebrand, Nathaniel}, \au{Dwivedi, Anubhav}, \au{Nichols,
  Joseph~W.}, \au{Jovanovi{\'{c}}, Mihailo~R.} \& \au{Candler, Graham~V.}}
  \yr{2018{\natexlab{{\em a\/}}}}  \at{{Simulation and stability analysis of
  oblique shock-wave/boundary-layer interactions at Mach 5.92}}.  \jt{Physical
  Review Fluids}  \bvol{3}~(1),  \pg{1--23}.

\bibitem[Hildebrand {\em et~al.\/}(2018{\natexlab{{\em b\/}}})Hildebrand,
  Nichols, Candler \& Jovanovic]{Hildebrand2018_2}
{\sc \au{Hildebrand, Nathaniel~J}, \au{Nichols, Joseph~W}, \au{Candler,
  Graham~V} \& \au{Jovanovic, Mihailo}} \yr{2018{\natexlab{{\em b\/}}}}
  \at{{Transient growth in oblique shock wave/laminar boundary layer
  interactions at Mach 5.92}}.  \bt{In {\em 2018 Fluid Dynamics Conference\/}}.
   \publ{American Institute of Aeronautics and Astronautics}.

\bibitem[Jacobs {\em et~al.\/}(2017)Jacobs, Jammy \& Sandham]{JACOBS201712}
{\sc \au{Jacobs, Christian~T.}, \au{Jammy, Satya~P.} \& \au{Sandham, Neil~D.}}
  \yr{2017}  \at{{OpenSBLI: A framework for the automated derivation and
  parallel execution of finite difference solvers on a range of computer
  architectures}}.  \jt{Journal of Computational Science}  \bvol{18},  \pg{12
  -- 23}.

\bibitem[Katzer(1989)]{Katzer1989}
{\sc \au{Katzer, Edgar}} \yr{1989}  \at{{On the lengthscales of laminar
  shock/boundary- layer interaction}}.  \jt{Journal of Fluid Mechanics}
  \bvol{206}~(1989),  \pg{477--496}.

\bibitem[Lusher {\em et~al.\/}(2018)Lusher, Jammy \& Sandham]{LUSHER201817}
{\sc \au{Lusher, David~J.}, \au{Jammy, Satya~P.} \& \au{Sandham, Neil~D.}}
  \yr{2018}  \at{Shock-wave/boundary-layer interactions in the automatic
  source-code generation framework opensbli}.  \jt{Computers \& Fluids}
  \bvol{173},  \pg{17 -- 21}.

\bibitem[Morajkar \& Gamba(2016)]{Morajkar2016}
{\sc \au{Morajkar, Rohan~R.} \& \au{Gamba, Mirko}} \yr{2016}  \at{{Swept shock
  corner flow interacions}}.  \jt{54th AIAA Aerospace Sciences Meeting}
  ~(January),  \pg{1--13}.

\bibitem[{Perry} \& {Hornung}(1984)]{1984ZFl}
{\sc \au{{Perry}, A.~E.} \& \au{{Hornung}, H.}} \yr{1984}  \at{{Some aspects of
  three-dimensional separation. II - Vortex skeletons}}.  \jt{Zeitschrift fur
  Flugwissenschaften und Weltraumforschung}  \bvol{8},  \pg{155--160}.

\bibitem[Quadros \& Bernardini(2018)]{Quadros2018}
{\sc \au{Quadros, Russell} \& \au{Bernardini, Matteo}} \yr{2018}
  \at{{Numerical Investigation of Transitional Shock-Wave/Boundary-Layer
  Interaction in Supersonic Regime}}.  \jt{AIAA Journal}  \bvol{56}~(7),
  \pg{2712--2724}.

\bibitem[Reguly {\em et~al.\/}(2014)Reguly, Mudalige, Giles, Curran \&
  McIntosh-Smith]{Reguly:2014:ODS:2691166.2691173}
{\sc \au{Reguly, Istv\'{a}n~Z.}, \au{Mudalige, Gihan~R.}, \au{Giles,
  Michael~B.}, \au{Curran, Dan} \& \au{McIntosh-Smith, Simon}} \yr{2014}
  \bt{{The OPS Domain Specific Abstraction for Multi-block Structured Grid
  Computations}}.  \pg{pp. 58--67}.  \publ{IEEE Press}.

\bibitem[Sansica {\em et~al.\/}(2013)Sansica, Sandham \& Hu]{Sansica2013}
{\sc \au{Sansica, Andrea}, \au{Sandham, Neil} \& \au{Hu, Zhiwei}} \yr{2013}
  \at{{Stability and Unsteadiness in a 2D Laminar Shock-Induced Separation
  Bubble}}.  \bt{In {\em 43rd Fluid Dynamics Conference\/}}.  \publ{American
  Institute of Aeronautics and Astronautics}.

\bibitem[Sansica {\em et~al.\/}(2016)Sansica, Sandham \& Hu]{Sansica2016}
{\sc \au{Sansica, Andrea}, \au{Sandham, Neil~D.} \& \au{Hu, Zhiwei}} \yr{2016}
  \at{{Instability and low-frequency unsteadiness in a shock-induced laminar
  separation bubble}}.  \jt{Journal of Fluid Mechanics}  \bvol{798},
  \pg{5--26}.

\bibitem[Sivasubramanian \& Fasel(2015)]{Sivasubramanian2015}
{\sc \au{Sivasubramanian, Jayahar} \& \au{Fasel, Hermann~F}} \yr{2015}
  \at{{Numerical Investigation of Shock-Induced Laminar Separation Bubble in a
  Mach 2 Boundary Layer}}.  \jt{45th AIAA Fluid Dynamics Conference}
  \bvol{2641},  \pg{1--36}.

\bibitem[Tobak \& Peake(1982)]{Tobak1982}
{\sc \au{Tobak, M} \& \au{Peake, D~J}} \yr{1982}  \at{{Topology of
  Three-Dimensional Separated Flows}}.  \jt{Annual Review of Fluid Mechanics}
  \bvol{14}~(1),  \pg{61--85}.

\bibitem[Wang {\em et~al.\/}(2015)Wang, Sandham, Hu \&
  Liu]{wang_sandham_hu_liu_2015}
{\sc \au{Wang, Bo}, \au{Sandham, Neil.~D.}, \au{Hu, Zhiwei} \& \au{Liu,
  Weidong}} \yr{2015}  \at{Numerical study of oblique shock-wave/boundary-layer
  interaction considering sidewall effects}.  \jt{Journal of Fluid Mechanics}
  \bvol{767},  \pg{526–561}.

\bibitem[White(2006)]{white2006viscous}
{\sc \au{White, Frank}} \yr{2006} {\em Viscous fluid flow\/}.  \publ{New York,
  NY: McGraw-Hill}.

\bibitem[Xiang \& Babinsky(2019)]{Xiang2019}
{\sc \au{Xiang, X} \& \au{Babinsky, H}} \yr{2019}  \at{{Corner effects for
  oblique shock wave/turbulent boundary layer interactions in
  rectangular channels}}.  \jt{Journal of Fluid Mechanics}  \bvol{862},
  \pg{1060--1083}.

\end{thebibliography}

\end{document}